# All Photonic Isolator using Atomically Thin (2D) Bismuth Telluride ($Bi_2Te_3$)

*Saswata Goswami, Bruno Ipaves, Juan Gomez Quispe, Caique Campos de Oliveira, Surbhi Slathia, Abhijith M.B, Varinder Pal, Christiano J.S. de Matos, Samit K. Ray, Douglas S. Galvao, Pedro A. S. Autreto* and Chandra Sekhar Tiwary**

S. Goswami, S. Slathia
School of Nano Science and Technology, Indian Institute of Technology, Kharagpur, West Bengal-721302, India

B. Ipaves, J.G. Quispe, C. C. Oliveira, P. A. S. Autreto
Center for Natural and Human Sciences (CCNH)
Federal University of ABC
Rua Santa Adélia 166, Santo André 09210-170, Brazil.
Email: pedro.autreto@uafbc.edu.br

Abhijit M.B
Materials Science Centre, Indian Institute of Technology, Kharagpur, West Bengal 721302, India

C. J. S. Matos
Mackenzie Presbyterian University
Rua da Consolação 896, São Paulo 01302-907, SP – Brasil

S. K. Ray
Department of Physics, Indian Institute of Technology Kharagpur, West Bengal 721302, India

D. S. Galvao
Department of Applied Physics and Center for Computational Engineering and Sciences, State University of Campinas, Campinas, 13083-859, SP, Brazil

V. Pal, C. S. Tiwary
Department of Metallurgical and Materials Engineering, Indian Institute of Technology Kharagpur, West Bengal 721302, India
E-mail: chandra.tiwary@metal.iitkgp.ac.in

**S.G and B.I contributed equally to this work**

## Abstract

This work has demonstrated that two-dimensional (2D) $Bi_2Te_3$ exhibits a robust light–matter interaction, enabling a broadband Kerr nonlinear optical response. This characteristic is advantageous for nonreciprocal light propagation in passive photonic isolators. The self-induced diffraction patterns generated at various wavelengths (650 nm, 532 nm, and 405 nm) in the far field are investigated to calculate the nonlinear refractive index ($n_2$) and third-order nonlinear susceptibility $\chi^{(3)}_{total}$ of the synthesized 2D $Bi_2Te_3$ using SSPM (Spatial Self-Phase Modulation) Spectroscopy method. 2D-$Bi_2Te_3$ exhibits a significant nonlinear refractive index on the order of $\approx 10^{-4}$ $cm^2$ $W^{-1}$, which is higher than that of graphene. The laser-induced hole coherence effect accounts for the significant magnitude of third-order nonlinear susceptibility $\chi^{(3)}_{monolayer}$ (in order of $\times 10^{-7}$ e.s.u.). The surface engineering method is applied to realize a fast-response photonic system. Ab initio simulations are also carried out to gain further insights into the observed experimental behaviour. Leveraging the robust Kerr nonlinearity of 2D $Bi_2Te_3$, a nonlinear photonic isolator that disrupts time-reversal symmetry has been successfully demonstrated, enabling unidirectional excitation. This demonstration of the photonic isolator shows $Bi_2Te_3$ as a novel 2D material, expanding its potential applications across multiple photonic devices, including detectors, modulators, and switches.

## 1. Introduction

The primary objective of designing photonic isolators is to provide nonreciprocal light transmission, which has applications in various domains, including optical telecommunications and integrated photonics. [1] The traditional methods for achieving optical nonreciprocity include acousto-optically induced interband photonic transitions[2], optomechanical mechanisms[3], customized waveguides[4], micro-resonator structures [1, 5], and others. The most traditional method to obtain optical isolation is using the Faraday effect, with an applied magnetic field along the direction of the optical axis. An optical isolator, also known as an optical diode, is a component that allows the unidirectional passage of light. The functionality of traditional optical isolators is dependent on the Faraday effect, which is generated by the magneto-optic effect used in the primary component, the Faraday rotator. However, it suffers from certain disadvantages like limited isolation ratio, temperature sensitivity, magnetic field dependence, size and weight,and wavelength dependence. Nonetheless, integrated isolators that do not depend on magnetism, and have less temperature sensitivity and wavelength dependence have also been developed in recent years.

Over the past decade, the application of photonic diodes has remained relatively underexplored. Wu et al. reported a two-dimensional (2D) Te nanosheets (Ns)-based air-stable nonlinear photonic diode, attributed to its strong light–matter interaction in the visible-to-infrared band.[6] Furthermore, the authors documented the distinct Kerr nonlinearity exhibited by 2D graphdiyne, which facilitated the development of a nonreciprocal light propagation device. Recently, there has been a significant increasing interest in the utilization of SSPM (Spatial Self Phase Modulation) Spectroscopy for probing non-reciprocal light propagation in 2D materials such as graphene, 2D Te, $NbSe_2$, $MoS_2$, $NiTe_2$, $MoTe_2$, $MoSe_2$, SnS, and $WSe_2$, Black and Violet Phosphorus, Violet Phosphorus QD, Silver modified Violet Phosphorus NS, MOP Microparticles, Hybrid Bismuth Halide, Bismuthene, $Ti_3C_2T_x$ MXene, and Antimonene.[6-7] SSPM spectroscopy can measure the nonlinear refractive index coefficients of 2D materials. Studies show that the value of nonlinear refractive index $n_2$ of Graphene is estimated in the order of $10^{-5}$ $cm^2W^{-1}$, calculated using SSPM.[7j] Zhang et al. measured the nonlinear refractive index of Black Phosphorus to be in the range of $10^{-5}$ $cm^2W^{-1}$, derived using the SSPM Spectroscopy method. [8] Among various 2D materials, $Bi_2Te_3$ is known as a topological insulator, characterized by its unique energy band structure. The novel optical properties of topological insulators make them highly desirable for applications in nonlinear optics, light modulation, fiber lasers, and other related photonic fields.[9] A topological insulator (TI) is a novel kind of quantum matter characterized by a distinct bulk gap and an odd quantity of relativistic Dirac fermions present on its surface. [10] Unlike graphene, TIs possess distinct photonic and opto-electronic characteristics that arise from the combined influence of robust spin-orbit interactions and surface states shielded by time-reversal symmetry.[11] In addition to their distinct quantum properties, TIs exhibit a wide range of nonlinear optical (NLO) responses [12] that span from visible to terahertz frequencies. This NLO response has been confirmed in numerous experimental setups, including diverse mode-locked or Q-switched laser operations conducted by various research groups. [13] Furthermore, the Z-scan approach has determined that the third-order nonlinear refractive index of the TI $Bi_2Te_3$ nano-platelets is around $10^{-8}$ $cm^2W^{-1}$.[12d]

This study introduces an innovative method for achieving nonreciprocal light propagation in a nonlinear photonic isolator by utilizing the robust nonlinear optical response of 2D $Bi_2Te_3$ and the reverse saturable absorption behavior of 2D hBN. 2D $Bi_2Te_3$, which are ultrathin TIs, have been successfully synthesized using a liquid phase exfoliation method. Our work indicates that 2D-$Bi_2Te_3$ displays a narrow electronic bandgap, enabling a broadband nonlinear optical response upon the passage of a laser beam. The nonlinear optical characteristics and their coefficients, specifically $n_2$ (nonlinear refractive index) and $\chi^{(3)}_{total}$ (third-order nonlinear susceptibility), were determined using the SSPM spectroscopic method. The findings demonstrate that 2D $Bi_2Te_3$ has a large nonlinear refractive index ($\approx 10^{-4}$ $cm^2$ $W^{-1}$), indicating a robust nonlinear optical response. This suggests that 2D $Bi_2Te_3$ may serve as an exceptional optical material for photonic isolators and all photonic devices.

The temporal evolution of the SSPM pattern was analyzed. The distortion in the SSPM pattern was also analyzed, and the variation in the refractive index was evaluated. Solvent exchange is utilized to identify the most efficient system capable of delivering rapid responses under laser beam exposure. The progression of the SSPM pattern under different wavelengths and solvents was determined. Ab initio computer simulations were also carried out to further investigate the interaction of 2D $Bi_2Te_3$ with solvents. A conclusion is drawn upon determining the solvent- 2D $Bi_2Te_3$ system, which yields a rapid response. 2D-hBN demonstrates reverse saturable absorption (RSA) with a significant electronic bandgap of 5.28 eV; however, exciting the diffraction rings presents challenges. The coupling of the two materials to form a hybrid 2D-$Bi_2Te_3$/2D-hBN structure results in the occurrence of propagational symmetry-breaking between the forward (2D-$Bi_2Te_3$/2D-hBN) and reverse (2D-hBN/2D-$Bi_2Te_3$) directions, facilitating the excitation of unidirectional diffraction rings. Results confirm that the suggested photonic isolator facilitates nonreciprocal light transmission in optical telecommunications or integrated photonics. In contrast to bulky traditional optical isolators, the proposed method utilizes the SSPM technique, leveraging 2D materials, specifically 2D-$Bi_2Te_3$ and 2D-hBN.

## 2. Results and Discussion

### 2.1 Synthesis and Characterization of 2D-$Bi_2Te_3$ nanostructure:

The synthesis and characterization are discussed in Supporting Information Section S1.

### 2.2 Experimental Section: SSPM Setup

**Figure 1a** depicts a method utilized to characterize the nonlinear optical response in a spatial self-phase modulation (SSPM) experiment. This experiment utilized three lasers with distinct continuous wavelengths of 650, 532, and 405 nm. The laser beams were directed through a convex lens with a focal length of 20 cm and then onto a cuvette, which included the sample. The SSPM effect is observable as the laser beam traverses the cuvette, producing a diffraction pattern on the far field screen, which is subsequently captured by a CCD camera. The underlying principle of the nonlinear optical characteristics of two-dimensional material based on SSPM can be explained by the Kerr nonlinearity.

### 2.3 Basic Understanding of SSPM Phenomenon and Calculation of the Nonlinear Optical Coefficients $n_2$ , $\chi^{(3)}_{total}$ and $\chi^{(3)}_{monolayer}$

This study explores the nonlinear Kerr effect, which is essential for understanding the nonlinear optical responses of the 2D $Bi_2Te_3$. The Kerr nonlinear effect is defined in Equation 1, which establishes the correlation between the intensity of the incident laser beam (I) and the refractive index (n).

$$n = n_0 + n_2 I \quad \text{................ (1)}$$

The terms "$n_0$" and "$n_2$" represent the linear and nonlinear refractive indices of the material, respectively. [14]

In the outgoing Gaussian light, there are at least two distinct locations, $r_1$ and $r_2$, where the slopes of the distribution curve, represented by $\left(\frac{d\Delta\psi}{dr}\right)_{r=r_1}$ and $\left(\frac{d\Delta\psi}{dr}\right)_{r=r_2}$, are equal and have the same phase. This is evident due to the fact that nonlinear phase shift has a Gaussian distribution, as seen in Figure 1b. Thus, the output light intensity profile exhibits a consistent phase difference while maintaining the same slope points. Clearly, these two points satisfy the requirement for interference. When $\Delta\psi_0 \geq 2\pi$, the diffraction ring appears. The self-induced diffraction pattern is observed as rings, which can be either bright or dark, in the far field.

The phase shift, which is crucial for the SSPM effect, generates a self-diffraction pattern in the far field, as seen in Figure 1(b).

The nonlinear refractive index $n_2$ is expressed as,

$$n_2 = \left(\frac{\lambda}{2n_0 L_{eff}}\right) \cdot \frac{dN}{dI} \quad \text{................ (2)}$$

"$n_0$" and "$n_2$" represent the linear and nonlinear refractive indices of the material, respectively. The $dN/dI$ is an important parameter to evaluate the nonlinear refractive index of the two-dimensional material, which is defined as a change in the number of rings, with change in intensity of the incoming laser beam. $L_{eff}$ is the effective transmission length of the laser beam inside the cuvette. The third-order nonlinear susceptibility $\chi_{total}^{(3)}$ is employed to characterize the nonlinear optical characteristics of materials. [7d, 15] It can be written as,

$$\chi_{total}^{(3)} = \frac{cn_0^2}{12\pi^2} 10^{-7} n_2 \ (e.s.u) \ \ldots\ldots\ldots\ldots\ldots\ldots\ldots \ (3)$$

Here, c represents the speed of light in vaccum, $n_0$ is the linear refractive index of the IPA solvent, and $n_2$ defines the effective length that the laser beam propagates through the cuvette. However, the effective number of 2D material available in the cuvette has a direct effect on the value of the $\chi_{total}^{(3)}$. Hence, it is necessary to determine the value of third order nonlinear susceptibility caused by a single layer of two-dimensional flakes $\chi_{monolayer}^{(3)}$. Detailed calculation of $n_2$ and $\chi_{total}^{(3)}$ is written in Supporting Information Section S2. The relationship between the overall electric field strength $E_{total}$ and the electric field strength $E_{monolayer}$ traveling through the single layer of $Bi_2Te_3$ can be mathematically represented as, [7j, 16]

$$E_{total} = \sum_{j=1}^{N_{eff}} E_j \cong N_{eff} \, E_{monolayer} \ \ldots\ldots\ldots\ldots\ldots \ (4)$$

In this context, $N_{eff}$ represents the number of 2D $Bi_2Te_3$ layers present in the solution, through which the beam passes through. The relationship between $\chi_{total}^{(3)}$ and $\chi_{monolayer}^{(3)}$ can be expressed as, [7d, 7g, 16]

$$\chi_{total}^{(3)} = N_{eff}^2 \chi_{monolayer}^{(3)} \ \ldots\ldots\ldots\ldots\ldots\ldots \ (5)$$

Wu et al. [17] described the SSPM phenomenon as originating due to nonlocal and intraband ac electron coherence. Authors also concluded that, as nearly all of the light is diffracted along the direction of the laser beam propagation, and Rayleigh scattering is found to be low, the SSPM procedure is inherently a third order-nonlinear optical process. In the Supporting Information Section S7, further experiments are provided to clarify that the SSPM phenomenon has an electronic origin instead of a thermal one. The concentration of 2D $Bi_2Te_3$ is varied to quantify the light-matter interaction taking place, although 0.25 mg $mL^{-1}$ was selected for further experimental purpose.

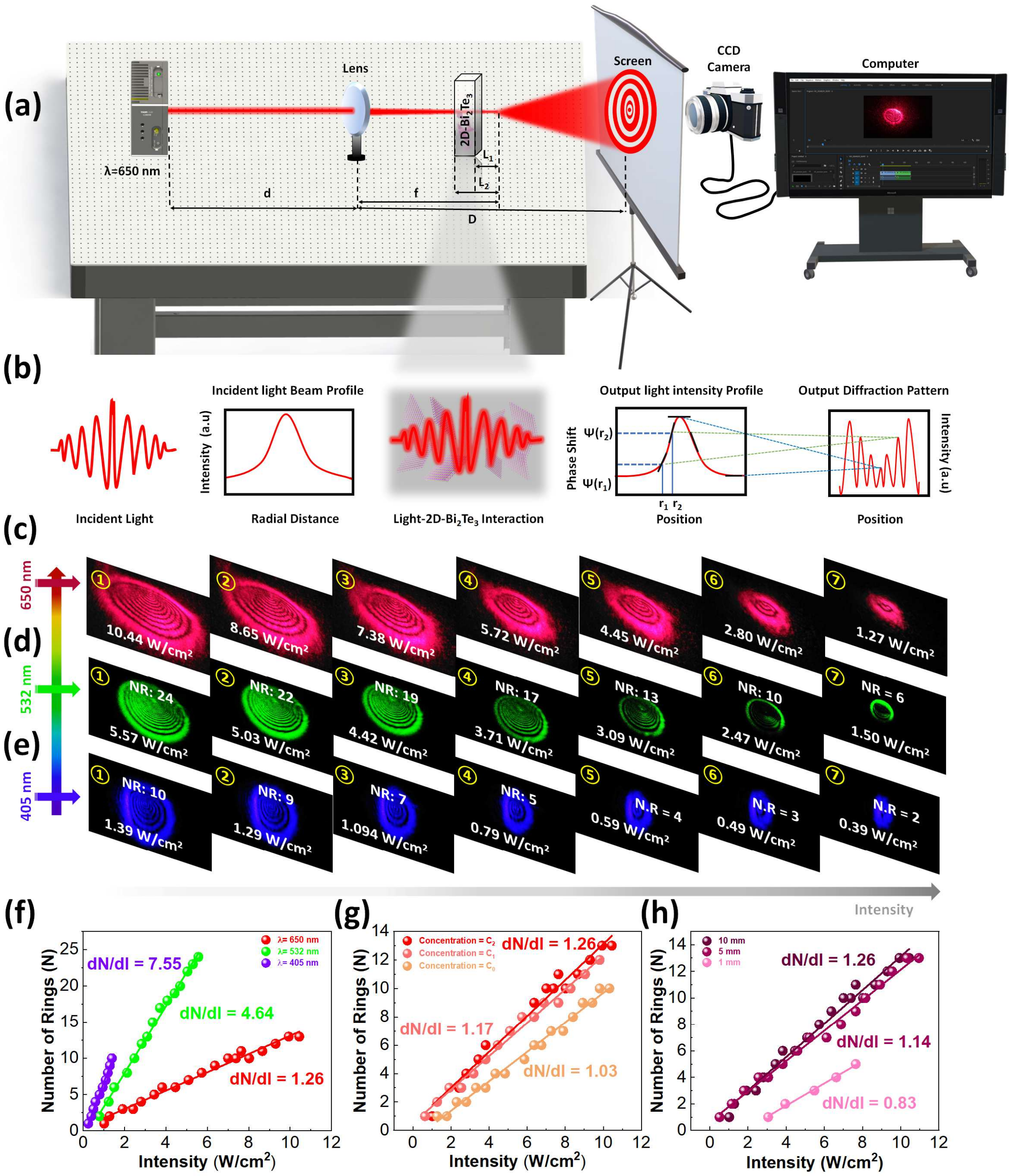

**Figure 1.** Experimental setup and SSPM spectroscopy-based observation in the far field. a) Diagram depicting the SSPM experimental setup. b) Schematic representation of the interaction between 2D $Bi_2Te_3$-light and the SSPM phenomenon by electronic coherence involvement with the laser beam. c-d-e) Progression of diffraction pattern on the far field with varying intensity at different wavelengths (λ = 650, 532, and 405 nm). f) The captured diffraction pattern on the far field for different wavelengths of 650, 532, and 405 nm at constant concentration $C_2$, and cuvette length of 10 mm. g) The variation in the number of diffraction rings in response to changes in laser intensity (λ= 650 nm) across different concentrations, while maintaining a constant cuvette length of 10 mm. h) Different cuvette lengths (L= 10 mm, 5 mm, and 1 mm) affect the number of diffraction rings with varying intensity at constant wavelength (λ= 650 nm) and concentration $C_2$.

The $N_{eff}$ calculation is presented in the Supporting Information Section S3. Figure 1c(①-⑦), Figure 1d(①-⑦), and Figure 1e(①-⑦) show the SSPM diffraction pattern recorded in the far screen for the wavelengths 650, 532, and 405 nm, which was recorded by a CCD camera. A linear increase in the number of rings is observed as the intensity is increased. The increment in intensity leads to a corresponding rise in both the horizontal and vertical diameter of the diffraction pattern. Figure 1f illustrates the relationship between intensity and the number of rings for laser beams with varying wavelengths (650, 532, and 405 nm). The $\frac{dN}{dI}$ calculated from the curve fitting is found to be 1.26 $cm^2W^{-1}$, 4.64 $cm^2W^{-1}$, and 7.55 $cm^2W^{-1}$ for $\lambda$= 650, 532, and 405 nm, respectively. A conclusion is made that when the wavelength reduces, the energy of photons rises, leading to an intensified SSPM effect. A control experiment was performed where the laser beam interacted with the solvent devoid of any 2D material, and no diffraction pattern was recorded at the far screen. The following experiment was explained in Supporting Information Section S9. Additional parameters that also contribute to the effective formation of the SSPM include the effective presence of 2D material in the solvent and the effective path length of the incoming laser beam within the cuvette. Figure 1g demonstrates the relationship between the intensity of laser light with a wavelength of 650 nm and the number of rings with varying concentrations of the 2D materials active in the solution. This correlation is seen while analyzing various concentrations of suspended two-dimensional material in the solution, as seen in Figure 1g. A linear relationship between the laser intensity and the number of diffraction rings was shown using curve fitting analysis. SSPM spectroscopy was conducted at specific concentrations of 0.0625 ($C_0$), 0.13 ($C_1$), and 0.25 ($C_2$) mg $mL^{-1}$, respectively. The corresponding slopes for $\frac{dN}{dI}$ were determined to be 1.03, 1.17, and 1.26 $cm^2W^{-1}$. As the effective number of 2D materials increases in the solution, light-matter interaction increases, leading to an increased number of rings in the diffraction pattern. Increasing the value of $\frac{dN}{dI}$ for the corresponding order of the concentration values shows enhanced light-matter interaction. The variation in the number of rings in relation to the intensity of the incoming laser beam for various cuvette lengths is seen in Figure 1h. It was observed that the changes in number of rings with respect to intensity ($\frac{dN}{dI}$) increase with increasing cuvette length. The slope $\frac{dN}{dI}$ is calculated to be 0.83, 1.14, and 1.26 $cm^2W^{-1}$ for cuvette thickness of 1 mm, 5 mm, and 10 mm at a constant concentration of 0.25 mg $mL^{-1}$ ($C_2$) and wavelength $\lambda$= 650 nm. With increasing the length of cuvette thickness, light-matter interaction increases, leading to the generation of more phase shifts and greater number of rings in the diffraction pattern. As the effective propagation length through the cuvette decreases, light-matter interaction decreases, resulting in lesser diffraction pattern for constant intensity at a fixed concentration. Experiment revealed that the value of $n_2$ is $2.7\times10^{-4}$, $8.14\times10^{-4}$, and $10.1\times10^{-4}$ cm$^2$W$^{-1}$ at the specified wavelengths of 650, 532, and 405 nm. The third order nonlinear susceptibility ($\chi^{(3)}_{total}$) is found to be 0.01569, 0.04741, and 0.05887 e.s.u., respectively, at the wavelengths of 650, 532, and 405 nm. An increase in both cuvette thickness and the concentration of suspended 2D $Bi_2Te_3$ within the cuvette leads to a corresponding increase in the number of rings, relative to intensity. The computed values of $n_2$ and $\chi^{(3)}_{total}$ obtained from the experiment described above are displayed in **Table 1**. The values of $n_2$ and $\chi^{(3)}_{total}$ increase significantly as the duration of travel inside the cuvette decreases and the quantity of active material in the cuvette increases. Measured values of $n_2$ and $\chi^{(3)}_{total}$ for 2D $Bi_2Te_3$ obtained through the SSPM approach are determined to be much higher than those of other transition metal dichalcogenides (TMDCs) (compared in between **Table 1** and Supporting Information **Table S1**). The electron experiences minimal or zero resistance while moving over the surface of the 2D $Bi_2Te_3$. This attribute allows for conduction even in the presence of many oxidized flaws on the surface of the TI material. The Fermi energy [18] of n-type $Bi_2Te_3$ higher than that of intrinsic $Bi_2Te_3$ and p-type $Bi_2Te_3$. Furthermore, the saturation density of photoexcited carriers in p-type $Bi_2Te_3$ ($n_p$) exceeds that in intrinsic $Bi_2Te_3$ ($n_i$) and n-type $Bi_2Te_3$.

**Table 1**. Experimental values of $\boldsymbol{n_2}$(nonlinear refractive index), $\boldsymbol{\chi^{(3)}_{total}}$ (third-order nonlinear susceptibiliy) and $\boldsymbol{\chi^{(3)}_{monolayer}}$

| Wavelength (nm) | Concentration (mg mL$^{-1}$) | L (mm) | Solvent | dN/dI (cm$^2$ W$^{-1}$) | $N_{eff}$ | $n_2$ (cm$^2$ W$^{-1}$) | $\chi^{(3)}_{total}$ (e.s.u) | $\chi^{(3)}_{monolayer}$ (e.s.u) |
|---|---|---|---|---|---|---|---|---|
| 650 | 0.25 | 10 | NMP | 1.26 | 361 | 2.7 × 10-4 | 0.01569 | 1.2 × 10$^{-7}$ |
| 650 | 0.25 | 10 | IPA | 1.29 | 361 | 3.038 × 10-4 | 0.01466 | 1.12 × 10$^{-7}$ |
| 532 | 0.25 | 10 | NMP | 4.64 | 361 | 8.14 × 10-4 | 0.04741 | 3.63 × 10$^{-7}$ |
| 405 | 0.25 | 10 | NMP | 7.55 | 361 | 10.1 × 10-4 | 0.05887 | 4.51 × 10$^{-7}$ |
| 650 | 0.13 | 10 | NMP | 1.17 | 188 | 2.51 × 10-4 | 0.01456 | 4.12 × 10$^{-7}$ |
| 650 | 0.0625 | 10 | NMP | 1.07 | 90 | 2.22 × 10-4 | 0.00129 | 1.59 × 10$^{-7}$ |
| 650 | 0.25 | 5 | NMP | 1.14 | 180 | 2.45 × 10-4 | 0.01419 | 8.76 × 10$^{-7}$ |
| 650 | 0.25 | 1 | NMP | 0.83 | 36 | 0.00178 | 0.10332 | 7.97 × 10$^{-7}$ |

$$n_{n,sat} = \int_{E_{Fn}}^{E_{photon}-E_{Fn}} D(E)f(E)dE < n_{i,sat} < \int_{E_{Fp}}^{E_{photon}-E_{Fp}} D(E)f(E)dE = n_{p,sat}$$

The symbol $D(E)$ represents the density of states (DOS) at the energy $E$. The function $f(E)$ represents the probability distribution of carriers. $E_{photon}$ is the energy of photon-excited electrons. $E_{Fn}$ and $E_{Fp}$ are the Fermi energies of the n-type and p-type $Bi_2Te_3$, respectively. The dopant type of transition metal (TI) changes the Fermi level without altering its energy bandgap. Thus, the photocarrier density and saturated photocarrier density vary between n-type and p-type materials. In this study, the 2D $Bi_2Te_3$ is found to be p-type in nature, which in turn may give rise to laser-induced hole-based ac coherence, as holes are more prominent electronic carriers compared to electrons in p-type materials.

The large third order nonlinear susceptibility coefficient $\chi^{(3)}_{monolayer}$ is observed at due to laser induced hole coherence effect.[19] The identification of high-lying bulk states positioned above the initial bulk state was determined to be the reason for this effect. However, here, the intensity value of the laser beam is maintained low. p-type-$Bi_2Te_3$ has a low-lying Fermi level compared to n-type-$Bi_2Te_3$, hence more capacity for excited carriers to be generated. These hole type carriers are involved in the laser-induced hole coherence that gives rise to high values of $n_2$, $\chi^{(3)}_{total}$, and $\chi^{(3)}_{monolayer}$ .[7k, 19]

### 2.4 Wind Chime Model: Diffraction Pattern Generation Under Different Viscosity Medium

#### 2.4.1 Time Evolution under Different Wavelength and Variable Intensity

In this analysis, we investigate the process of pattern formation and the fundamental mechanism of the SSPM. Wu et al. introduced a model to elucidate the mechanism of pattern formation in the SSPM process, encompassing the duration required for the pattern to emerge at a consistent level of intensity.[7d] This model suggests that at the initial interaction of the incoming laser beam with the 2D $Bi_2Te_3$, the angles between the 2D and the electric field of the laser beam are orientated in arbitrary directions. This incoming laser beam polarizes the suspended 2D $Bi_2Te_3$ in the solution. [16] Eventually, the polarized 2D $Bi_2Te_3$ aligns along with the external electric field and the polarization direction of the incident laser's electric field due to the energy relaxation process.[7d] As time progresses, the macroscopic angle between the 2D $Bi_2Te_3$ decreases, leading to a more precise alignment with the electric field, which results in a continuous increase in the number of diffraction rings. Once all 2D-$Bi_2Te_3$ that are in the path length of the incoming laser beam are all perfectly aligned, the number of rings reach its maximum value.

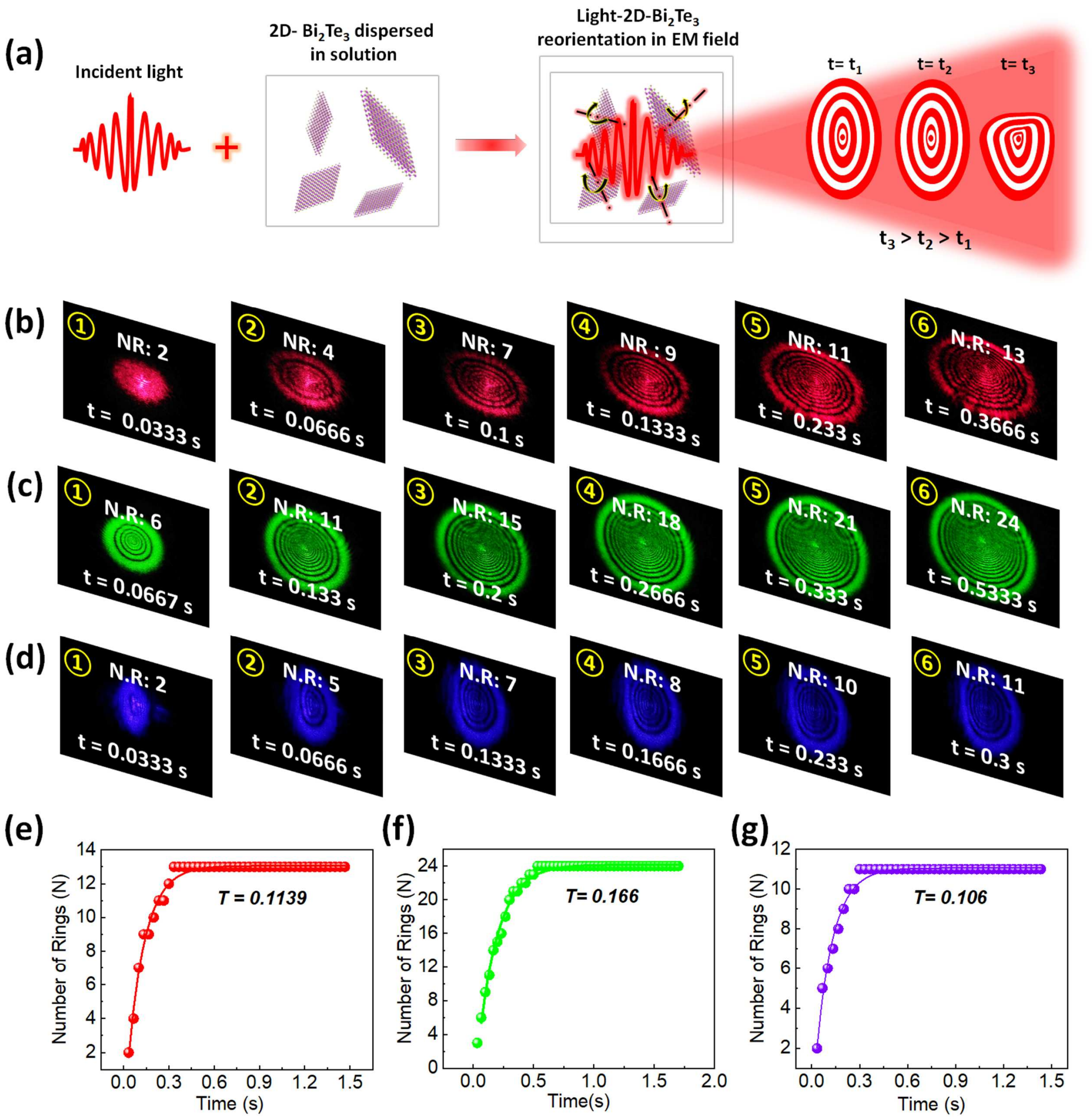

**Figure 2.** The Figure shows the far-field diffraction pattern progression over time. a) Depiction of wind chime model and progression of the SSPM pattern on the far-field. b-c-d) The far-field diffraction pattern over time at different wavelengths (λ= 650, 532, and 405 nm). e-f-g) The temporal variation in the number of diffraction rings for various wavelengths (λ= 650, 532, and 405 nm).

To validate this model and its dependence on the medium's viscosity, we performed an experiment to investigate the effect of the intense laser beam's polarisation on the suspended 2D $Bi_2Te_3$ in IPA and NMP. **Figure 2a** explains the windchime model for 2D $Bi_2Te_3$ in the NMP/IPA dispersion. At first, a point of outgoing Gaussian beam emerges in the far field, progressively transforming into a complete circular pattern of light diffraction when the suspended 2D $Bi_2Te_3$ align themselves with the electric field of the incoming laser beam. Figure 2b(①-⑥), 2c(①-⑥), and 2d(①-⑥) show the SSPM pattern's advancement under continuous- wave laser beams at λ = 650, 532, and 405 nm. The duration necessary for the 2D $Bi_2Te_3$ or domains to achieve full alignment with the electric field of the incident laser beam corresponds to the time required for the full formation of the diffraction ring pattern. An exponential model can explain the process of ring generation with time. [7f]

$$N = N_{max}\left(1 - e^{t/\tau_{rise}}\right) \text{ .......... (6)}$$

Here, $N$ is the quantity of rings observed in the diffracted pattern, $N_{max}$ represents the maximum number of rings that are created when the laser intensity remains constant, and the $\tau_{rise}$ is the duration required for pattern development.

The Wind Chime model properly represents the time ($\mathcal{T}$) required for the full-diameter diffraction pattern to appear. [7d, 7g]

$$\mathcal{T} = \frac{\epsilon_r \pi \eta \xi R_C}{1.72(\epsilon_r - 1) I h} \text{ .............. (7)}$$

here $\epsilon_r$ represents the relative dielectric constant of 2D $Bi_2Te_3$ understood to be 5[20], $\eta$ represents the viscosity coefficient of the solvent, and the values are $2.4 \times 10^{-3}$ Pa.s at 20°C and $1.65 \times 10^{-3}$ Pa.s at 25°C for IPA and NMP, respectively, and $\xi$ denotes the segment of the fluid sphere that is proximal to the 2D nanostructure. The variable $R_C$ represents the radius of the 2D $Bi_2Te_3$, specifically referring to the radius of its exposed surface that is in contact with the solvent, h represents the vertical height of the suspended 2D nanostructure, whereas $I$ denotes the intensity of the incident laser beam. The value of the domain radius ($R_C$) was obtained through atomic force microscopy is 50 nm, and the flake thickness (h) is 3.5 nm. Furthermore, the theoretically predicted values of $\mathcal{T}$ are determined to be 0.32 s, 0.478 s, and 0.263 s for the wavelengths of 650, 532, and 405 nm for NMP solvent. These values, obtained by curve fitting, are compatible with experimentally derived values of 0.333 s, 0.5333 s, and 0.3 s for wavelengths of 650, 532, and 405 nm, respectively, as shown in Figure 2e-g. The number of the rings is greater in the case of 532 nm compared to other wavelengths, indicating enhanced light-matter interaction; this causes prolonged time for diffraction pattern formation compared to others. The time taken for the diffraction pattern to reach maximum number of rings is shown in **Table 2** for different wavelengths and viscous medium.

**Table 2.** Experimental and theoretical value of time for the diffraction pattern to reach maximum number of rings.

| Sample | Solvent | Wavelength (nm) | C [mg/mL] | L [mm] | $\eta$ [mPa.s] | $\tau_c$ [s] | Experimental $\mathcal{T}$ [s] | $\mathcal{T}$ from Windchime model |
|---|---|---|---|---|---|---|---|---|
| 2D-$Bi_2Te_3$ | NMP | 650 | 0.25 | 10 | 1.65 | 0.1139 s | 0.333 s | 0.328 s |
| 2D-$Bi_2Te_3$ | IPA | 650 | 0.25 | 10 | 2.4 | 0.213 s | 0.5 s | 0.4 s |
| 2D-$Bi_2Te_3$ | NMP | 532 | 0.25 | 10 | 1.65 | 0.166 s | 0.5333 s | 0.6179 s |
| 2D-$Bi_2Te_3$ | NMP | 405 | 0.25 | 10 | 1.65 | 0.106 s | 0.3 s | 0.263 s |

### 2.4.2 Time Evolution in Different Viscous Medium

As discussed previously, here we examine the phenomenon of nonlocal electron coherence in a detailed manner. Within our sample, every individual flake-like nanostructure represents a domain. Initially, charge carriers, including photocarriers, excitons, free electron-hole pairs, and intrinsic electrons and holes, are distinctly out of phase, regardless of whether they exist in separate or in the same domains. Moreover, every domain possesses an arbitrary orientation.

Wu et al.[7d] employed a wind-chime model to describe the evolution of electron coherence in the nonlocal regions induced by SSPM. According to the authors, initially, there is no fixed orientation angle between a polarized 2D $Bi_2Te_3$ nanostructure and the electric field. The electric field induces an energy relaxation process, prompting the flakes or nanostructures to realign in such a manner that each domain takes an orientation parallel to the polarization of the external electric field. This configuration makes each domain appear suspended by a vertical filament. This above-mentioned domain consists of the nanostructure and also the fluid surrounding the nanostructure. A depiction resembling a wind chime model is presented in Figure 2a. Now this laser beam induces a torque on the flake/ nanostructure, although the viscous force of the fluid counters this torque.Viscous force experienced by the nanostructure through the boundary is found to be [7d]:

$$V = \eta \int_0^{\pi} \frac{dv}{dr}(R\ sin\varphi)(\xi.\, 2\pi R\ sin\varphi)(Rd\varphi) = \pi\eta\Omega\xi R^3 \quad \ldots\ldots\ldots\ldots\ldots (8)$$

In this context $\Omega$ represents the rotational velocity and $\xi$ denotes the segment of the fluid sphere that is close to the disc. The viscous force is directly dependent upon the value of $\eta$ (viscosity coefficient).

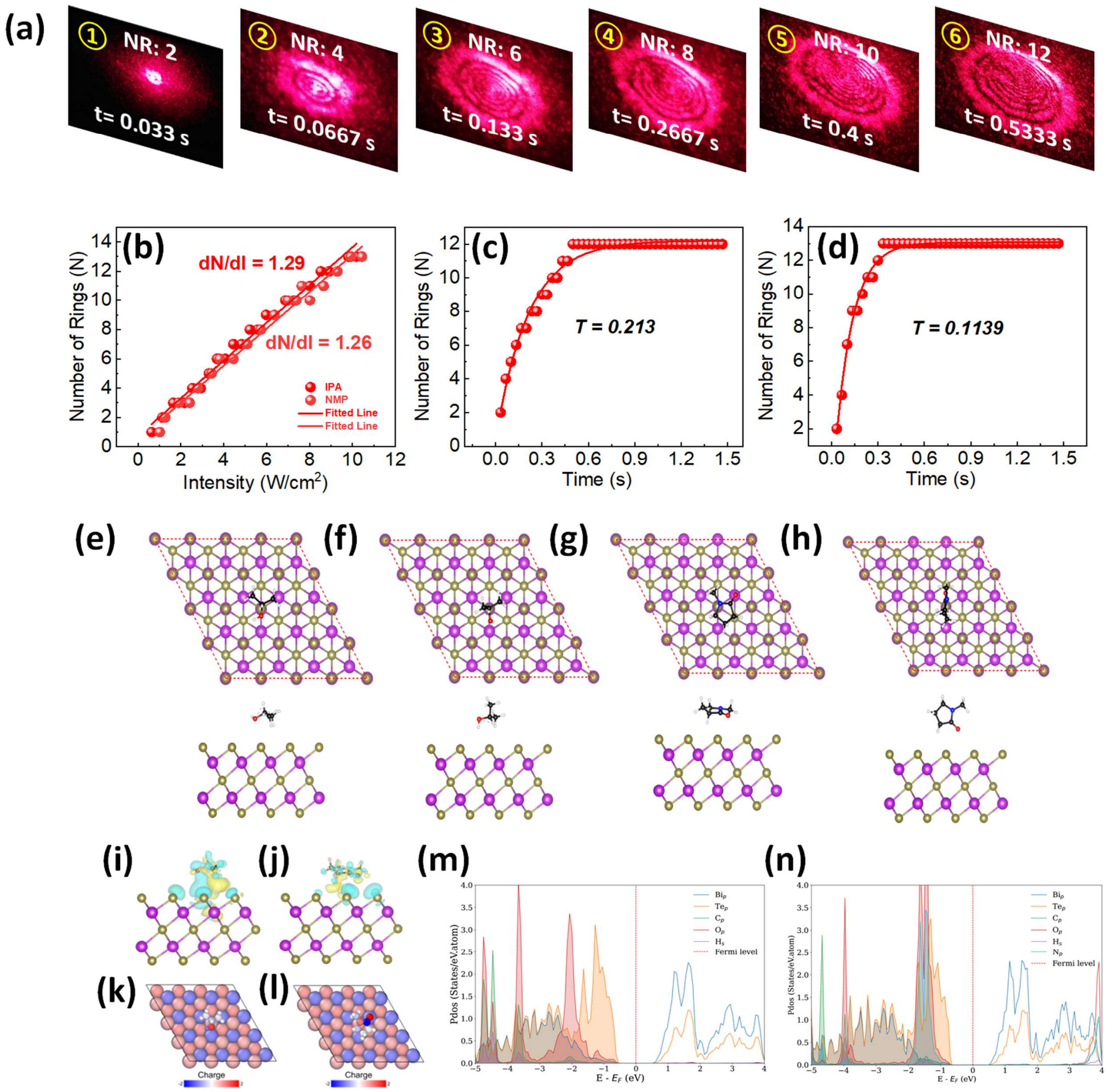

**Figure 3.** Comparative study of time evolution in IPA and NMP sample, a first principle angle of observation: a) Time evolution of diffraction pattern for IPA at wavelength 650 nm and concentration $C_2$. b) $\frac{dN}{dI}$ for IPA and NMP at wavelength 650 nm and concentration $C_2$. c-d) Rise time of the diffraction pattern for the IPA-2D $Bi_2Te_3$ and NMP- 2D $Bi_2Te_3$ solution. Top and side views of the molecules on top of the 2D-$Bi_2Te_3$ surface. (e) C1-IPA, (f) C2-IPA, (g) C1-NMP and (h) C2-NMP. The pink, yellow, black, blue, red, and white spheres represent the Bi, Te, C, N, O, and H atoms, respectively. The red dashed lines denote the simulation unit cell. Side views of the differential charge density plots for i) for C1-IPA (isovalue $1.3 \times 10^{-4}$ e $Bohr^{-3}$) and j) for C1-NMP (isovalue $2.5 \times 10^{-4}$ e $Bohr^{-3}$). Regions of charge accumulation and depletion are denoted by yellow and cyan colors, respectively. Top views of Bader charge analysis k) for C1-IPA and l) for C1-NMP. Projected density of states per atom for the valence states of m) for C1-IPA and n) for C1-NMP.

Although these nanostructures have a quasi-two-dimensional structure, having exposed surface area, this creates non-passivated bonds on the structure's surface.Thus, the probability of the interaction with a solvent molecule is high. It was found that in our experiment, IPA possesses a higher viscosity coefficient compared to NMP. Hence, the viscous force is considered to be low in case of NMP-2D $Bi_2Te_3$ compared to IPA-2D $Bi_2Te_3$. The time taken by the diffraction pattern to fully create IPA-2D $Bi_2Te_3$, is longer compared to NMP-2D $Bi_2Te_3$. However, employing the Wind Chime model yields a different conclusion when considering an ab initio calculation to investigate further interactions between NMP-2D $Bi_2Te_3$ and IPA-2D $Bi_2Te_3$ separately.

### 2.4.3 Ab Initio Study on 2D Material-Solvent Interaction and Effect on Time Evolution

**Figure 3a** shows the progression of the number of the rings with respect to time for the IPA-2D $Bi_2Te_3$ system. In Figure 3b, the calculated value of $\frac{dN}{dI}$ for both NMP- 2D $Bi_2Te_3$ and IPA-2D $Bi_2Te_3$ systems and the values were found to be very close, as the nonlinear coefficient value depends on the material itself. In Figure 3c and Figure 3d, the rise time of the IPA-2D $Bi_2Te_3$ and NMP-2D $Bi_2Te_3$ were calculated to be 0.213 s and 0.1139 s. Previous investigations show lower values of rise time for NMP- 2D $Bi_2Te_3$ compared to the IPA- 2D $Bi_2Te_3$ system. In Supporting Information **Table S2**, the SSPM formation times of the diffraction pattern are reported from recent literature works. Yet the time taken for the diffraction pattern with the maximum number of rings takes longer to appear for the case of NMP-2D $Bi_2Te_3$ when compared to the IPA-2D $Bi_2Te_3$.

We have carried out a series of computer simulations to understand the interactions between 2D-$Bi_2Te_3$ and the molecules (IPA or NMP). Computational details are provided in the Supporting Information Section S4. We have considered four configurations: two for IPA (C1-IPA and C2-IPA) and two for NMP (C1-NMP and C2-NMP), as illustrated in Figure 3e-f-g-h. All optimized configurations exhibited negative binding energies, indicating that both IPA and NMP can potentially bind to the 2D-$Bi_2Te_3$ surface. The calculated binding energies were computed using Eq. 12 (Supporting Information), and the obtained values were -0.30 eV, -0.29 eV, -0.56 eV, and -0.26 eV for C1-IPA, C2-IPA, C1-NMP, and C2-NMP, respectively. Since all binding energies are negative, all four configurations are viable. However, a more negative binding energy indicates a stronger interaction. Therefore, C1-NMP was identified as the most favorable configuration.

The vertical distance between the molecules and the surface ranged from 1.91 Å to 2.80 Å, with C1-NMP exhibiting the shortest distance, which is suggestive of a stronger interaction and is consistent with its most negative binding energy. **Table 3** summarizes binding energies and shortest vertical distances between the molecules and the 2D-$Bi_2Te_3$ surface.

**Table 3.** Binding energy values and shortest vertical distances of the optimized configurations.

| Configuration | Eb (eV) | Distance (Å) |
|---|---|---|
| C1-IPA | -0.30 | 2.50 |
| C2-IPA | -0.29 | 2.52 |
| C1-NMP | -0.56 | 1.91 |
| C2-NMP | -0.26 | 2.80 |

We have then analyzed the interactions of C1-IPA and C1-NMP using differential charge density analysis (Eq. 13 from Supporting Information) and Bader charge analysis.[21] As depicted in Figure 3i for C1-IPA and Figure 3j for C1-NMP, there is charge accumulation (highlighted in yellow) between Te and the molecules, suggesting a significant interaction between them. Moreover, as illustrated in Figure 3k for C1-IPA and Figure 3l for C1-NMP, Bader charge analyses indicate that there are no significant changes on the surface, i.e., there is only a slight charge transfer between the molecules and 2D $Bi_2Te_3$, as we have mainly van-der Waals (vdW) interactions.

Furthermore, in Figures 3m-n, we have examined the Projected Density of States (pDOS) of the C1-IPA and C1-NMP systems to obtain deeper insights. Figures 3m and 3n display the pDOS projected onto the s states of the H atoms and the p states of the Bi, Te, C, O, and N atoms for the C1-IPA and C1-NMP systems, respectively. Both systems exhibit an electronic band gap value of approximately 1.30 eV. Additionally, the pDOS indicates a stronger interaction for C1-NMP compared to C1-IPA, as evidenced by the observed hybridization in the C1-NMP case between the more prominent peak of the Te atom's p orbital and the p orbitals of the O and N atoms near the valence band maximum (VBM).

These results align with the experimental findings (see **Table 2**). The rise time for NMP-2D $Bi_2Te_3$ is shorter than that of IPA-2D $Bi_2Te_3$ due to the higher polarization observed in NMP-2D $Bi_2Te_3$. This increased polarization enhances the likelihood of electron-hole pair generation under the same laser illumination intensity. Consequently, the torque generated under an identical electric field is more pronounced for NMP-2D $Bi_2Te_3$ compared to IPA-2D $Bi_2Te_3$. The time duration needed to attain the maximum number of rings is longer for NMP-2D $Bi_2Te_3$ than for IPA-2D $Bi_2Te_3$. The wind chime model explained this behavior, where the 2D $Bi_2Te_3$ and its solvent interface form a rotating domain. For NMP-2D $Bi_2Te_3$, this rotation takes more time due to stronger molecular interactions with the solvent, which increases the effective viscosity of the system. In contrast, the IPA-2D $Bi_2Te_3$ system exhibits weaker molecular interactions, thus allowing faster rotation in a longer period of time. The conclusion drawn can be stated as follows: before the laser beam hits the solvent-2D $Bi_2Te_3$ solution and induces polarization, this polarization is higher in the case of NMP-2D $Bi_2Te_3$, as the number of generated hole pairs is also higher. The initial torque experienced by the NMP-2D $Bi_2Te_3$ is higher compared to IPA-2D $Bi_2Te_3$, thus the shorter rise time observed in the case of NMP-2D $Bi_2Te_3$. As time progresses on the polarized NMP-2D $Bi_2Te_3$ structure, it also polarizes the surrounding NMP molecules. Consequently, it starts to experience a stronger dragging force compared to IPA-2D $Bi_2Te_3$. Hence, NMP-2D $Bi_2Te_3$ in the cuvette solution takes a longer time to align itself with the electric field and to exhibit a complete AC electric coherence.

### 2.5 Intensity Dependent Dynamic Collapse of Diffraction Pattern and Variation in Nonlinear Refractive Index with Different Solvent

The previous section discussed the time evolution of diffraction patterns at different laser beam wavelengths and solvents.. Supporting Information Section S5 shows the intensity-dependent dynamic collapse of the diffraction pattern and change in the nonlinear refractive index for different wavelengths ($\lambda$ = 650, 532, and 405 nm). Here, the dynamics of the collapse phenomenon at different solvents are discussed. **Figure 4a**(①-②) shows the pre and post-collapse images of the diffraction pattern for the 2D $Bi_2Te_3$-IPA system at wavelength 650 nm, concentration $C_2$, and L= 10 mm. Figure 4b shows the progression of the vertical and horizontal diameters with time at a wavelength of $\lambda$= 650 nm, in IPA solvent. The Figures 4c and 4d show $\frac{\Delta n_2}{n_2}$% distortion for the IPA and NMP solvents, respectively.

The absorbed laser energy leads to a gradual increase in temperature within the solvent, attributed to thermal effects. The laser heating induces a convection process, where the upward convection current noted in the central area of the beam assists in flattening out the temperature gradient profile in the upper section of the heated region.[22] This results in a reduction of 2D nanostructure in the upper section of the cuvette volume, which subsequently decreases the light-matter interactions in the upper area relative to the lower one, acccroding to the laser beam path length. The upper portion of the diffraction pattern is distorted compared to the lower region. Consequently, thermal action induces variations in density that depend on the temperature within the liquid sample. This density gradient generates the spatial distribution of the buoyant force inside the solvent. The movement of the liquid layers leads to the transfer of heat from the region of the concentrated light beam to the surrounding area.

While some authors have previously proposed the concept, there is currently no extant experimental evidence supporting such an experiment.[23] Previously, it was concluded that the IPA-2D $Bi_2Te_3$ interface experiences less frictional force compared to the NMP-2D $Bi_2Te_3$ one. This less frictional force should influence the convection current to be faster in the case of IPA than in NMP. In contrast, the whole collapse process for IPA takes 0.9666 s, as described in Figure 4b, while the time taken by the collapse process for NMP is only 0.6666 s. Thus, it is observed that fluid convection is more significant in the case of NMP-2D $Bi_2Te_3$ compared to IPA-2D $Bi_2Te_3$. As the amount of volume of 2D $Bi_2Te_3$ is smaller than the volume of the solvent, the characteristics of the solvent become dominant in the convection process. The temperature gradient was monitored using an Optris PI 640i infrared camera in the optical setup shown in Figure 4e. The thermal convection flow within the cuvette is anticipated to follow the behavior depicted in Figure 4f. The analysis indicates that the heat-releasing process is more significant for the NMP method, resulting in a more uniform temperature gradient at the upper section of the solvent. Figure 4g depicts a shift in the vertical temperature difference before and after laser illumination in the IPA-2D $Bi_2Te_3$ solution. Figure 4h(①-②) shows the temperature map before and after laser illumination at 73 mW, at the wavelength (λ) of 650 nm. Figure 4i illustrates the vertical temperature profile change before and after laser illumination in NMP-2D $Bi_2Te_3$ solution. Figure 4j(①-②), taken at 71 mW, shows the temperature profile of the surface for the NMP-2D $Bi_2Te_3$ solution. A higher temperature gradient is observed compared to the IPA-2D $Bi_2Te_3$ solution, as shown in Figure 4j(①-②). Figures 4h(①-②) and 4j(①-②) capture the snapshot of the temperature profile of the surface, although where the beam is projected in the solvent, the temperature profile is different. Figure 4j(①-②) shows a higher temperature gradient than Figures 4h(①-②). The same is seen in Figures 4i-g. IPA[24] [161.2 (J $mol^{-1}K^{-1}$)] has a lower specific heat compared to NMP[25] [412.4 (J $mol^{-1}K^{-1}$)], hence the temperature at the beam focal point is higher for IPA-2D $Bi_2Te_3$ solvent than NMP-2D $Bi_2Te_3$. Figure 4j(②) shows a different temperature profile map compared to Figure 4h(②). In contrast, the temperature where the beam is focused is higher for the IPA-2D $Bi_2Te_3$ system compared to the NMP-2D $Bi_2Te_3$ one. From these observations, we can conclude that the heat exchange capacity of the NMP-2D $Bi_2Te_3$ is more significant than IPA-2D $Bi_2Te_3$, as a smaller temperature difference between the focal point and the top surface of the liquid is observed for the latter case.

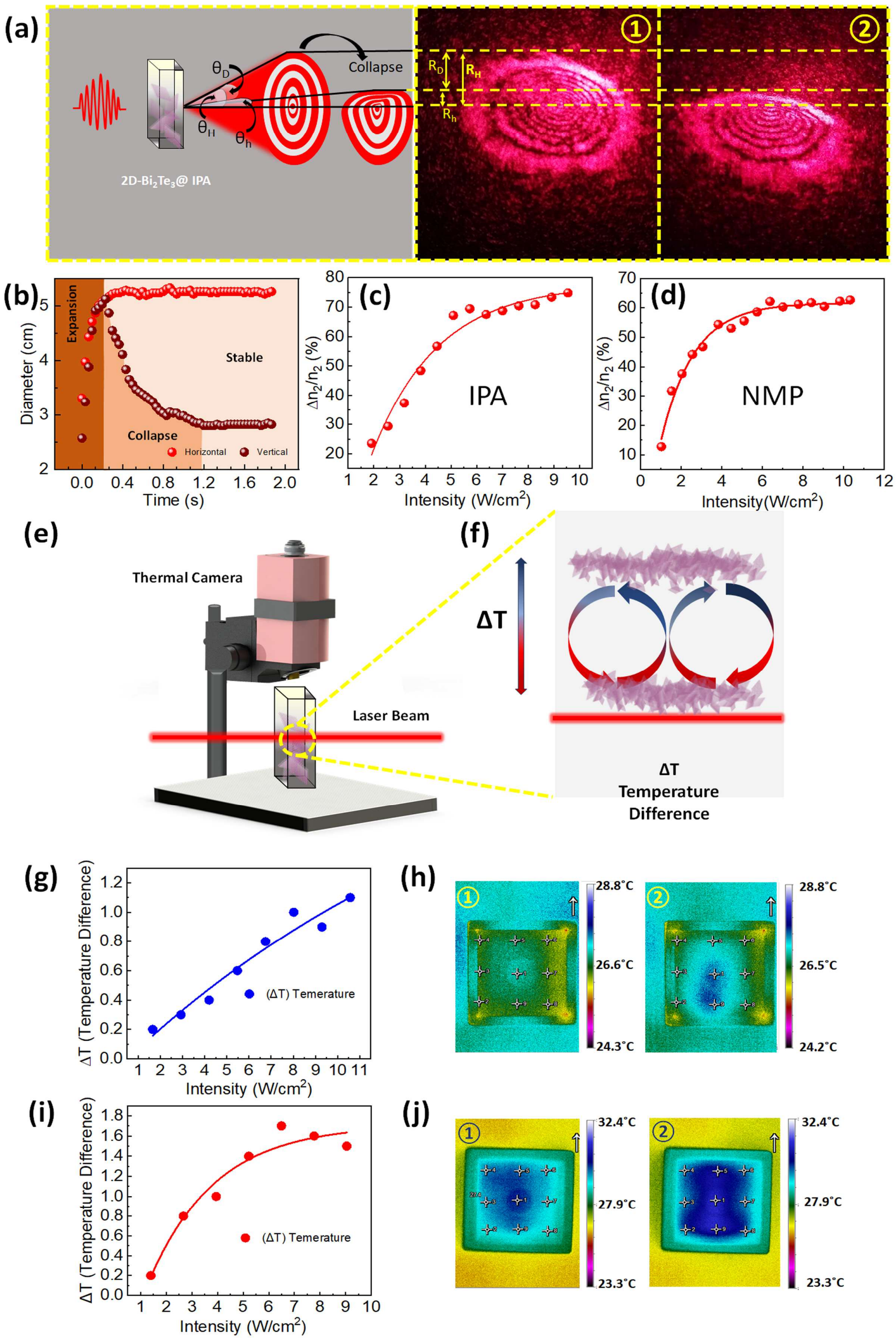

**Figure 4.** a) Illustration of the collapse phenomenon of the SSPM diffraction pattern featuring a semi-cone and distortion angle for 2D $Bi_2Te_3$ in IPA solvent. b) The progression of the vertical diameter and horizontal diameter with time at a wavelength ($\lambda$) of 650 nm) in IPA solvent. c-d) The distortion of the refractive index $(\Delta n_2/n_2)\%$ as recorded in the far field for 650 nm in IPA and NMP. e) Thermal camera setup to visualize the temperature profile of the top surface. f) Thermal convection flow created due to laser heating of the solvent. g) Temperature difference before and after laser impact of the laser beam with respect to intensity for IPA solvent. h) ①-② Temperature profile of the top surface of solvent in the cuvette before and after of laser beam impact for IPA solvent at intensity of 73 mW. i) Temperature difference before and after the impact of the laser beam with respect to intensity for NMP solvent. j) ①-② Temperature profile of the top surface of solvent in the cuvette before and after of the laser beam impact for NMP solvent at intensity of 71 mW.

The convection velocity is linearly dependent on ΔT, as the fluid has a tendency to flow from the high-density/hot region to the cold region/low-density region, which gives rise to convection velocity. NMP has lower viscosity compared to the IPA solvent, which aids the NMP system in this convection process (here, the 2D $Bi_2Te_3$ -NMP system is not mentioned; only solvent characteristics are highlighted). This convection velocity causes the distortion, as seen in Figure 4a(①-②). From the experiment, it was verified that $\frac{\Delta n_2}{n_2}$ % is more pronounced in the case of IPA-2D $Bi_2Te_3$ solution, compared to NMP-2D $Bi_2Te_3$. Figure 4c shows the variation of $\frac{\Delta n_2}{n_2}$ % with varying intensity for NMP-2D $Bi_2Te_3$ solution, we can see a saturating behavior after a threshold intensity (6.36 $Wcm^{-2}$). The distortion can also be compared with Figures 4b and S5b, where for the same laser beam intensity IPA- 2D $Bi_2Te_3$ system shows more collapse in the vertical diameter. The same conclusion can be drawn by observing the top surface temperature profile in Figure 4i. This leads to the conclusion that the convection velocity becomes fixed after a certain threshold. However, IPA-2D $Bi_2Te_3$ does not show any saturating behavior, as observed in Figure 4g. High viscosity leads to more distortion in diffraction patterns and nonsaturating behavior, as seen in Figure 4g. The thermal conductivity of NMP is higher than that of IPA, allowing the fluid to achieve a more uniform and stable condition faster.

The total pattern formation time for IPA is 1.2 s, and for NMP is 0.9333 s. Also, the calculated time for the maximum number of rings to appear is documented in **Table 4**. Two different solvents are used to realize a fast response all-photonic isolator, although it is seen in the end that viscosity plays a more important role in the pattern formation.

Total Pattern Formation Time = Time taken for Full Diameter Rings to appear + Time required for the diffraction rings to reach stable condition

Or, Total Pattern Formation Time ≈ Time needed for the Maximum Number of Rings to appear + Time required for the diffraction rings to reach stable condition

As the volume of the 2D $Bi_2Te_3$ is small compared to the volume of the solvent, by comparing under the same wavelength (650 nm) of the laser beam at a particular intensity and constant concentration of 2D $Bi_2Te_3$-NMP system, it takes the shortest time for the stable diffraction pattern formation. Thus, NMP was selected as the solvent for the realization of the all-photonic isolator under three different wavelengths. The 2D structure of the nanoparticles contributes in improved Rise time compared to other microparticles like TaAs, MoP.[19, 26]

**Table 4**. Rise Time, time needed for the maximum number of rings to appear, Time taken for Full Vertical Diameter Rings to appear, Collapse Time for Full Vertical Distortion observed in the Diffraction Pattern, and Time required for the diffraction rings to reach stable condition based on Solvent type, Wavelength, and Intensity of the laser beam.

| Sample | Wavelength (λ nm) | Intensity ($W/cm^2$) | Rise Time Fitted ( $\tau_c$ [s]) | Time needed for Maximum number of Rings to appear (s) | Time taken for Full Vertical Diameter Rings to appear (s) | Collapse Time for full vertical distortion (s) | Time required for the diffraction rings to reach stable condition (s) |
|---|---|---|---|---|---|---|---|
| NMP-2D-Bi2Te3 | 650 nm | 10.44 | 0.1139 s | 0.3333 s | 0.3666 s | 0.5666 s | 0.93333 s |
| IPA-2D-Bi2Te3 | 650 nm | 10.44 | 0.213 s | 0.5 s | 0.5 s | 0.7 s | 1.2 s |
| NMP-2D-Bi2Te3 | 532 nm | 5.3 | 0.1666 s | 0.5333 s | 0.5333 s | 1.2 s | 1.7333 s |
| NMP-2D-Bi2Te3 | 405 nm | 1.29 | 0.106 s | 0.3 s | 0.3333 s | 1.3 s | 1.6333 s |

### 2.6 2D $Bi_2Te_3$/2D hBN Based Nonlinear All-Optical Isolator

To further investigate the photonic isolators exploiting the nonlinear optical (NLO phenomena, a unique photonic nonlinear isolator has been created by utilizing a hybrid structure of 2D $Bi_2Te_3$ and 2D hBN. The 2D-hBN material was synthesized using the LPE process, and it exhibits a larger optical bandgap of 5.38 eV compared to 2D-$Bi_2Te_3$. The approach used to estimate the 2D-hBN optical band gap is described in the Supporting Information Section S8. Consequently, this combination of 2D $Bi_2Te_3$ and 2D hBN could be used in the field of all-photonic isolator applications.[15a, 16] The application of the all-photonic isolator has been realized through the use of laser beams at wavelengths of 650, 532, and 405 nm, employing SSPM Spectroscopy. When the forward-biased configuration 2D $Bi_2Te_3$/2D hBN is realized, as shown in **Figure 5a**, the diffraction pattern is generated. This pattern can be observed in Figures 5c (①-⑧), 5e (①-⑧), and 5g (①-⑦) for the wavelengths 650, 532, and 405 nm, respectively.

In the second configuration illustrated in Figure 5b, where the 2D-hBN solution is added prior to the 2D $Bi_2Te_3$ one, the beam profile consistently displays a Gaussian shape. This is due to the reverse saturation property of 2D-hBN, which causes a decrease in the strength of the laser beam that is transmitted through the cuvette. This laser beam of low intensity does not produce the diffraction pattern as it does not exceed the threshold level. The 2D-hBN material exhibits reverse saturable absorption, which is utilized to create an all-optical isolator configuration.[27] The Gaussian beam profile may be seen in Figure 5d (①-⑧), 5f (①-⑧), and 5h (①-⑤) for wavelengths of 650, 532, and 405 nm, respectively. The calculated values of $\frac{dN}{dI}$ for the different wavelengths (650, 532, and 405 nm) are found to be 1.14, 4.34, and 7.46 $cm^2W^{-1}$, respectively, for forward bias condition. The observed results are similar to those obtained from a single cuvette containing a 2D $Bi_2Te_3$-NMP solution. Figures 5i-k displays the linear regression graphs for the forward and reverse bias conditions. Asymmetric light propagation is achieved by employing three distinct laser types, each with different wavelengths of λ= 650, 532, and 405 nm. Each of these wavelengths possesses photon energy that surpasses the bandgap of the 2D $Bi_2Te_3$, which is measured at 0.9 eV. Thus, the laser photons have the ability to induce the band-to-band transitions. In the forward configuration illustrated in Figure 5l, the photons from the incoming laser beam excite the valence band electrons, prompting their transition to the conduction band. Subsequently, the electrons that have gained energy decay to their original state, emitting a photon in the process. The photon, possessing a distinct phase, interacts with the laser beam, resulting in the emergence of a diffraction pattern.[7g, 15a] The electrons in the conduction band will undergo oscillations in a direction opposite to the electric field of the laser beam, resulting in the generation of charges with opposite polarities in the suspended material.[7d] Upon interaction with the incoming laser beam, the polarized flakes will align themselves with the electric field axis of the laser beam in order to minimize their interaction energy. The nonlinear optical response of 2D $Bi_2Te_3$ is enhanced, leading to the observation of the optical Kerr effect.[7g, 14] The band-to-band transition in the 2D hexagonal boron nitride (2D-hBN) cannot be induced by laser beams with wavelengths of λ= 650, 532, and 405 nm. The electrons undergo energy dissipation during an intraband transition. The reverse saturable absorption feature of 2D-hBN causes a decrease in the incoming beam intensity below a certain threshold. This prevents the generation of a diffraction pattern from the 2D $Bi_2Te_3$-NMP solution, as seen in Figure 5m. The claimed all-optical isolator exhibits operation under different applicable wavelengths. This all-optical diode allegedly has an extensive operational wavelength range. A similar technique may be used to determine the value of $n_2$ for the 2D $Bi_2Te_3$-based photonic diode, as explained in Supporting Information Section S10. In the Supporting Information, Section S6 the relationship between $\chi^{(3)}_{monolayer}$ on $m^*$ and $\chi^{(3)}_{monolayer}$ on $\mu$ is obtained, and the identification between N-type and P-type $Bi_2Te_3$ is explored.

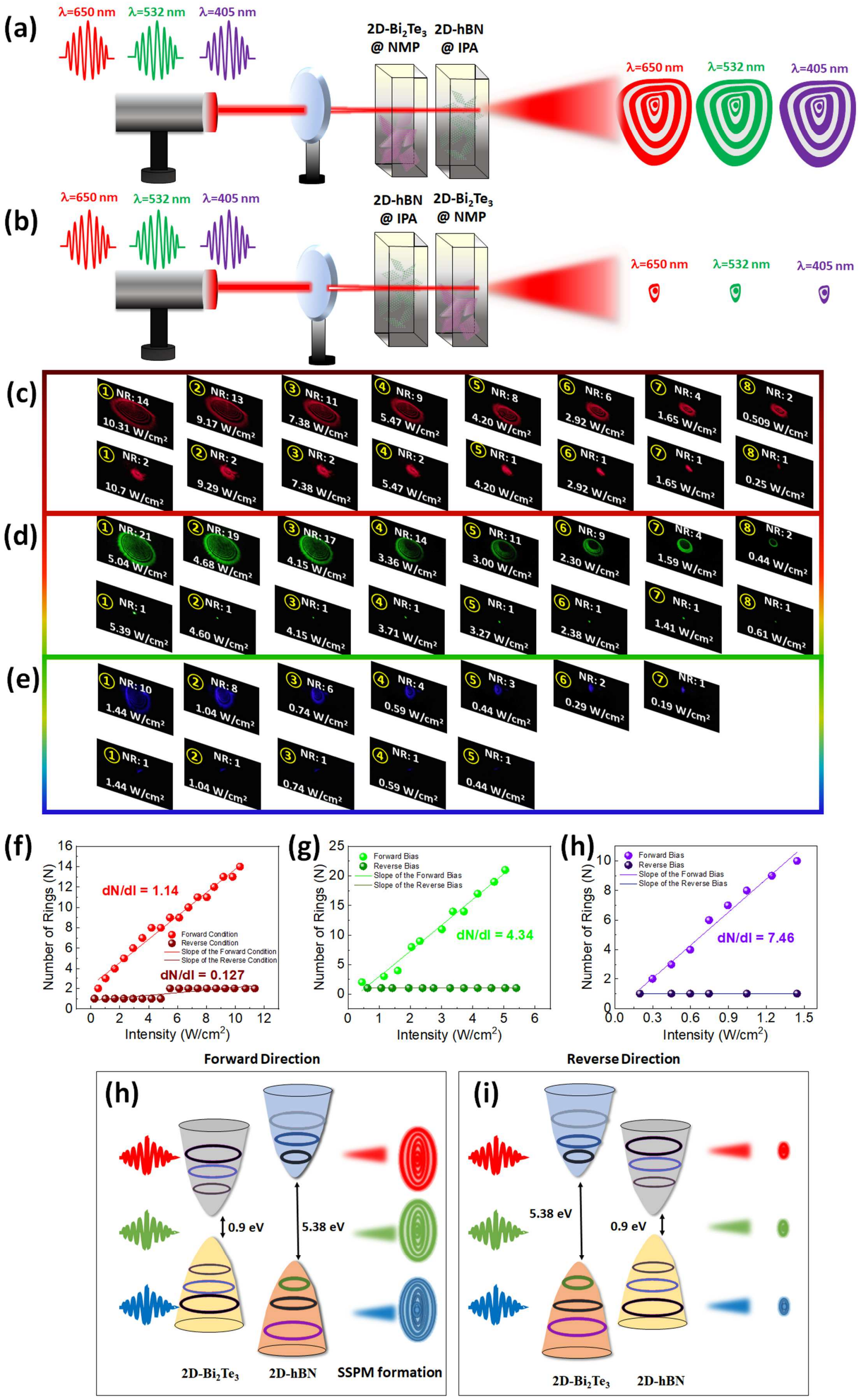

**Figure 5.** Depiction of All-Photonic isolator. a-b) Figure showing forward and reverse biased analogy for the all-photonic diode. c-e-g) Figures showing the diffraction rings vs intensity for forward bias condition for different wavelengths (λ= 650 nm, 532 nm, and 405 nm). d-f-h) Figures showing the diffraction rings vs intensity for reverse bias condition for different wavelengths (λ= 650 nm, 532 nm, and 405 nm). i-j-k) Experimental values of $\frac{dN}{dI}$ obtained for asymmetric light propagation for different wavelengths (λ= 650 nm, 532 nm, and 405 nm). l-m) Diagrams illustrating the mechanism of all-photonic isolator using band structure information for forward and reverse conditions.

## 3. Conclusion

This work presents the synthesis of high-quality 2D $Bi_2Te_3$ via a liquid phase exfoliation technique. Furthermore, we have explored the mechanisms and underlying physical principles of SSPM pattern formation through the use of 2D $Bi_2Te_3$. We have examined the observed phenomena using the Wind Chime model while considering the electronic coherence theorem. The nonlinear optical coefficient values, such as $n_2$ and $\chi^{(3)}_{total}$ were computed. The value of $n_2$ was found to be $2.7\times10^{-4}$, $8.14\times10^{-4}$, and $10.1\times10^{-4}$ $cm^2W^{-1}$ at the specified wavelengths of 650, 532, and 405 nm. The diffraction pattern formation time is accounted for different wavelengths while maintaining other variables constant. In this study, the computed value of third-order nonlinear susceptibility for monolayer ( $\chi^{(3)}_{monolayer}$ ) are $1.2 \times 10^{-7}$ e.s.u, $3.63\times 10^{-7}$ e.s.u, and $4.51\times 10^{-7}$ e.s.u. for the wavelengths 650, 532, and 405 nm. The calculated value of $\chi^{(3)}_{total}$ shows a comparably high value compared to other 2D materials discussed in the literature. Time evolution of the diffraction patterns for different wavelengths (650, 532, and 405 nm) and solvents were also calculated. The study of diffraction ring formation under SSPM, utilizing two solvents, demonstrated that the rise time escalated with increased solvent viscosity, aligning with the Wind Chime model, whereas the time to reach the maximum number of rings presented inconsistencies with the Wind Chime model. First-principles calculations utilizing Density Functional Theory (DFT) were carried out to examine the interaction mechanisms between the molecules and 2D $Bi_2Te_3$, as well as, the contributions to the inconsistent experimental results. NMP shows stronger interaction with 2D $Bi_2Te_3$, which could be related to the increase in the effective viscosity. On the other hand, IPA exhibited weaker molecular interactions, leading to faster rotation under the same applied electric field. The distortion in the vertical direction of the SSPM pattern was quantified, and the collapse time for each solvent was evaluated. Two different solvents were used to realize a fast-operating photonic isolator. The derived experimental and theoretical conclusions were used to fabricate solution-2D nanostructure system capable of showing a fast response, while maintaining a wide wavelength range response. The all-photonic isolator demonstrated in the literature shows ${dN}/{dI}$ values 1.14, 4.34, and 7.46 $Wcm^{-2}$ for wavelengths 650, 532, and 405 nm while demonstrating forward bias condition. This work presents a novel configuration of a nonlinear photonic isolator designed to facilitate nonreciprocal light propagation. To further differentiate this effect from the thermal lens effect, the dependency of $\chi^{(3)}_{monolayer}$ on $m^*$ and $\chi^{(3)}_{monolayer}$ on $\mu$ is obtained, resembling the pertinent curves documented by other authors. The high values of $n_2$ and $\chi^{(3)}_{total}$ are predicted to result from laser-induced hole coherence.

**Data Availability Statement**

Data is available from corresponding author upon request.

**References**

[1] L. Del Bino, J. M. Silver, M. T. Woodley, S. L. Stebbings, X. Zhao, P. Del'Haye, *Optica* **2018**, 5, 279.

[2] a)X. Huang, S. Fan, *Journal of lightwave technology* **2011**, 29, 2267; b)M. S. Kang, A. Butsch, P. S. J. Russell, *Nature Photonics* **2011**, 5, 549.

[3] a)Z. Shen, Y.-L. Zhang, Y. Chen, C.-L. Zou, Y.-F. Xiao, X.-B. Zou, F.-W. Sun, G.-C. Guo, C.-H. Dong, *Nature Photonics* **2016**, 10, 657; b)K. Fang, J. Luo, A. Metelmann, M. H. Matheny, F. Marquardt, A. A. Clerk, O. Painter, *Nature Physics* **2017**, 13, 465.

[4] a)K. Gallo, G. Assanto, K. R. Parameswaran, M. M. Fejer, *Applied Physics Letters* **2001**, 79, 314; b)L. Feng, M. Ayache, J. Huang, Y.-L. Xu, M.-H. Lu, Y.-F. Chen, Y. Fainman, A. Scherer, *Science* **2011**, 333, 729.

[5] L. Fan, J. Wang, L. T. Varghese, H. Shen, B. Niu, Y. Xuan, A. M. Weiner, M. Qi, *Science* **2012**, 335, 447.

[6] L. Wu, W. Huang, Y. Wang, J. Zhao, D. Ma, Y. Xiang, J. Li, J. S. Ponraj, S. C. Dhanabalan, H. Zhang, *Advanced Functional Materials* **2019**, 29, 1806346.

[7] a)Y. Liao, C. Song, Y. Xiang, X. Dai, *Annalen der physik* **2020**, 532, 2000322; b)L. Wu, X. Yuan, D. Ma, Y. Zhang, W. Huang, Y. Ge, Y. Song, Y. Xiang, J. Li, H. Zhang, *Small* **2020**, 16, 2002252; c)Y. Jia, Y. Liao, L. Wu, Y. Shan, X. Dai, H. Cai, Y. Xiang, D. Fan, *Nanoscale* **2019**, 11, 4515; d)Y. Wu, Q. Wu, F. Sun, C. Cheng, S. Meng, J. Zhao, *Proceedings of the National Academy of Sciences* **2015**, 112, 11800; e)X. Li, R. Liu, H. Xie, Y. Zhang, B. Lyu, P. Wang, J. Wang, Q. Fan, Y. Ma, S. Tao, *Optics Express* **2017**, 25, 18346; f)L. Hu, F. Sun, H. Zhao, J. Zhao, *Optics Letters* **2019**, 44, 5214; g)K. Sk, B. Das, N. Chakraborty, M. Samanta, S. Bera, A. Bera, D. S. Roy, S. K. Pradhan, K. K. Chattopadhyay, M. Mondal, *Advanced Optical Materials* **2022**, 10, 2200791; h)L. Wu, Z. Xie, L. Lu, J. Zhao, Y. Wang, X. Jiang, Y. Ge, F. Zhang, S. Lu, Z. Guo, *Advanced Optical Materials* **2018**, 6, 1700985; i)Y. Jia, Y. Shan, L. Wu, X. Dai, D. Fan, Y. Xiang, *Photonics Research* **2018**, 6, 1040; j)R. Wu, Y. Zhang, S. Yan, F. Bian, W. Wang, X. Bai, X. Lu, J. Zhao, E. Wang, *Nano letters* **2011**, 11, 5159; k)S. Goswami, C. C. de Oliveira, B. Ipaves, P. L. Mahapatra, V. Pal, S. Sarkar, P. A. S. Autreto, S. K. Ray, C. S. Tiwary, *Laser & Photonics Reviews*, n/a, 2400999; l)K. Sk, B. Das, N. Chakraborty, M. Samanta, S. Bera, A. Bera, D. S. Roy, S. K. Pradhan, K. K. Chattopadhyay, M. Mondal, *Advanced Optical Materials* **2022**, 10, 2200791; m)X. Xu, M. Wang, Y. Zhang, Q. Li, W. Niu, Y. Yang, J. Zhao, Y. Wu, *Laser & Photonics Reviews* **2024**, 18, 2300930; n)Y. Gao, C. Ling, D. Weng, G. Rui, J. He, B. Gu, *Laser & Photonics Reviews* **2024**, 18, 2301062; o)X. Xu, Z. Cui, Y. Yang, Y. Zhang, Q. Li, L. Tong, J. Li, X. Zhang, Y. Wu, *Laser & Photonics Reviews* **2025**, 19, 2401521; p)D. Weng, C. Ling, Y. Gao, G. Rui, L. Fan, Q. Cui, C. Xu, B. Gu, *Laser & Photonics Reviews* **2025**, 19, 2401587; q)Z. Xu, H. Wang, W. Niu, Y. Guo, X. Zhai, P. Li, X. Zeng, S. Gull, J. Liu, J. Cao, X. Xu, G. Wen, G. Long, Y. Wu, J. Li, *Laser & Photonics Reviews* **2025**, 19, 2401929; r)L. Lu, Z. Liang, L. Wu, Y. Chen, Y. Song, S. C. Dhanabalan, J. S. Ponraj, B. Dong, Y. Xiang, F. Xing, D. Fan, H. Zhang, *Laser & Photonics Reviews* **2018**, 12, 1700221; s)L. Wu, X. Jiang, J. Zhao, W. Liang, Z. Li, W. Huang, Z. Lin, Y. Wang, F. Zhang, S. Lu, Y. Xiang, S. Xu, J. Li, H. Zhang, *Laser & Photonics Reviews* **2018**, 12, 1800215.

[8] J. Zhang, X. Yu, W. Han, B. Lv, X. Li, S. Xiao, Y. Gao, J. He, *Optics letters* **2016**, 41, 1704.

[9] a)C. Zhao, Y. Zou, Y. Chen, Z. Wang, S. Lu, H. Zhang, S. Wen, D. Tang, *Optics express* **2012**, 20, 27888; b)X. Wu, Q. Wang, Y. Guo, D. Wang, Y. Wang, D. Meng, *Materials Letters* **2015**, 159, 80; c)P. Seifert, K. Vaklinova, K. Kern, M. Burghard, A. Holleitner, *Nano Letters* **2017**, 17, 973.
[10] a)X.-L. Qi, S.-C. Zhang, *Reviews of modern physics* **2011**, 83, 1057; b)H. Zhang, C.-X. Liu, X.-L. Qi, X. Dai, Z. Fang, S.-C. Zhang, *Nature physics* **2009**, 5, 438; c)Y. Xia, D. Qian, D. Hsieh, L. Wray, A. Pal, H. Lin, A. Bansil, D. Grauer, Y. S. Hor, R. J. Cava, *Nature physics* **2009**, 5, 398.
[11] J. E. Moore, *Nature* **2010**, 464, 194.
[12] a)S. Chen, C. Zhao, Y. Li, H. Huang, S. Lu, H. Zhang, S. Wen, *Optical Materials Express* **2014**, 4, 587; b)X. Zhang, J. Wang, S.-C. Zhang, *Physical Review B—Condensed Matter and Materials Physics* **2010**, 82, 245107; c)S. Lu, C. Zhao, Y. Zou, S. Chen, Y. Chen, Y. Li, H. Zhang, S. Wen, D. Tang, *Optics express* **2013**, 21, 2072; d)V. V. Kim, A. Bundulis, V. S. Popov, N. A. Lavrentyev, A. A. Lizunova, I. A. Shuklov, V. P. Ponomarenko, J. Grube, R. A. Ganeev, *Optics Express* **2022**, 30, 6970.
[13] a)Y.-H. Lin, S.-F. Lin, Y.-C. Chi, C.-L. Wu, C.-H. Cheng, W.-H. Tseng, J.-H. He, C.-I. Wu, C.-K. Lee, G.-R. Lin, *Acs Photonics* **2015**, 2, 481; b)F. Bonaccorso, Z. Sun, *Optical Materials Express* **2013**, 4, 63; c)Z. Luo, Y. Huang, J. Weng, H. Cheng, Z. Lin, B. Xu, Z. Cai, H. Xu, *Optics express* **2013**, 21, 29516; d)Z.-C. Luo, M. Liu, H. Liu, X.-W. Zheng, A.-P. Luo, C.-J. Zhao, H. Zhang, S.-C. Wen, W.-C. Xu, *Optics letters* **2013**, 38, 5212; e)C. Zhao, H. Zhang, X. Qi, Y. Chen, Z. Wang, S. Wen, D. Tang, *Applied Physics Letters* **2012**, 101.
[14] G. Wang, S. Zhang, X. Zhang, L. Zhang, Y. Cheng, D. Fox, H. Zhang, J. N. Coleman, W. J. Blau, J. Wang, *Photonics Research* **2015**, 3, A51.
[15] a)L. Wu, Y. Dong, J. Zhao, D. Ma, W. Huang, Y. Zhang, Y. Wang, X. Jiang, Y. Xiang, J. Li, *Advanced Materials* **2019**, 31, 1807981; b)Y. Shan, L. Wu, Y. Liao, J. Tang, X. Dai, Y. Xiang, *Journal of Materials Chemistry C* **2019**, 7, 3811.
[16] Y. Liao, Y. Shan, L. Wu, Y. Xiang, X. Dai, *Advanced Optical Materials* **2020**, 8, 1901862.
[17] Y. Wu, Q. Wu, F. Sun, C. Cheng, S. Meng, J. Zhao, *Proceedings of the National Academy of Sciences* **2015**, 112, 11800.
[18] a)Y.-H. Lin, S.-F. Lin, Y.-C. Chi, C.-L. Wu, C.-H. Cheng, W.-H. Tseng, J.-H. He, C.-I. Wu, C.-K. Lee, G.-R. Lin, *ACS Photonics* **2015**, 2, 481; b)P. Roushan, J. Seo, C. V. Parker, Y. S. Hor, D. Hsieh, D. Qian, A. Richardella, M. Z. Hasan, R. J. Cava, A. Yazdani, *Nature* **2009**, 460, 1106.
[19] Y. Huang, H. Zhao, Z. Li, L. Hu, Y. Wu, F. Sun, S. Meng, J. Zhao, *Advanced Materials* **2023**, 35, 2208362.
[20] M. Shalaby, N. Yousif, L. Wahab, H. Hashem, *Materials Science and Engineering: B* **2021**, 271, 115246.
[21] W. Tang, E. Sanville, G. Henkelman, *Journal of Physics: Condensed Matter* **2009**, 21, 084204.
[22] C. M. Vest, M. Lawson, **1972**.
[23] R. Karimzadeh, *Journal of optics* **2012**, 14, 095701.
[24] G. Roux, D. Roberts, G. Perron, J. E. Desnoyers, *Journal of Solution Chemistry* **1980**, 9, 629.
[25] W. Steele, R. Chirico, A. Nguyen, I. Hossenlopp, N. Smith, National Inst. for Petroleum and Energy Research, Bartlesville, OK (United …, 1991.
[26] D. Weng, C. Ling, Y. Gao, G. Rui, L. Fan, Q. Cui, C. Xu, B. Gu, *Laser & Photonics Reviews* **2025**, 19, 2401587.
[27] P. Kumbhakar, A. K. Kole, C. S. Tiwary, S. Biswas, S. Vinod, J. Taha‐Tijerina, U. Chatterjee, P. M. Ajayan, *Advanced Optical Materials* **2015**, 3, 828.

**Supporting Information**

Supporting Information is available from the Wiley Online Library or from the author.

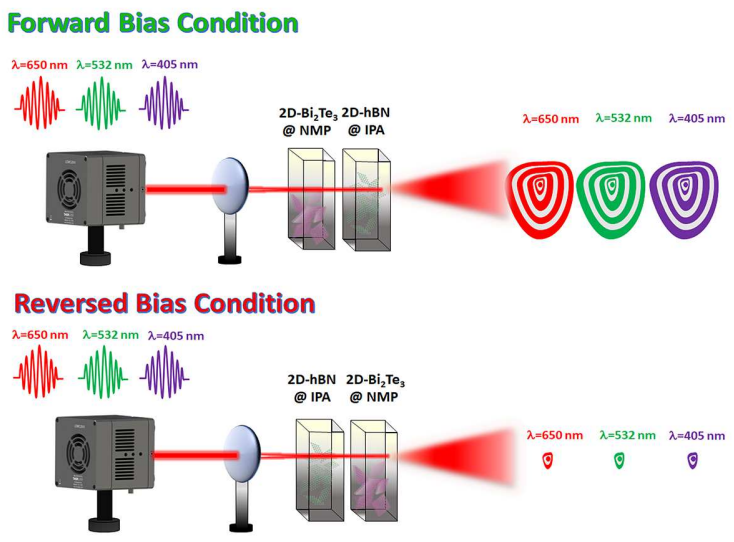

A light-rectifying Fast Response diode: The nonlinear optical properties of the 2D-$Bi_2Te_3$ were examined via calculating the third order nonlinear susceptibility and the nonlinear refractive index by SSPM spectroscopy. The evolving and distorted nature of the SSPM's diffraction pattern is investigated for different solvent to generate a nonlinear photonic diode using 2D-$Bi_2Te_3$/ 2D-hBN heterostructure.

**All Photonic Isolator using Atomically Thin (2D) Bismuth Telluride ($Bi_2Te_3$)**

*Saswata Goswami, Bruno Ipaves, Juan Gomez Quispe, Caique Campos de Oliveira, Surbhi Slathia, Abhijith M.B, Varinder Pal, Christiano J.S. de Matos, Samit K. Ray, Pedro A. S. Autreto* and Chandra Sekhar Tiwary**

**Section S1. Synthesis and Characterization of 2D-$Bi_2Te_3$ nanostructure**

Pure Bismuth (99.99%) and Tellurium (99.99%) were combined in a 2:3 molar ratio to produce crystals of Bismuth Telluride ($Bi_2Te_3$). The synthesis of bulk $Bi_2Te_3$ was achieved by subjecting the mixture to flame melting at a temperature of 850°C±50°C for a duration of 10 minutes in a sealed argon environment. The material was then cooled using a furnace. The bulk material was subsequently subjected to annealing at a temperature of 450°C in the presence of $N_2$ gas within a sealed quartz tube using a muffle furnace. The bulk crystal was pulverised into powder form with a mortar and pestle. This powder was subsequently probe sonicated in isopropyl alcohol (IPA) for a duration of 4 hours using a Rivotek SM250PS ultrasonic probe sonicator. Subsequent to a 12-hour relaxation, the suspended solution underwent centrifugation at 2000 revolutions per minute for 30 minutes utilising a REMI PR-24 centrifuge apparatus. The 2D $Bi_2Te_3$ was subsequently extracted from the suspension, rendering this synthesis technique for 2D $Bi_2Te_3$ nanostructure straightforward and brief.

**X-Ray Diffraction:** The XRD analysis was performed using the Bruker D8 advance with Cu-$K_\alpha$ radiation ($K_\alpha$=1.54 Å). X-ray diffraction (XRD) analysis reveals that identical peaks are seen in both the bulk and 2D structure of the material.[1] **Figure S1a** demonstrates that the structure remains intact during the exfoliation process. The reflections along (006), (015), (222), (1010), (0015), (1016), (0210), (0120) and (1115) exhibit sharp and strong characteristics, with corresponding 2θ values of 17.29°, 27.59°, 37.81°, 44.432°, 53.97°, 57.04°, 62.16° and 65.80° respectively. Figure S1a presents the X-ray diffraction (XRD) pattern for both the bulk and 2D $Bi_2Te_3$, revealing rhombohedral $R\bar{3}m$ space group with $D_{3d}^5$ point symmetry crystal structure. All peaks of the XRD pattern matched with rhombohedral phase of $Bi_2Te_3$. (ICDD PDF card No. 00-015-0863, a= 4.386 Å, c= 30.497 Å). Rietveld Refinement[2] is performed to refine the lattice parameters, and the results converged to the weighted and expected residual factors of $R_{wp}$ = 0.0806 and $R_p$ = 0.0596, to yield a structure in rhombohedral space group $R\bar{3}m$ with lattice parameters of a = 4.3963 Å and c = 30.52322 Å. The pseudo-voigt function is chosen to fit the peak profile during refinement process, this secures a goodness of fit $\chi^2 = (\frac{R_{wp}}{R_{exp}})^2 = 1.82$. Fitted graph is shown Supporting information Figure S2, Rietveld refinement is done using Topas software.

**TEM:** A high-resolution transmission electron microscope (NEOARM/JEM-ARM200F) was employed to image the exfoliated nanostructure of 2D $Bi_2Te_3$. Figure S1b illustrates the enlarged depiction of the 2D flake like nanostructure. Figure S1b depicts translucent sheets arranged in a stacked formation, with the grid and the sample clearly discernible. A 3D map is presented in the Figure S1c to illustrate the staircase-like layered structure in terms of intensity variation. As depicted in Figure S1d cluster of 2D flake-type nanostructures of 2D $Bi_2Te_3$ is visible. The Top inset of Figure S1d illustrates the Fast Fourier

Transform (FFT) SAD pattern corresponding to the (110) rhombohedral plane. The determined value of d spacing for the (110) plane was 0.2074 nm (shown in the bottom inset of the Figure S1d). Figure S1d illustrates the enlarged depiction of the 2D flake. The upper inset of Figure S1d presents the selected area electron diffraction (SAED) pattern corresponding to the region marked in red. The lower inset shows the d spacing derived from the IFFT of the SAED pattern.

**AFM:** An atomic force microscope (Agilent Technologies, 5500) was used to evaluate the thickness and lateral width of the two-dimensional $Bi_2Te_3$ (shown in Figure S1e). Figure S1f illustrates the width profile of the 2D flakes, which ranges from 60 to 140 nm. The average width is considered to be 100 nm. Figure S1g demonstrates that the thickness distribution of the 2D $Bi_2Te_3$ spans roughly 0 to 8 nm, with certain particles displaying markedly larger thickness. Average value of thickness is considered to be 3.5 nm.

**SEM:** Scanning electron microscopy was done using JSM-IT300HR JEOL microscope. Figure S1h shows SEM image and Figure S1i and S1j shows the EDX mapping image of Bi and Te elements. And schematic indicates that elements are uniformly distributed across the nanostructure. The EDX spectrum shown in the Supporting information Figure S3 indicates that the Bi / Te atomic ratio of the nanostructure is found to be near 2:3. Although the presence of excess Bi atomic percentage indicates P-type behavior in the nanostructure. Due to low melting point of the Te, some amount Te might have evaporated during melting process, which induces P-type behavior in the nanostructure. EDAX spectrum of 2D $Bi_2Te_3$ is shown Supporting Information Figure S4.

**UV-Vis Spectroscopy:** Figure S1n illustrates that UV-Visible Spectroscopy was used to assess the optical absorbance of 2D $Bi_2Te_3$ in the IPA solvent. Figure S1o shows the Tauc plot that was used to estimate the direct bandgap of the 2D $Bi_2Te_3$. Tauc plot is described in the following expression,

$$(\alpha h\nu)^{1/n} = A(h\nu - E_g) \qquad (1)$$

Here α is the optical absorption coefficient, $h\nu$ is the discrete photon energy, $E_g$ is the bandgap of the material and n is the order of the polynomial, n is considered as $1/2$, as the material has direct bandgap nature. The value of the bandgap is calculated to be 0.9 eV.

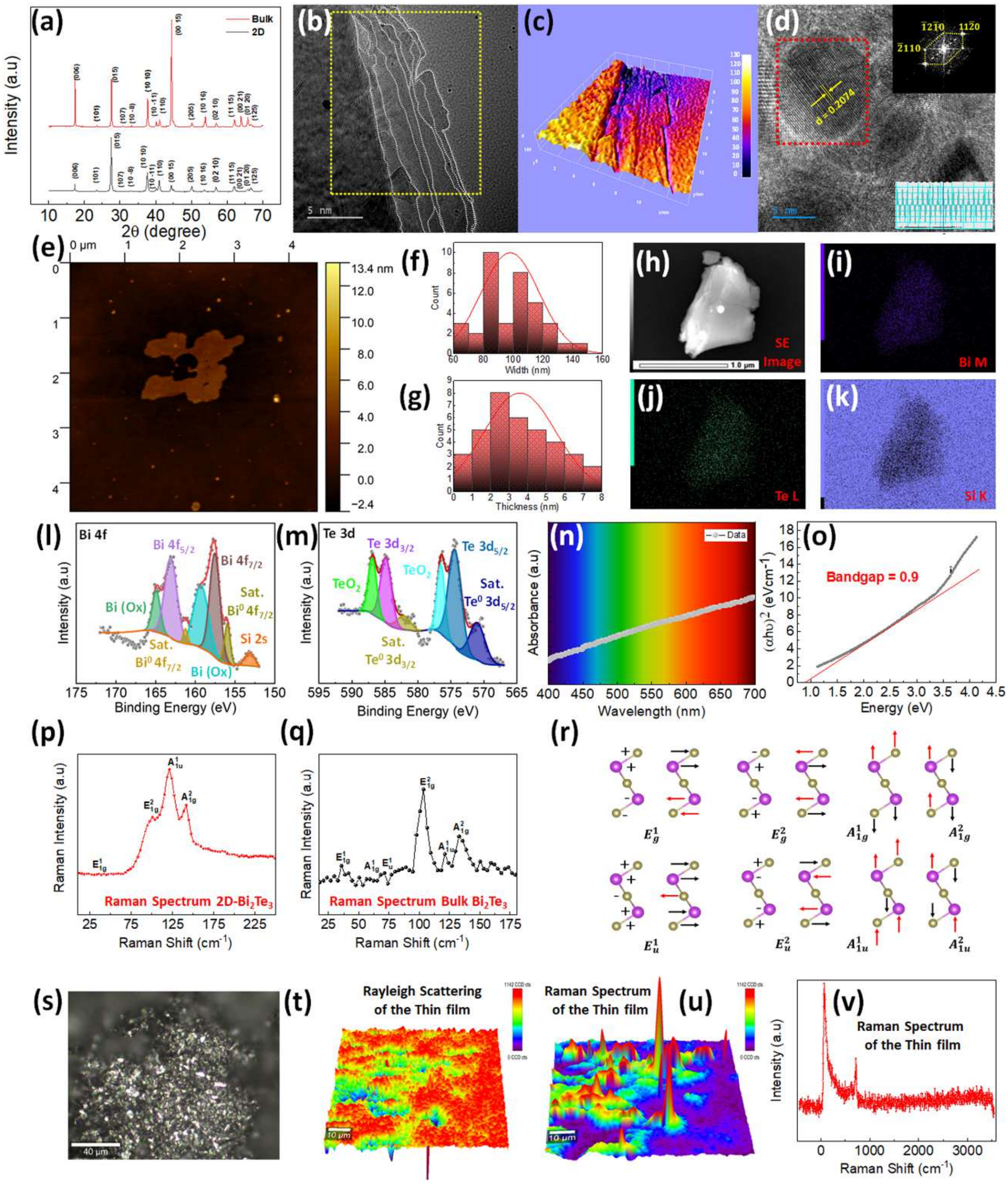

**Figure S1.** Characterization of 2D $Bi_2Te_3$. a) Bulk and 2D $Bi_2Te_3$ XRD comparison. b) Transmission electron microscopy image of the 2D $Bi_2Te_3$ showing two-dimensional morphology. Dotted lines showing one flake on top of another. c) Figure showing 3D map of pixel intensity profile done using ImageJ software, showing flakes stacked over each other. d) Figure showing a single 2D $Bi_2Te_3$ flake like nanostructure. Inset showing FFT pattern of (110) plane matched with plane orientation showing (110) plane. Bottom inset showing calculated value of d spacing. e) AFM image of the flake like nanostructure distributed on the surface. Figure f) and g) showing the width (lateral dimension) and height (thickness) of the exfoliated 2D-$Bi_2Te_3$. h) SEM image of the exfoliated nanostructure showing layered morphology. Elemental mapping on the exfoliated nanostructure shows i) Bi, j) Te, and k) Si substrate. The high-definition XPS spectrum of l) Bi 4f and m) Te 3d orbitals. Figure n) showing absorbance spectrum derived from UV-vis Spectroscopy. Figure o) showing determined value of direct bandgap using Tauc plot. Figure

p) and q) showing Raman spectrum of 2D $Bi_2Te_3$ and bulk form respectively. r) Schematic showing different vibrational modes of 2D $Bi_2Te_3$. s) Figure showing thin film of 2D $Bi_2Te_3$. Raman mapping taken from the film shown in the previous Figure shows t) Rayleigh, u) Raman Signal of the corresponding sample. v) Raman spectrum of the thin film.

**XPS:** The IPA solution containing 2D $Bi_2Te_3$ was then drop casted on a Si wafer for chemical analysis. PHI 5000 VERSA PROBE III ULVAC PHI (Physical Electronics) was used for the XPS analysis. Supporting Information Figure S4 displays the chemical characterization data obtained by XPS (X Ray Photoelectron Spectroscopy, the XPS survey spectrum spans the binding energy range from 0 to 1100 eV. The exhibited peaks found in the survey spectrum were determined to correspond to the elements Te, Bi, C, and O. Figure S1l displays the high-resolution XPS scan of Bi 4f doublet peaks located at about 158.5 and 164 eV, which correspond to Bi $4f_{7/2}$ and Bi $4f_{5/2}$ energy levels, respectively. The peaks may be separated into two distinct signals located at 157.6 eV and 163.12 eV. These signals are linked to the presence of Bi-Te bonds, providing confirmation of the fabrication of the $Bi_2Te_3$ phase.[3] In addition, two peaks located at 159.3 eV and 165 eV correspond with the peaks seen in bismuth oxide ($BiO_x$), indicating the potential development of a surface oxidation phase.[3b, 4] The redox signals that reflect the oxidation and reduction potential of Bi oxide species show a relationship with the oxidation of $Bi^0$ films and the reduction of $Bi^{3+}$ ions. The double peaks at 156 eV and 165 eV corresponds to the $Bi^0$ peak or the elemental bismuth present in the sample, this can be caused due to the exposed Bi atoms on the surface. In addition, Figure S1m displays the Te 3d peaks, which consist of two doublets can be separated into three component peaks each. The peaks at 573.8 eV and 584.1 eV are characteristic peaks of the 2D $Bi_2Te_3$ semiconductor phase. [5] Similarly, we see two peaks at 574 eV and 586.87 eV that are associated with the $Te^{4+}$ state, maybe caused by the existence of the $TeO_2$ phase. The peaks at 577.2 and 587.6 eV correspond to the $Te^0$, which is also probably a feature of exposed Te atom surface.[3c, 6] The $Te^{2+}$ atoms are predominantly located on the surface of the material. The $TeO_2$ peaks arise from the reaction between $Te^{4+}$ atoms distributed on the surface of the sample and oxygen present in the surrounding air.[6b] Supporting Information Figure S5a indicates the presence of C in the sample, deconvolution of the peaks reveals two distinct sub peak corresponding to C-O-C and C-C.[7] Both of this peaks are located at 283.62 eV and 285.62 eV. The presence of C peaks is most likely attributed to the sample preparation conditions.[8] In contrary, the oxygen peaks confirm the inherent inclination of $Te^{4+}$ state to interact with the surrounding air, hence causing surface oxidation. This is supported by the double peaks seen for Bi and Te in Figure S5b.[9]

In particular, the O 1s spectrum seen in Supporting Information Figure S5b displays three distinct bonding types at energy levels of 528.87 eV, 531.25 eV, and 532 eV. The dominant signal at 531.25 eV in the Bi-Te samples may be linked to lattice oxygen resulting from metal-oxide interactions.[3b] In conclusion, the small intensity peaks seen at 532 eV and 528.87 eV are associated with the presence of dangling bonds $O^-$ and $O^{2-}$, as well as surface adsorbed oxygen $O^{2-}$, respectively. Peak located at 532 eV may suggest long chain organic compound present in the sample. This may have happened due to the laboratory condition.

**Raman Spectroscopy:** Raman spectroscopy measurement is performed using WiTec alpha 300 system equipped with a 600 lines/nm grating. The excitation was performed using 532 nm diode laser with a spot size of 500 nm. The data analysis is carried out using the WiTec project software.

Raman Spectroscopy was used to determine the vibrational modes, investigating the chemical bonds would provide a wealth of information about the structure and stability of these compounds. According to the TEM and XRD results, few-QL $Bi_2Te_3$ has a hexagonal nanostructure. Based on group theory, this unit cell belongs to the space group $D_3^{d3}$ $p\overline{3}$m1. In a unit cell each layer consists of five monoatomic planes of $Te^{(1)}$–Bi–$Te^{(2)}$–Bi–$Te^{(1)}$[10], this five atoms arranged in vertical axis occupy ~1nm is measured to be 1 QL. At the Brillouin zone centre (q=0), five atoms produce four Raman-active modes: the in-plane vibrational modes $E_g^1$ and $E_g^2$, and the out-of-plane vibrational modes $A_{1g}^1$ and $A_{1g}^2$. Group theory classification indicates that there are 12 optical modes represented as $2A_{1g}$ +$2E_g$ +$2A_{1u}$ +$2E_u$.[11] The phonon modes in this material are considered to be Raman or infrared (IR) active because of the crystal's inversion symmetry. The $A_{1g}$ and $E_g$ modes demonstrate 2-fold degeneracy.

The atoms oscillate within the same plane and perpendicular to it in diverse phonon modes. In $E_g$ modes, the atoms oscillate inside the basal plane, but in $A_{1g}$ modes, the atoms oscillate along the CH direction. Crystals have inversion symmetry, indicating that the IR-active modes ($A_{1u}$) must demonstrate odd parity, whereas the Raman-active modes ($E_g$, $A_{1g}$) must exhibit even parity under inversion.

$A_{1g}^1$ mode is the vibration mode that moves towards higher frequencies in the out-of-plane direction. This change relates to alterations in the interlayer van der Waals force along the (0 0 1)-direction, induced by dislocations resulting from lattice mismatch and residual internal stress. The Fröhlich electron-phonon interaction stimulates the longitudinal optical $A_{1g}^1$ vibration, indicating that the blueshift of the $A_{1g}^1$ mode signifies an increase in internal stress due to electron-phonon coupling.[12] According to group theory, the ~121 $cm^{-1}$ peak is attributed to the longitudinal optical phonon mode.[12-13] The same can be observed in Figure S1p.

Figure S1p illustrates that the Raman intensity of $A_{1u}^1$ out-of-plane vibrations along the c axis surpasses that of the a-b in-plane lattice vibrations $E_{1g}^2$, suggesting that the 2D $Bi_2Te_3$ consist of multiple QLs in thickness.[14] The Raman spectra of the thin bulk exhibit a notable result, characterized by an additional peak of lower intensity at approximately 121 cm$^{-1}$, as illustrated in Figure S1q. $Bi_2Te_3$ demonstrates a notable level of symmetry and crystallization order in its 2D nanostructure formed from stoichiometric atoms. The crystal structure significantly influences the IR-active Raman $A_{1u}^{1}$ mode. The peak is identified as the $A_{1u}^{1}$ mode, comprising longitudinal optical (LO) phonons situated near the boundary of the Brillouin Zone (BZ), particularly at the Z point. The $A_{1u}^{1}$ mode is identified as an infrared active mode; however, it has been observed that there is no Raman active mode present in bulk $Bi_2Te_3$ crystal.[15] Thus, we may rationally attribute the formation of the IR active $A_{1u}^{1}$ mode in thin nanostructures to the violation of crystal symmetry resulting from the occurrence of two surfaces.
This mode has an odd parity and is not allowed in bulk crystals because of their inversion symmetry.[16] The presence of the strongest $A_{1u}^{1}$ peak can be attributed to the disruption of crystal symmetry on nanoplates with extremely thin thickness. Therefore, the Raman spectra suggest that the thickness of the nanostructure created in their original state is not consistent. As seen in Figure S1p the signal of $A_{1u}^{1}$ is not prominent, conveying the same conclusion, that it shows the Raman signature of 2D $Bi_2Te_3$. However, the $A_{1u}^{1}$ is infrared active mode (IR), the influence of temperature enhances the oscillation of bonded atoms and the collision between electrons and phonons. The temperature effect amplifies the oscillation of bonded atoms and the interaction between electrons and phonons. Figure S1r illustrates the Raman vibrational modes of the 2D $Bi_2Te_3$. The image Figure S1s show the thin film taken for Raman mapping. Raman mapping derived from the film depicted in the preceding picture illustrates t) Rayleigh, u) Raman Signal of the corresponding sample. Figure S1v shows the Raman spectrum of the thin film.

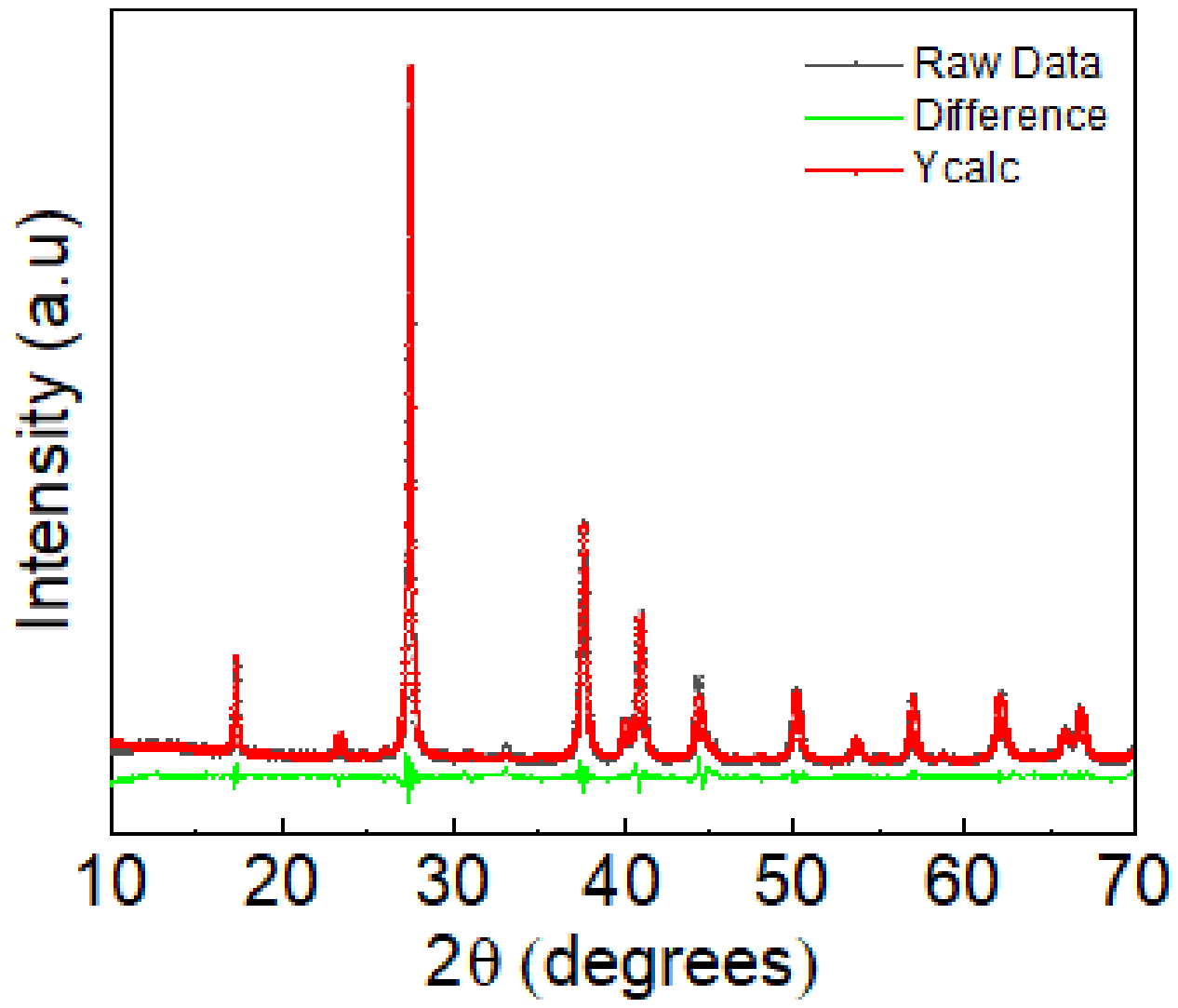

**Figure S2**: Fitted XRD pattern obtained through Topas software.

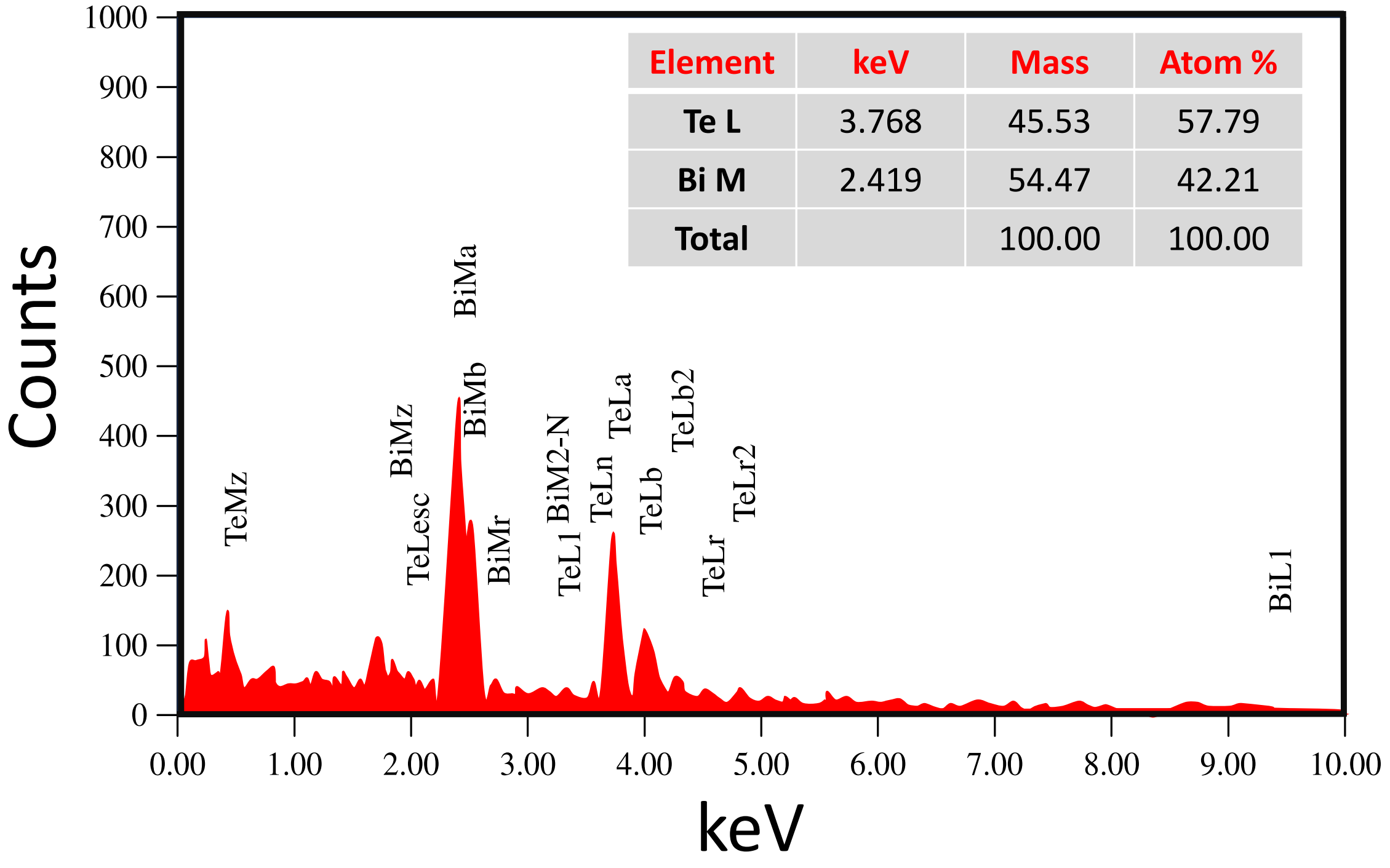

**Figure S3**: EDX Spectrum of 2D-$Bi_2Te_3$ nanostructure.

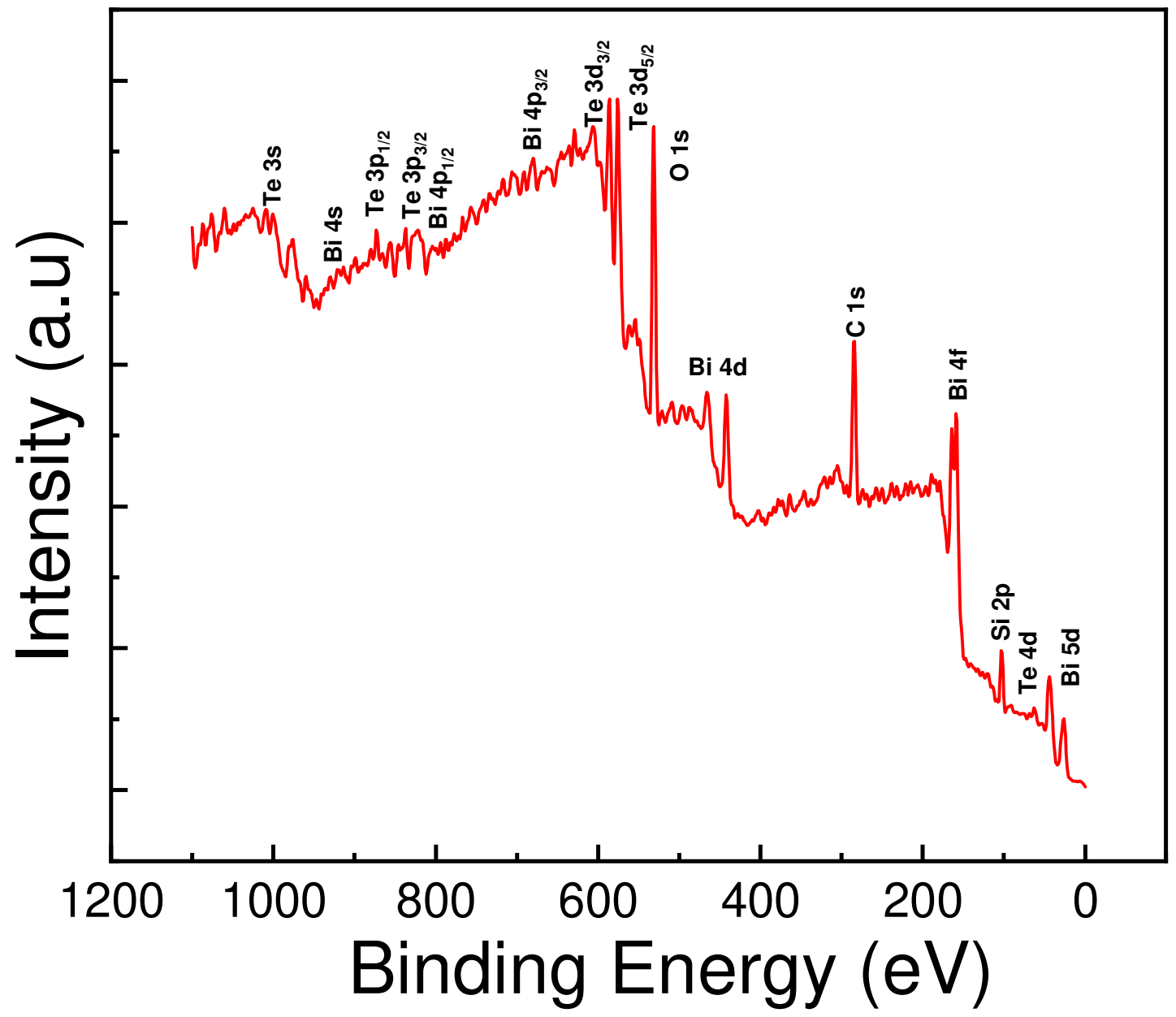

**Figure S4**: The XPS survey spectrum covering the binding energy range of 0 to 1100 eV.

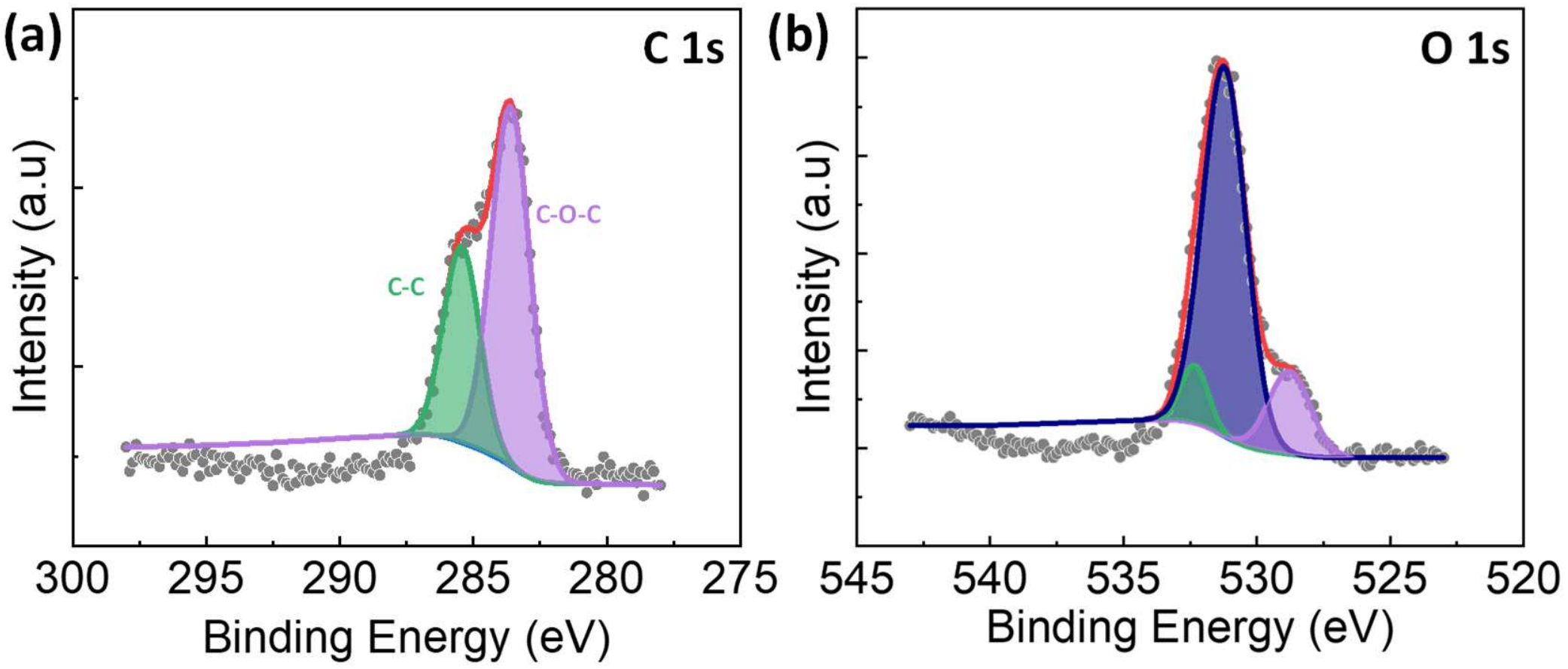

**Figure S5**: (a) XPS Spectrum of the C 1S. (b) XPS Spectrum of the O 1S.

**Section S2: Theoretical Estimation of the Formula of $n_2$ and $\chi^{(3)}_{total}$**

In this work we focus on investigating the nonlinear Kerr effect, which plays a crucial role in determining the nonlinear optical responses of the 2D-$Bi_2Te_3$. Equation 2 defines the Kerr nonlinear effect by establishing a relationship between the incident laser beam intensity (I) and the refractive index (n).

$$n = n_0 + n_2 I \quad \ldots\ldots\ldots\ldots\ldots\ldots (2)$$

The terms "$n_0$" and "$n_2$" represent the linear and nonlinear refractive indices of the material, respectively. [17]

Moreover, a modification in the refractive index of the medium will cause the light wave to undergo a mandatory phase shift. As a result, the light wave transmitted through the medium, experiences its own intensity modulation, known as self-phase modulation (SPM). In general, the SSPM phenomenon is excited using the fundamental mode Gaussian beam ($TEM_{00}$). A Gaussian beam traverses a nonlinear medium of length L along the z-axis, and its electric field distribution may be characterized as following:

$$E(r,z) = E(0,z)\frac{\omega_0}{\omega(z)}\exp\left(-\frac{r^2}{\omega(z)^2}\right)\times\exp\left(-i\left[kz - arctan\frac{z}{z_0} + \frac{kn_0r^2}{2R(z)}\right]\right)\ldots\ldots\ldots\ldots\ldots\ldots (3)$$

Where λ is the wavelength of the laser, $k = \frac{2\pi}{\lambda}$ is the wave vector, $z_0 = {\pi\omega_0^2}/{\lambda}$ is the Rayleigh length (the length of the beam that propagates like wave), $R(z) = z\,[1 + (\frac{z}{z_0})^2]$ is the radius of curvature of the wavefront, $\omega_0$ is the beam waist radius at $z$. Using the radius of the laser beam waist as the starting point for beam propagation, the electric that is incident on the 2D material can be defined as,

$$E(r, z_0) = E(0, z_0)\exp\left(-\frac{r^2}{\omega(z_0)^2}\right) \times \exp\left(-i\frac{kn_0r^2}{2R(z)}\right) \ldots\ldots\ldots\ldots\ldots\ldots (4)$$

The light intensity distribution can be expressed as,

$$I(r, z) = I_0(1 + z^2/z_0^2)^{-1}\exp\left(-{2r^2}/{\omega_0^2}\right) \ldots\ldots\ldots\ldots\ldots\ldots (5)$$

Where $I_0 = 2P_{ave}/[\pi\ \omega(z)^2]$ is the central intensity of the Gaussian Beam, $P_{ave}$ is the average power of the laser. Therefore, the refractive index can be expressed as,

$$n(r) = n_0 + n_2 I(r) \ldots\ldots\ldots\ldots\ldots\ldots (6)$$

Assuming thin sample the output can be expressed as,

$\Delta\phi = \Delta\psi_L(r) + \Delta\psi_{NL}(r)$, where, $\Delta\psi_L(r) = kn_0r^2/[2R(z)]$ is the linear phase shift, and $\Delta\psi_{NL}(r)$ is a nonlinear phase shift related to the light intensity is described as,

$$\Delta\psi_{NL}(r) = \frac{2\pi n_0}{\lambda}\int_0^L n_2\, I(r, z)dz \ldots\ldots\ldots\ldots\ldots\ldots (7)$$

The above-mentioned nonlinear phase shift " $\Delta\psi$ ", as shown in Equation 7 is induced by potent Kerr Nonlinear Effect caused by incoming Gaussian laser beam

The transverse propagation wave vector is thus expressed as,

$$\delta k(r) = \frac{d\Delta\psi_{NL}}{dr} = \frac{-8krn_2I(r,z)\exp\left(-2r^2/\omega_0^2\right)}{n_0} \ldots\ldots\ldots\ldots\ldots\ldots (8)$$

The character λ denotes the wavelength of the laser, while $L_{eff}$ refers to the effective transmission length of the laser as it travels through the cuvette, $r \in [0, +\infty)$ is the radial coordinate and $I(r, z)$ is the radial intensity distribution.[18]

In the outgoing Gaussian light, there are at least two distinct locations, $r_1$ and $r_2$, where the slopes of the distribution curve, represented by $(d\Delta\psi/dr)_{r=r_1}$and $(d\Delta\psi/dr)_{r=r_2}$, are equal and have the same phase. This is evident due to the fact that nonlinear phase shift has a Gaussian distribution as seen in Figure 2b. Thus, the output light intensity profile exhibits a consistent phase difference while maintaining the same slope points. Clearly, these two points meet the requirement for interference. When $\Delta\psi_0 \geq 2\pi$, the diffraction ring appears. The self-induced diffraction pattern is observed as rings, which can be either bright or dark, in the far field. The positioning of these rings is determined by the Equation 9.

$$\Delta\psi_{r_1} - \Delta\psi_{r_2} = 2M\pi \text{ .................. (9)}$$

In this context, M is used to represent an integer. The terms "dark field " and "bright field" are achieved at the odd and even values of M, respectively.[19] The phase shift, which is crucial for the SSPM effect, generates a self-diffraction pattern in the far field, as seen in Figure 2(b). The calculation of the effective transmission length ($L_{eff}$) using Equation 10 is essential for finding the nonlinear refractive index ($n_2$) as described in Equation 14.[20]

$$L_{eff} = \int_{L_1}^{L_2}(1+\frac{z^2}{z_0^2})^{-1}\,dz = z_0\arctan\left(\frac{z}{z_0}\right)\Big|_{L_1}^{L_2},\ z_0 = \frac{\pi\omega_0^2}{\lambda} \text{ .................... (10)}$$

$L_2$ and $L_1$ denote the distance between the focus point of the laser beam and the boundaries of the quartz cuvette in this particular situation. The thickness of the quartz cuvette is calculated to be $L_2 - L_1$. The intensity profile of the transmitted beam follows a Gaussian distribution, with the highest intensity at the center, $I(0,z) = 2I$. Here $I$ represent the mean intensity of the incoming laser beam $z_0$ is the diffraction length and $\omega_0$ denotes the beam radius ($\frac{1}{e^2}$).

The Light Intensity is included in the expression of $\Delta\psi_{NL}(r)$,

$$\Delta\psi(r) = \frac{2\pi n_0 n_2}{\lambda} I_o L_{eff} e^{-2r^2/\omega_0^2} \text{ ................ (11)}$$

Also, $\Delta\psi(r_1) - \Delta\psi(r_2) = M\pi$, here brightness of the diffraction ring decided upon phases of radial distances on two beams ($r_1$) and ($r_2$). The center and infinity position of the gaussian beam is satisfied as,

$$\Delta\psi(0) - \Delta\psi(\infty) = 2N\pi \text{ ................ (12)}$$

Where N is the number of diffraction ring, and

$$\Delta\psi(\infty) = 0, \Delta\psi(0) = \frac{2\pi n_0 n_2}{\lambda} I_0 L_{eff} \text{................ (13)}$$

The nonlinear refractive index $n_2$ is expressed as,

$$n_2 = (\frac{\lambda}{2n_0 L_{eff}}).\frac{dN}{dI} \text{ ................ (14)}$$

The ${dN}/{dI}$ is an important parameter to evaluate the nonlinear refractive index of the two-dimensional material. The third-order nonlinear susceptibility $\chi^{(3)}_{total}$ is employed to characterize the nonlinear optical characteristics of materials.[21]

It can be written as

$$\chi^{(3)}_{total} = \frac{cn_0^2}{12\pi^2} 10^{-7} n_2 \ (e.s.u) \text{ ..................... (15)}$$

Here, c represents the speed of light in free space, $n_0$ is the linear refractive index of the IPA solvent, and $n_2$ defines the effective length that the laser beam propagates through the cuvette. However, the effective number of 2D material available in the cuvette has a direct effect on the value of the $\chi^{(3)}_{total}$. Hence it is necessary to determine the value of third order nonlinear susceptibility caused by a single layer of two-dimensional sheets $\chi^{(3)}_{monolayer}$. The relationship between the overall electric field strength $E_{total}$ and the electric field strength $E_{monolayer}$ traveling through the single layer of $Bi_2Te_3$ may be mathematically represented as,[19, 22]

$$E_{total} = \sum_{j=1}^{N_{eff}} E_j \cong N_{Eff} \, E_{monolayer} \text{ ............... (16)}$$

In this context, $N_{Eff}$ represents the number of 2D-$Bi_2Te_3$ layers present in the solution, through which the beam passes through. The relationship between $\chi^{(3)}_{total}$ and $\chi^{(3)}_{monolayer}$ can be expressed as, [21c, 22-23]

$$\chi^{(3)}_{total} = N_{eff}^2 \chi^{(3)}_{monolayer} \text{ .................. (17)}$$

**Section S3: Calculation of effective number of 2D-$Bi_2Te_3$ nanostructure present in the solution**

The molecular weight of $Bi_2Te_3$ is found to be 800.7608 g.mol$^{-1}$. The concentration can be expressed as $3.122\times10^{-4}$ mol. L$^{-1}$. The volume of the cuvette is considered to be $4.5\times10^{-3}$ L. Total number of molecules of $Bi_2Te_3$ in the solution is M = $\rho \times V \times N_A$ ($N_A$ is the Avogadro's constant). The space group of the $Bi_2Te_3$ is $R\bar{3}m$, which is a hexagonal crystal system, and lattice constants are found to be a = 4.386 Å and c = 30.49 Å, respectively. Thus, a single effective layer contains number of molecules $m = 1 \times 4.5\ cm^2 / (Sin\ 90°) \times (4.386)^2 = 2.079 \times 10^{15}$ molecules. The number of layers of the nanoflakes can be calculated as the following, n = $M/m$ = 361.

**Table S1.** The value of $n_2$ and $\chi^{(3)}_{total}$ calculated using SSPM Spectroscopy in recent literature.

| 2D Material | Type of laser | $n_2$ | $\chi^{(3)}_{total}$ | $\chi^{(3)}_{monolayer}$ | References |
|---|---|---|---|---|---|
| Graphene | CW 532 nm | $2.5 \times 10^{-9}$ m$^2$.W$^{-1}$ | $1 \times 10^{-3}$ | $1 \times 10^{-7}$ | [19] |
| $MoS_2$ | CW 478 nm | $10^{-7}$ | $1.44 \times 10^{-4}$ | $1.6 \times 10^{-9}$ | [21a] |
| $Ti_3C_2Ti_x$ | CW 457 nm/ 532nm / 671 nm | $11 \times 10^{-4}$ / $4.75 \times 10^{-4}$/ $4.72 \times 10^{-4}$ | ----- | $4.34 \times 10^{-7}$/ $1.68 \times 10^{-7}$/ $0.15 \times 10^{-7}$ | [24] |
| 2D Te NS | CW 457 nm/ 532nm / 671 nm | $6.14 \times 10^{-5}$/ $6.202 \times 10^{-5}$/ $7.37 \times 10^{-5}$ | ------ | ------- | [25] |
| Sb FS/QD | CW 532 nm / 633 nm | FS - $2.88 \times 10^{-5}$ / $0.979 \times 10^{-5}$ QD- $1.91 \times 10^{-5}$/ $0.719 \times 10^{-5}$ | FS - $3.98 \times 10^{-9}$ / $1.74 \times 10^{-9}$ QD- $2.87 \times 10^{-5}$/ $1.29 \times 10^{-5}$ | ----- | [26] |
| $Bi_2Te_3$ | CW 1070 nm | $2.91\times10^{-9}$ m$^2$.W$^{-1}$ | $10^{-3}$ | $10^{-8}$ | [27] |
| Graphene Oxide | CW 532 nm CW 671 nm | $3.57 \times 10^{-6}$ cm$^2$W$^{-1}$ $1.1 \times 10^{-6}$ cm$^2$W$^{-1}$ | $1.7 \times 10^{-6}$ $5.32 \times 10^{-6}$ | ------ | [28] |
| $MoTe_2$ | CW 473 nm /532 nm/ 750 nm/ 801 nm | ------ | ------- | $1.88 \times 10^{-9}$ e.s.u $1.3 \times 10^{-9}$ e.s.u $1.14 \times 10^{-9}$ e.s.u $0.98 \times 10^{-9}$ e.s.u | [29] |
| $NbSe_2$ | 532 nm / 671 nm / | $1.352 \times 10^{-5}$ m$^2$.W$^{-1}$ / $2.0 \times 10^{-5}$ m$^2$.W$^{-1}$/ $1.07 \times 10^{-5}$ m$^2$.W$^{-1}$ | $1.352 \times 10^{-5}$ / $9.354 \times 10^{-6}$ / $5.03 \times 10^{-6}$ | $3.34 \times 10^{-9}$/$2.59 \times 10^{-9}$ /$3.39 \times 10^{-9}$ | [30] |
| $Bi_2Se_3$ | 350 nm/ 600nm/ 700nm/ 1160 nm | $1.16 \times 10^{-8}$ / $3.53 \times 10^{-9}$ / $2.5 \times 10^{-9}$/ $1.65 \times 10^{-9}$ (m$^2$.W$^{-1}$) | $5.76 \times 10^{-3}$ /$1.82 \times 10^{-3}$ / $1.29 \times 10^{-3}$ /$8.53 \times 10^{-4}$ (e.s.u) | $10^{-8}$/ $10^{-8}$ / $10^{-9}$/ $10^{-9}$ | [31] |
| Black Phosphorus | Pulsed Laser 350-1160 nm | $10^{-5}$ cm$^2$.W$^{-1}$ | $10^{-8}$ (e.s.u) | ----- | [20] |
| $WSe_2$ | CW 532nm / 671 nm/ 457 nm/ | $2.94 \times 10^{-6}$ / $8.66 \times 10^{-6}$ / $6.402 \times 10^{-6}$ | $1.371 \times 10^{-6}$/ $4.04 \times 10^{-6}$/ $2.98 \times 10^{-6}$ | $8.14 \times 10^{-10}$/ $8.44 \times 10^{-11}$ /$3.69 \times 10^{-9}$ | [18] |
| $TaS_2$ | CW 532nm / 671 nm/ 457 nm | $1.14 \times 10^{-5}$, $0.88 \times 10^{-5}$, $0.69 \times 10^{-5}$ cm$^2$W$^{-1}$ | ------ | $1.2 \times 10^{-6}$/ $0.9 \times 10^{-6}$/ | [32] |
| | | | | $0.7 \times 10^{-6}$ | |
| $TaSe_2$ | 532 nm / 671 nm | $8.0 \times 10^{-7}$/ $3.3 \times 10^{-7}$ ($cm^2.W^{-1}$) | 1.37 x $10^{-7}$ /1.58 X $10^{-7}$ | $3.1 \times 10^{-10}$/$1.64 \times 10^{-10}$ | [21c] |
| GeSe | 532 nm | $4.841 \times 10^{-6}$ ($cm^2.W^{-1}$) | $2.258 \times 10^{-6}$ | $2.945 \times 10^{-10}$ | [33] |
| Boron NS | CW 457 nm / 532nm /671 nm | $1.25 \times 10^{-5}$ / $3.43 \times 10^{-6}$ / $9.45 \times 10^{-6}$ ($cm^2.W^{-1}$) | $1.75 \times 10^{-7}$ / 0.64 X $10^{-6}$/ 0.48 X $10^{-6}$ (e.s.u) | $4 \times 10^{-9}$/ $1.8 \times 10^{-9}$/ $1.8 \times 10^{-9}$ (e.s.u) | [34] |
| SnS NS | CW 532 nm/ 633 nm | $4.531 \times 10^{-5}$/ $0.323 \times 10^{-5}$ ($cm^2.W^{-1}$) | $2.317 \times 10^{-5}$/$0.165 \times 10^{-5}$ (e.s.u) | $6.995 \times 10^{-10}$ / $2.037 \times 10^{-10}$ (e.s.u) | [35] |
| $Bi_2S_3$ | CW 457nm/ 532 nm/ 671 nm | $3.34 \times 10^{-5}$/ $1.26 \times 10^{-6}$/ $1.62 \times 10^{-7}$($cm^2.W^{-1}$) | ------ | ------- | [36] |
| $MoSe_2$ | CW 532nm | $3.24 \times$ 10-10 $W.m^{-2}$ | -------- | $1.1\times10^{-9}$ (e.s.u) | [37] |
| TaAs | 405 nm/ 532 nm/ 671 nm/ 841 nm | ------- | $6.06 \times 10^{-4}$ / $5.68 \times 10^{-4}$ / $5.30 \times 10^{-4}$/ $4.65 \times 10^{-4}$ (e.s.u) | $10.50 \times 10^{-9}$ / $9.86 \times 10^{-9}$ /$9.19 \times 10^{-9}$ / $8.07 \times 10^{-9}$ (e.s.u) | [38] |
| $NiTe_2$ | CW 650 nm/ 532 nm / 405 nm | $3.22 \times10^{-5}$/ $6.15 \times10^{-5}$ / $7.68 \times 10^{-5}$ ($W.cm^{-2}$) | $1.56 \times 10^{-3}$ / $2.99 \times 10^{-3}$ / $3.76 \times 10^{-3}$ | $4.06 \times 10^{-9}$/ $7.79 \times 10^{-9}$/ $9.89\times 10^{-9}$ (e.s.u.) | [39] |
| Violet Phosphorus NS | CW 405 nm/ 473 nm / 532 nm/ 671 nm/ 721 nm | -- | -- | $3.54\times 10^{-8}$/ $1.65 \times 10^{-8}$/ $9.64 \times 10^{-9}$/ $4.63 \times 10^{-9}$/ $2.31 \times 10^{-9}$ | [40] |
| Hybrid Bismuth Halide (PPA)3BiI6 | CW 405 nm/ 473 nm / 532 nm | -- | -- | $1.77 \times 10^{-8}$/ $1.26\times 10^{-8}$/ $7.74\times 10^{-8}$/ | [41] |
| MoP Microparticles | CW 532 nm | $1.91 \times 10^{-5}$ ($W/cm^2$) | -- | -- | [42] |
| $Fe_{3-x}GeTe_2$, $Fe_{4-x}GeTe_2$, $Fe_{5-x}GeTe_2$ | 671 nm / 532 nm | $2.49 \times 10^{-5}$ ($W.cm^{-2}$)/ $2.16 \times 10^{-5}$ ($W.cm^{-2}$), $1.23 \times 10^{-5}$ ($W.cm^{-2}$)/ $1.09 \times 10^{-5}$ ($W.cm^{-2}$), $0.93 \times 10^{-5}$ ($W.cm^{-2}$)/ $0.92 \times 10^{-5}$ ($W.cm^{-2}$) | -- | $2.75 \times 10^{-8}$/ $2.37 \times 10^{-8}$, $1.57 \times 10^{-8}$ /$1.39 \times 10^{-8}$,$1.3 \times 10^{-8}$ /$1.29 \times 10^{-8}$ | [43] |
| $Bi_2Te_3$ (Our work) | CW 650 nm/ 532 nm / 405 nm | $2.7 \times10^{-4}$/ $8.14 \times10^{-5}$ / $10.1 \times 10^{-5}$ ($W.cm^{-2}$) | $1.56 \times 10^{-3}$ / $4.741\times 10^{-3}$ / $5.88 \times 10^{-3}$ | $1.2 \times 10^{-7}$/ $3.63 \times 10^{-7}$ / $4.51\times 10^{-9}$ (e.s.u.) | |

**Table S2.** SSPM formation time of the diffraction pattern as reported in the relevant literature.

| Material | Laser Specification | Solvent | Intensity | Formation time ($T$) | Ref. |
|---|---|---|---|---|---|
| $MoTe_2$ | 473/ 532/ 750 nm (CW) | NMP | 252 $W.cm^{-2}$ | 0.45 s/ 0.6 s/ 0.62 s | [29] |
| $MoSe_2$ | 671 nm (CW) | NMP/ Acetone | 12 $W.cm^{-2}$ | 0.41 s/ 0.22 s | [30] |
| TaAs | 589 nm/ 532 nm/ 473 nm (CW) | NMP | 90 $W.cm^{-2}$ | 2.5 s/ 2.5 s/ 2.3 s | [38] |
| Graphene Oxide | 532 nm (CW) | IPA | ---- | 0.43 s | [44] |
| Black Phosphorus | 700 nm (CW) | NMP | 18.9 $W.cm^{-2}$ | 0.7 s | [20] |
| Violet Phosphorus | CW 405 nm/ 473 nm / 532 nm/ 671 nm/ 721 nm | NMP | 1.55 $W.cm^{-2}$ / 3.55 $W.cm^{-2}$ / 9.80 $W.cm^{-2}$ / 19.65 $W.cm^{-2}$ / 18.64 $W.cm^{-2}$ | 0.3 s/ 0.47 s/ 0.7s/ 0.83 s/ 0.77 s | [45] |
| MoP Microparticles | CW 532 nm | NMP | 219.8 $W.cm^{-2}$ | 0.4 s | [42] |
| $Fe_{3-x}GeTe_2$, $Fe_{4-x}GeTe_2$, $Fe_{5-x}GeTe_2$ | 671 nm / 532 nm | NMP | 15 $W.cm^{-2}$ | 0.28/ 0.31, 0.16/ 0.23, 0.15/0.18 | [43] |
| $Bi_2Te_3$ | 650 nm/ 532 nm/ 405 nm | NMP | 10.318 $W.cm^{-2}$, 5.39 $W.cm^{-2}$, and 1.244 $W.cm^{-2}$ | 0.266 s, 0.5 s ,0.3999 s | |

## Section S4: Computational Details

We investigated the interaction between the molecules isopropyl alcohol (IPA)[46] and N-methyl-2-pyrrolidone (NMP)[47] with the host material 2D-$Bi_2Te_3$[48] using density functional theory (DFT)[49]. The calculations were conducted with the SIESTA software[50], employing a plane-wave basis set with an energy cutoff of 350 eV. In SIESTA, the Kohn-Sham orbitals were expanded using a double-zeta basis set of numerical pseudoatomic orbitals of finite range, augmented with polarization orbitals. We used optimized pseudopotentials and atomic bases from the SIMUNE database[51] with the Perdew-Burke-Ernzerhof (PBE) approximation for the exchange-correlation functional[52]. The Dion et al. scheme[53], optimized by Klimes et al. (optB88-vdW)[54], was employed to describe dispersive van der Waals (vdW) interactions accurately. Bader charge analysis was conducted to quantify the charge transfer between the IPA and NMP molecules and the host material, 2D-$Bi_2Te_3$.[55]

The total energy convergence threshold for electronic calculation was $1 \times 10^{-4}$ eV. Geometry optimizations were carried out using the conjugate gradient (CG) algorithm, ensuring that the magnitude of the forces acting on each ion was minimized to less than 0.01 eV/Å for the primitive cell of 2D-$Bi_2Te_3$ and less than 0.02 eV/Å for a 4×4×1 supercell. The irreducible Brillouin zone was sampled with a 5×5×1 k-point mesh for the primitive cell and a 2×2×1 k-point mesh for the 4×4×1 supercell[56] .

Before constructing the heterostructure composed of IPA or NMP and 2D-$Bi_2Te_3$, we optimized the primitive cell of 2D-$Bi_2Te_3$ using SIESTA. The unit cell dimensions for the 2D-$Bi_2Te_3$ were found to be a=b=4.40 Å, which are in good agreement with the literature[48], while the lattice parameter perpendicular to the sheets (z-axis) was fixed at c= 30 Å. Subsequently, a 4×4×1 supercell of 2D-$Bi_2Te_3$ was optimized with lattice parameters a = b= 17.62 Å and c = 30 Å. To construct the 2D-$Bi_2Te_3$ and molecules heterostructure, we minimized the total energy concerning the distance between the molecules and the surface in the z-direction. Initially, the molecules were placed approximately 1.5 Å from the 2D-$Bi_2Te_3$ surface, and the distance was incrementally increased by 0.5 Å. After determining the most favorable distance, we optimized the systems by allowing only the molecules to move.

We have also calculated the binding energy (Eb) using the following Equation[57]:

$$E_b = E_{Bi_2Te_{3+Mol}} - (E_{Bi_2Te_3} + E_{Mol}) \text{ ............. (18)}$$

In this expression, $E_{Bi_2Te_{3+Mol}}$ represents the total energy of the system with the molecule (IPA or NMP) adsorbed onto the 2D-$Bi_2Te_3$ surface, $E_{Bi_2Te_3}$ is the total energy of the pristine 2D-$Bi_2Te_3$, and $E_{Mol}$ denotes the energy of the isolated molecule (IPA or NMP).

To investigate charge redistribution during adsorption, we performed a charge density difference analysis (CDD). This involved calculating the charge density of the pristine 2D-$Bi_2Te_3$ surface, the charge density of the surface with the adsorbed molecule ($\rho_{Bi_2Te_3+Mol}$), and the charge density of the isolated molecule (IPA or NMP), using the following Equation[57b, 58] :

$$\Delta\rho = \rho_{Bi_2Te_3+Mol} - (\rho_{Bi_2Te_3} + \rho_{Mol}) \text{............... (19)}$$

Regions where $\Delta\rho > 0$ indicate charge accumulation, while the ones where $\Delta\rho < 0$ indicate charge depletion.

**Section S5. Intensity Dependent Dynamic Collapse of Diffraction Pattern and Variation in Nonlinear Refractive Index with Different Solvent**

Through further examination of the SSPM phenomenon, it becomes clear that the self-diffraction pattern displays a phenomenon of distortion as time progresses. At some point when the diffraction pattern exhibits the maximum vertical diameter, the upper portion of the pattern begins to converge towards the middle region of the diffraction pattern. Vertical distortion is more prominent than horizontal distortion as seen in **Figure S6a**. Initially documented by Wang et al.[59] this distortion of the pattern can be described by non-axis thermal convection, which elucidates the underlying mechanism of the collapse process. The values $R_H$ and $\theta_H$ are designated as the maximum radius and the associated half-cone angle, respectively before the collapse process.[35, 60] The relationship can be represented as,

$$\theta_H = \frac{R_H}{D} \quad \text{.....................} \quad (20)$$

The Equation 20 is valid only if D >> $R_H$. The parameter D defines the distance from the cuvette to the screen. Over time, the thermal convection effect causes distortion in the diffraction pattern, resulting in changes to both the maximum diffraction radius and the half-cone angle, that can be denoted as $R_h$ and $\theta_h$. If $D >> R_h$, therefore, the relationship may be represented as,

$$\theta_h = \frac{R_h}{D} \quad \text{..............................} \quad (21)$$

Hence, the measurement of distortion is expressed in terms of the distortion radius ($R_D$) and the distortion angle ($\theta_D$). The relationship might be characterized as,

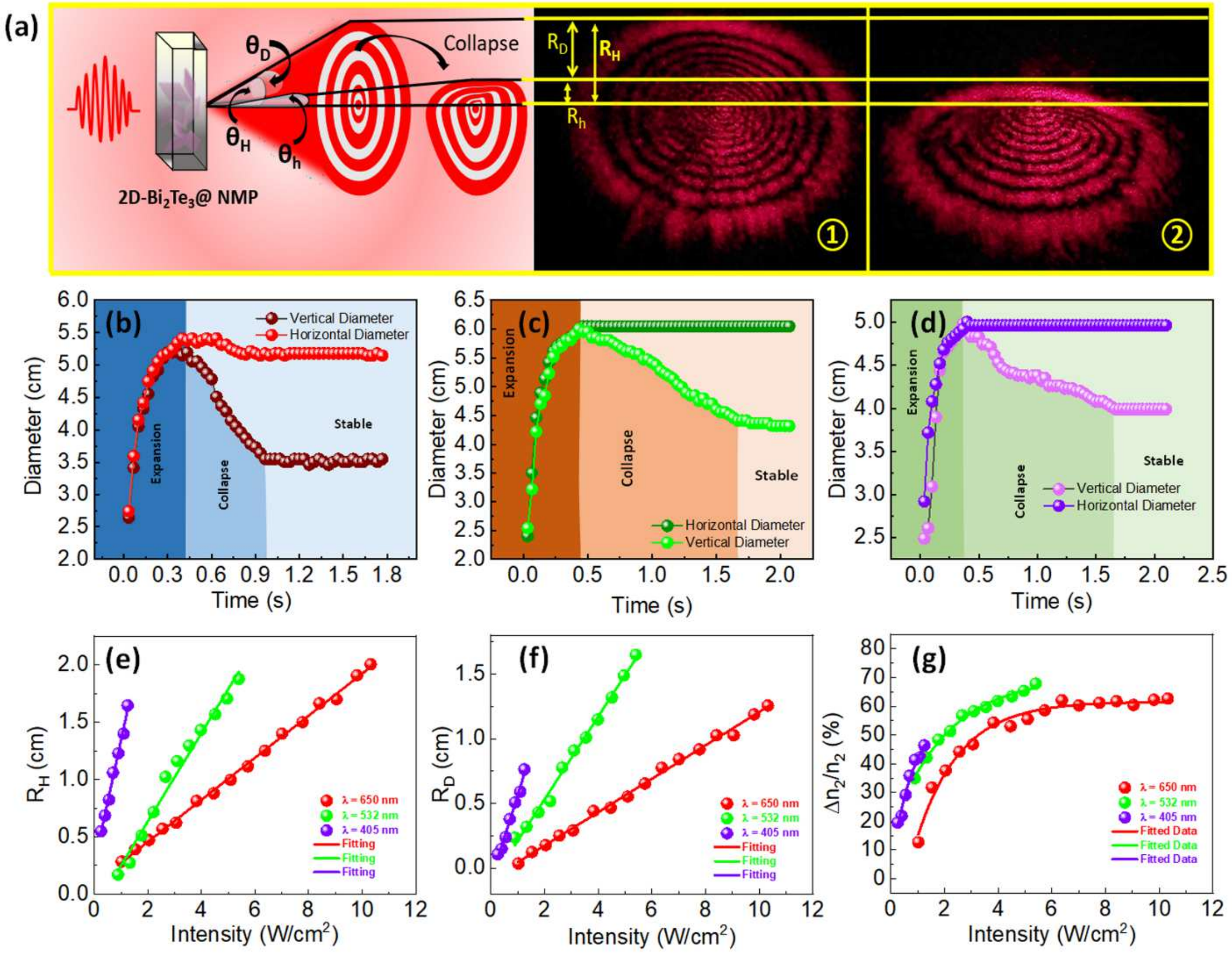

**Figure S6.** Figurative depiction of the dynamic collapse process. a) Graphical depiction of the collapse phenomenon of the SSPM diffraction pattern featuring a semi-cone and distortion angle for 2D $Bi_2Te_3$ in NMP solvent. b-c-d) The temporal evolution of the vertical and horizontal diameters over various wavelengths (λ = 650, 532, and 405 nm) in NMP solvent. e) The variation in maximum radius ($R_H$) with intensity across different wavelengths (λ = 650, 532, and 405 nm) in NMP solvent. f) The change in the distorted radius ($R_D$) with changing intensity for different wavelengths (λ = 650, 532, and 405 nm) in NMP solvent. g) The analysis of the variation of $n_2$ (nonlinear refractive index) with varying intensity across different wavelengths (λ = 650, 532, and 405 nm) in NMP solvent.

$$\theta_H - \theta_h = \frac{R_D}{D} \quad (22)$$

Furthermore, the half-cone angle may be approximated as,

$$\theta_H = \frac{\lambda}{2\pi}\left(\frac{d\psi}{dr}\right)_{max}, r\in[0,+\infty] \quad (23)$$

Here r represents the transverse position of the beam. For a Gaussian beam, the above Equation can be $\theta_H$ can also be expressed as,[61]

$$\theta_H = n_2 IC \quad (24)$$

Where $C = [-\frac{8rL_{eff}}{\omega_0^2}\exp(-\frac{2r^2}{\omega_0^2})\ ]$ is constant when $r\in[0,+\infty)$. Figure S6a illustrates the evolution of the diffraction pattern for the wavelength 650 nm. The variance in distortion may be quantified by analyzing the angle of distortion that occurs between the sample and the screen, with the degree of distortion being dependent on the intensity.[23] Consequently, we may determine the collapse angle as follows,

$$\theta_D = \theta_H - \theta_h = (n_2 - n_2')IC = \Delta n_2 IC \quad (25)$$

The distortion observed in the top region of the diffraction pattern is a result of the non-uniform fluid motion produced by heat, which lacks symmetry along the axis and causes the collapse of the diffraction pattern.[59] The thermal convection occurs as a result of the laser beam's propagation, causing the solvent to heat up due to its absorption coefficient is constant.[23, 62] Due to indirect effect of laser heating, a temperature gradient is created at a right angle to the axis of the laser point, which enhances the process of 'thermal convection'. Based on the information provided in Equation 24 and Equation 25, it can be inferred that,

$$\frac{\Delta n_2}{n_2} = \frac{\theta_D}{\theta_H} = \frac{R_D}{R_H} \quad (26)$$

Using Equation 26, we may determine the relative change of the nonlinear refractive index by observing the dynamic change( $\frac{\theta_D}{\theta_H}$) or ( $\frac{R_D}{R_H}$) in the distortion of the diffraction ring. The value of $\Delta n_2/n_2$ is influenced by various factors such as laser wavelength, temperature, and duration.[63] Although the value of $\Delta n_2/n_2$ is primarily dependent on the incoming laser intensity.[33, 61] The value of $\Delta n_2/n_2$ can be determined using mathematical computation.

Figure S6a(①-②) displays the image captured by the CCD camera. The diffraction pattern exhibits the maximum number of rings with the greatest vertical diameter and the condition of equilibrium following the collapse phenomena, when the laser beam of the wavelength of 650 nm. Figure S6b–d illustrates the temporal development of the diffraction ring pattern with progression of time for three laser beams with wavelengths (λ) values of 650, 532, and 405 nm, respectively, at their maximum laser intensity. All three lasers exhibited a progressive rise in horizontal diameters with respect to time until reaching their maximum diameter, after which they stayed constant subsequently. Simultaneously, the vertical diameters attain their peak magnitude and thereafter gradually diminish over time to reaching certain constant value. After reaching a certain value of ring number the thermal convection process becomes prominent and distortion in the diffraction pattern starts to appear. This distortion in the vertical direction becomes prominent with time. Experimentally determined maximum vertical radius during pre-collapse and distorted value of radius are shown in Figure S6e-f respectively. The observed relative shift in the SSPM diffraction ring was attributed to the heat convection mechanism. The quantification of this distortion is achieved in terms of the nonlinear refractive index, as described by Equation 26. This distortion index expressed as $\Delta n_2/n_2$, increases with the increment of the value of intensity. Distortion index ($\Delta n_2/n_2$) is quantified in percentage for three laser beams with wavelengths (λ) values of 650, 532, and 405 nm as shown in Figure S6g. The distortion or relative nonlinear refractive index is found to be 62.16 % (650 nm), 66.41% (532 nm), and 46.48% (405 nm) for intensities 10.318 $Wcm^{-2}$, 5.39 $Wcm^{-2}$, and 1.244 $Wcm^{-2}$.

The 532 nm laser induces more distortion compared to the 650 nm laser at similar intensity levels, due to each photon in the 532 nm laser beam possessing greater energy than those in the 650 nm laser. The extent of distortion caused by 405 nm is observed to be the most significant. However, the increase in distortion attains its peak once the laser beam's intensity surpasses a specific threshold.

## Section S6. Electronic Relationship Between $\chi^{(3)}_{monolayer}$, Mobility ($\mu$), and Effective Mass ($m^*$)

In addition, an experiment is conducted to determine the source of the SSPM pattern generation by establishing a correlation between $\chi^{(3)}$ and mobility, as well as $\chi^{(3)}$ and effective mass. Hu et. al. introduced a technique to establish a relationship between the electronic coherence phenomena and the carrier mobility and effective mass.[29] Carrier mobility denotes the ability of a carrier to react to an external electric field through movement. The dynamics of the carrier within the ab plane of the unit cell are influenced by the effective mass and the dispersion characteristics of the carrier. The high value of $\chi^{(3)}_{monolayer}$ is a consequence of the reduced presence of electron scattering in the 2D material. Therefore, based on this reasoning, the optical coefficient $\chi^{(3)}$ is associated with electrical characteristics such as effective mass ($m^*$) and carrier mobility ($\mu$). In this study, we have investigated a correlation between the value of $\chi^{(3)}$ acquired from the SSPM experiment and the values of mobility and effective mass of 2D $Bi_2Te_3$ as reported in the relevant literature.[64] The value of other included materials are taken from relevant literature, also the value of the μ and $m^*$ are derived Supporting information in Table S3. The reported value of $\chi^{(3)}$ ($10^{-7}$ e.s.u) and its relationship to electronic activity is better than that of Graphene, as seen in **Figure S7a** and Figure S7b. It may be inferred from Figure S6b that the carrier type of the semiconductor nanostructure can be identified using Equation 28. The relationship in the termed can be expressed as, $\chi^{(3)}$ vs $\mu$ and $\chi^{(3)}$ vs $m^*$,[30] i.e.

$$\chi^{(3)} = 8.00/\sqrt{m^*} \quad \text{.............................. (27)}$$

And,

$$\chi^{(3)} = 0.146 \times \sqrt{\mu} \quad \text{......................... (28)}$$

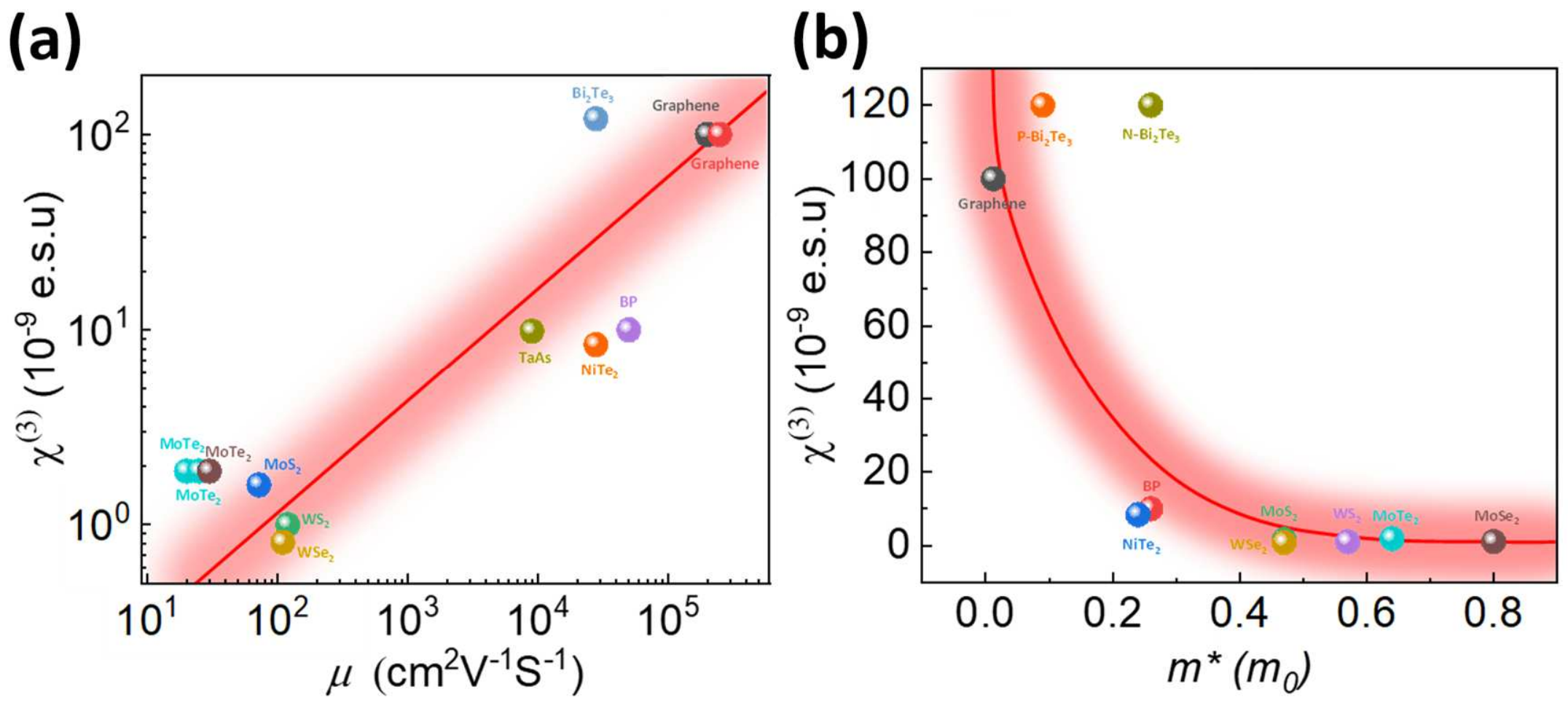

**Figure S7.** Comparative study between optical and electronic coefficients. a) $\chi^{(3)}(10^{-9}\ e.s.u)$ versus $\mu$. b) $\chi^{(3)}(10^{-9}\ e.s.u)$ versus effective mass ($m^*$) of several 2D materials.

**Table S3:** The values of $\chi^{(3)}$, Mobility (μ) & Effective Mass ($m^{*}$)

| Material & Corresponding Wavelength | $\chi^{(3)}_{monolayer}$ (Third-order nonlinear susceptibility) | Mobility (μ) & Effective Mass ($\boldsymbol{m}^{*}$) | Reference |
|---|---|---|---|
| Graphene | $1 \times 10^{-3}$ (e.s.u) | [65] | [19, 59] |
| BP | $10^{-5}$ (e.s.u) | [66] | [20] |
| $MoS_2$ | $1.44 \times 10^{-4}$ (e.s.u) | [67] | [21a] |
| $WSe_2$ | $1.371 \times 10^{-6}$/ $4.04 \times 10^{-6}$/$2.98 \times 10^{-6}$ (e.s.u) | [68] | [18] |
| TaAs (405 nm/ 532 nm/ 671 nm/ 841 nm) | $6.06 \times 10^{-4}$ / $5.68 \times 10^{-4}$ / $5.30 \times 10^{-4}$/ $4.65 \times 10^{-4}$ (e.s.u)/ $6.06 \times 10^{-4}$ / $5.68 \times 10^{-4}$ / $5.30 \times 10^{-4}$/ $4.65 \times 10^{-4}$ (e.s.u) | [38] | [38] |
| $MoSe_2$ (532 nm) | $1.76 \times 10^{-4}$ (e.s.u) | [69] | [23] |
| $WS_2$ | $8.14 \times 10^{-10}$/ $8.44 \times 10^{-11}$ / $3.69 \times 10^{-9}$ (e.s.u) | [70] | [61] |
| $MoTe_2$ | $1.88 \times 10^{-9}$ e.s.u<br>$1.3 \times 10^{-9}$ e.s.u<br>$1.14 \times 10^{-9}$ e.s.u<br>$0.98 \times 10^{-9}$ e.s.u<br>(CW 473 nm /532 nm/ 750 nm/ 801 nm) | [71] | [29] |

## Section S7:

This study focuses on the nonlinear optical characteristics of 2D $Bi_2Te_3$ through the application of SSPM Spectroscopy. The light source employed in the SSPM experiment is a continuous wave laser. According to recent findings, the SSPM does not arise from the thermally induced response. The following paragraphs will elaborate on the supporting arguments. In the Supporting Information Section S6 Electronic Relation between $\boldsymbol{\chi^{(3)}_{monolayer}}$), Mobility (μ), and Effective Mass ($\boldsymbol{m}^{*}$), we have examined the relationship between effective mass and carrier mobility. The mobility of the carrier is characterized by its capacity to traverse in the presence of an external field. This relies on the effective mass and the scattering behavior of the carriers. In the scenario of third-order nonlinear optical response, the mobility of carriers within the coherent laser field establishes the ac nonlocal electronic coherence. Hu et al.[29] anticipated that $\boldsymbol{\chi^{(3)}_{monolayer}}$ shows a positive correlation with carrier mobility. Both parameters indicate the ability to store charge, reflecting the observed energy storage mechanism, while $\boldsymbol{\chi^{(3)}_{monolayer}}$ shows a negative correlation with the effective mass of the carriers. The effective mass represents a dissipative characteristic, highlighting the mechanism through which energy is released. The light field improves the motion of the carriers, resulting in reduced scattering during their movement, which leads to a higher value of $\boldsymbol{\chi^{(3)}_{monolayer}}$ . The elevated value of $\chi^{(3)}_{monolayer}$ rises as the wavelength decreases, corresponding to an increase in photon energy with shorter wavelengths. The main manuscript presents a summary of the values of $\chi^{(3)}_{monolayer}$ assessed using the SSPM Spectroscopy method in relation to carrier mobility μ and effective mass $\boldsymbol{m}^*$. $\chi^{(3)}_{monolayer}$ represents an optical parameter, μ denotes an electronic property, and $\boldsymbol{m}^*$ pertains to the electronic structure in the excited state. The relationship between optical properties and electronic properties reinforces the phenomenon of laser-induced coherence.

In addition to the Kerr nonlinearity, fluctuations in the medium's temperature caused by a powerful laser beam can modify the refractive index, leading to effects akin to self-phase modulation as documented earlier. Dabby et al.[72] characterized this phenomenon as the "thermal lens effect." The study we have conducted focuses on the "Wind Chime Model," which relies on the polarization of suspended nanostructures and their subsequent reorientation when subjected to a laser beam. The thermal lens effect demonstrates a linear optical response.[73] The mechanical chopper is utilized to validate the previously stated assertion. the phenomenon of coherence. Figure S8a illustrates the optical configuration that includes a mechanical chopper. The chopper is first operated within a frequency range of 50 to 200 Hz (50 Hz interval) at an intensity of 11.45 W/cm². At the same intensity, the chopper was run at a higher frequency of 0.5-3.75 kHz (250 Hz interval) to examine the variation in the number of rings; however, no significant change was seen. For high chopping frequency from 0.5-3.75 kHz no change in number of rings is observed hence indicating similar level of electronic excitation. This same experiment was performed for lower frequencies 50-200 Hz (with 25 Hz step increment). Even in lower frequencies of mechanical chopper no change in the number of rings of the diffraction pattern was observed with increasing frequency. Therefore, we may conclude that the phenomena of electronic coherence, which induces a nonlinear optical response, predominates over the thermal lens effect.

The thermal camera was utilized to record the thermal profile under the same intensity in the configuration illustrated in Figure S9a at different frequency. The thermal interaction of the laser with the solvent is measured at various chopping frequencies, where the emergence of the "Thermal lens effect" is anticipated. The temperature difference is measured before and after laser impact at frequencies of 0 kHz, 1 kHz, and 2 kHz. Figure S9b①, c①, d①, and e① illustrate the upper view of the cuvette prior to laser impact at frequencies of 0 kHz, 1 kHz, and 2 kHz, respectively. Figures S9b②, c②, d②, and e② illustrate the upper view of the cuvette following laser impact at frequencies of 0, 0.5, 2, and 3.5 kHz, respectively. Figure S9g presents the measurement of temperature difference prior to and following laser impact, incorporating chopper frequency. As the chopping frequency increases, the interaction of the laser with the solvent diminishes, resulting in reduced localized heating of the solvent. A decrease in temperature is observed with an increase in chopping frequency. It was observed that after each increase of 0.25 kHz chopping frequency, the decrease in temperature difference is observed. The number of rings remains constant despite the increase in chopping frequency, as illustrated in Figure S8b ①-⑮. The initial reduction in the number of rings is attributed to the diminished intensity of the laser beam, resulting in decreased electronic coherence and a lower quantity of rings produced. The decrease in temperature did not alter the ring number, as illustrated in Figure S9f, indicating that the SSPM phenomenon is not thermally induced. The aforementioned statements offer a comprehensive rationale indicating that SSPM is not a thermally induced phenomenon resulting from the "thermal lens effect".

Similar experiment was performed using 532 nm, to observe whether previously drawn conclusion can be applied here also. The thermal interaction of the laser with the solvent is assessed at different chopping frequencies, where the occurrence of the "Thermal lens effect" is expected. The temperature variation is assessed before to and after to laser application at frequencies of 0 kHz, 1 kHz, and 2 kHz. Figures S11a①, b①, c①, and d① show the top view of the cuvette before laser exposure at frequencies of 0 kHz, 1 kHz, and 2 kHz, respectively. Figures S11a②, b②, c②, and d② show the superior perspective of the cuvette subsequent to laser exposure at frequencies of 0, 0.5, 2, and 3.5 kHz, respectively. Figure S11f illustrates the temperature differential measured before and after laser application, including chopper frequency. As the chopping frequency escalates, the contact between the laser and the solvent lessens, leading to a decrease in localized heating of the solvent. Although a decrease in temperature difference was observed but the number of rings stayed same as seen in Figure S11e. The number of rings with variable chopping frequency is presented in Figure S9, with chopping frequency documented in between 0-3.75 kHz with interval of 250 Hz.

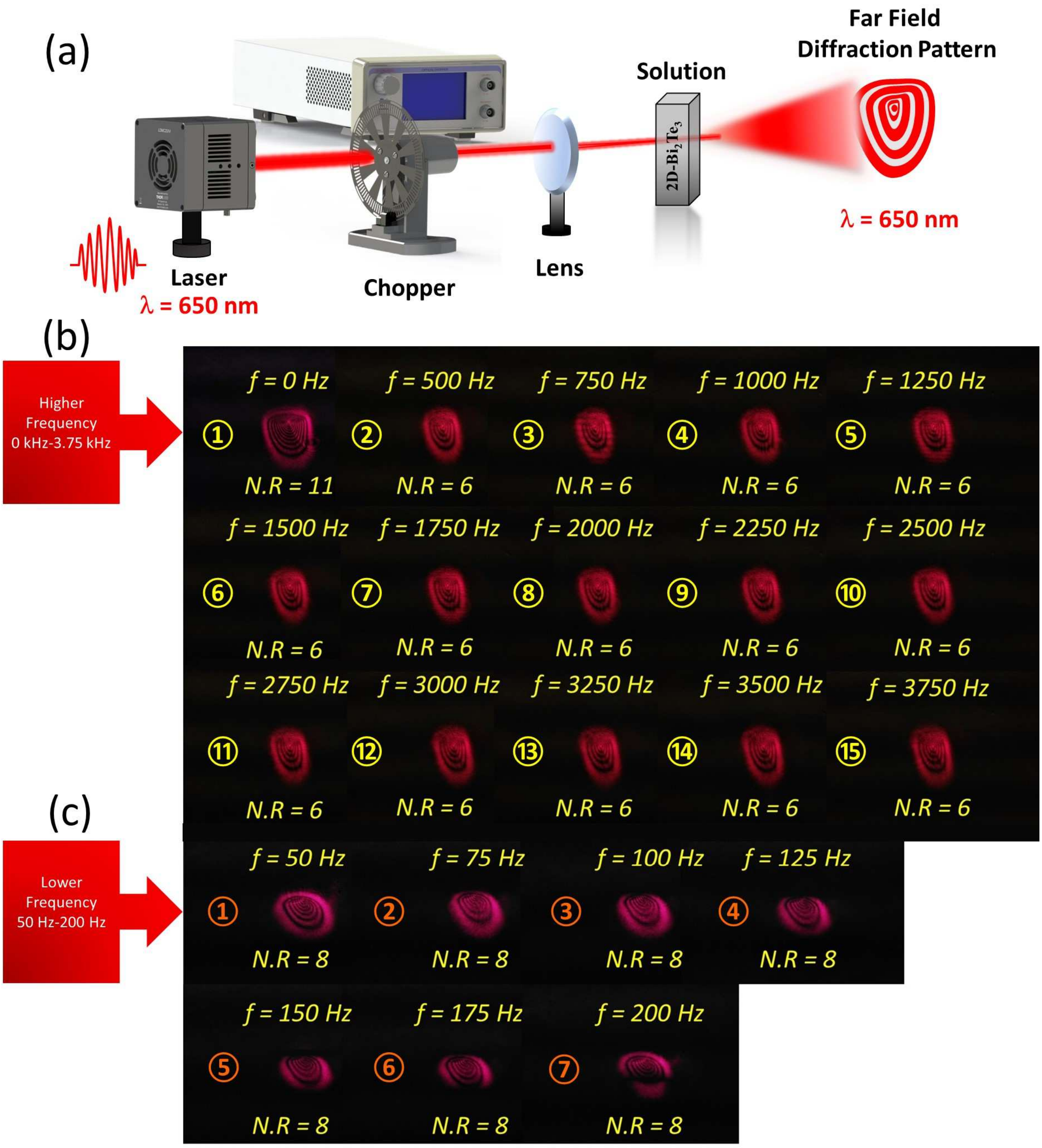

Figure S8. a) Schematic representation of the SSPM setup. b-c) mechanical chopper operating between high Frequency (0.5-3.75 kHz) and low frequency (0-200 Hz).

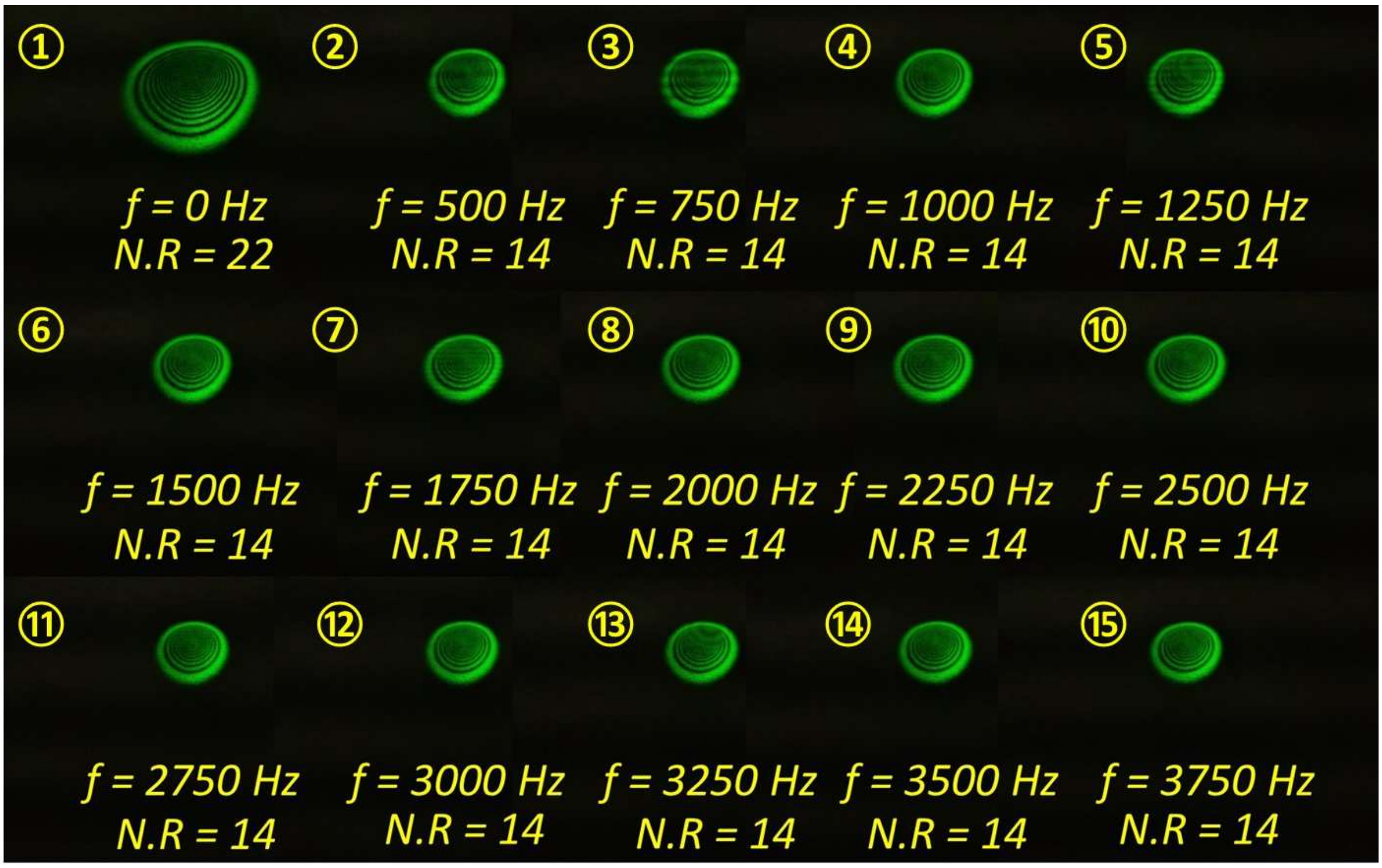

Figure S9. Mechanical chopper operating between high Frequency for 532 nm. (0-3.75 kHz).

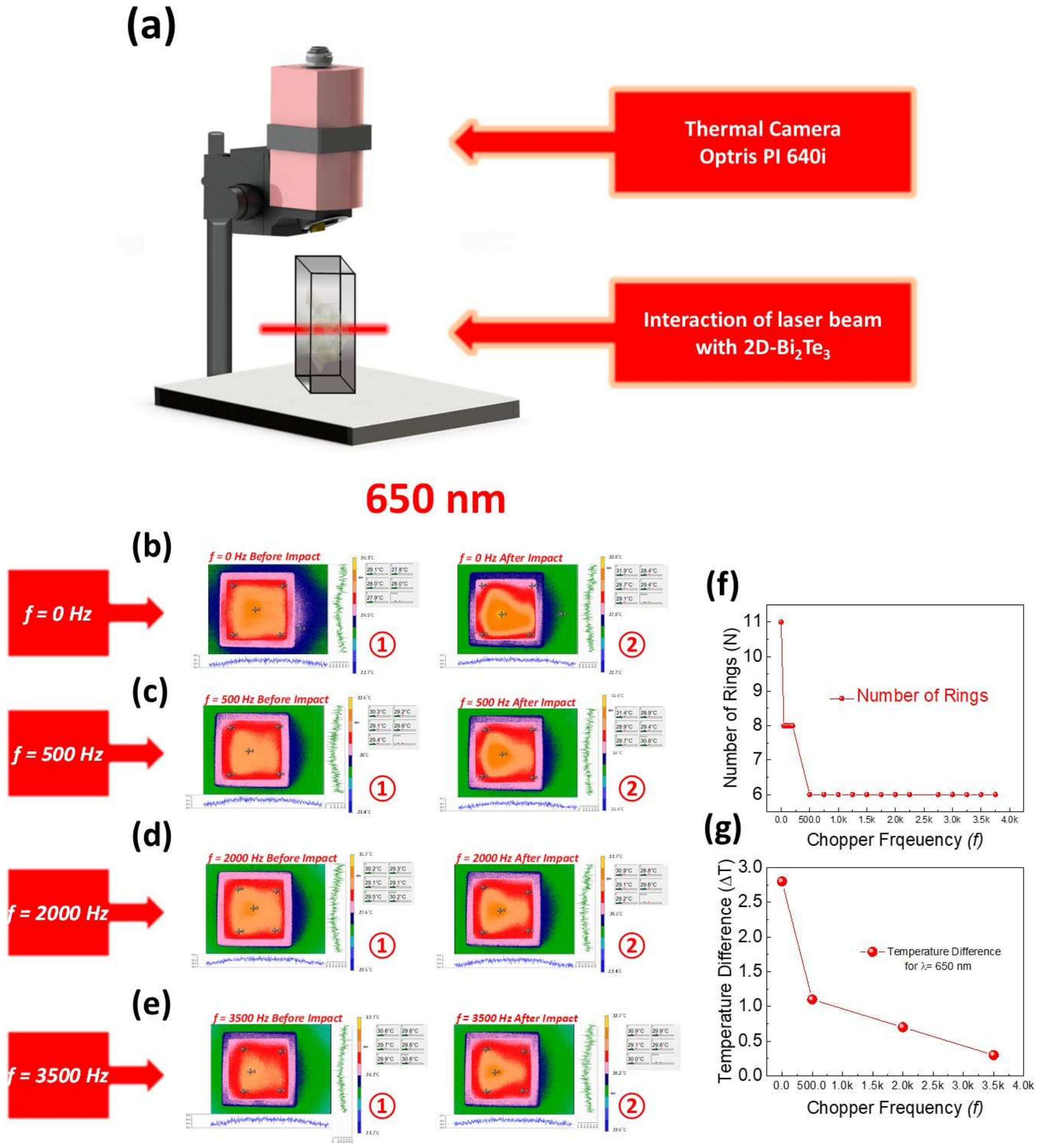

Figure S10. (a) Thermal camera setup for measurement of the upper view of the thermal profile of the cuvette. (b)- (c) -(d)- (e) Thermal profile of the cuvette before and after the laser impact for frequencies 0 kHz, 500 Hz, 2 kHz, and 3.5 kHz at wavelength 650 nm. (f) Number of Rings vs frequency of the mechanical chopper. (g) The temperature difference before and after laser impact of the incoming laser beam.

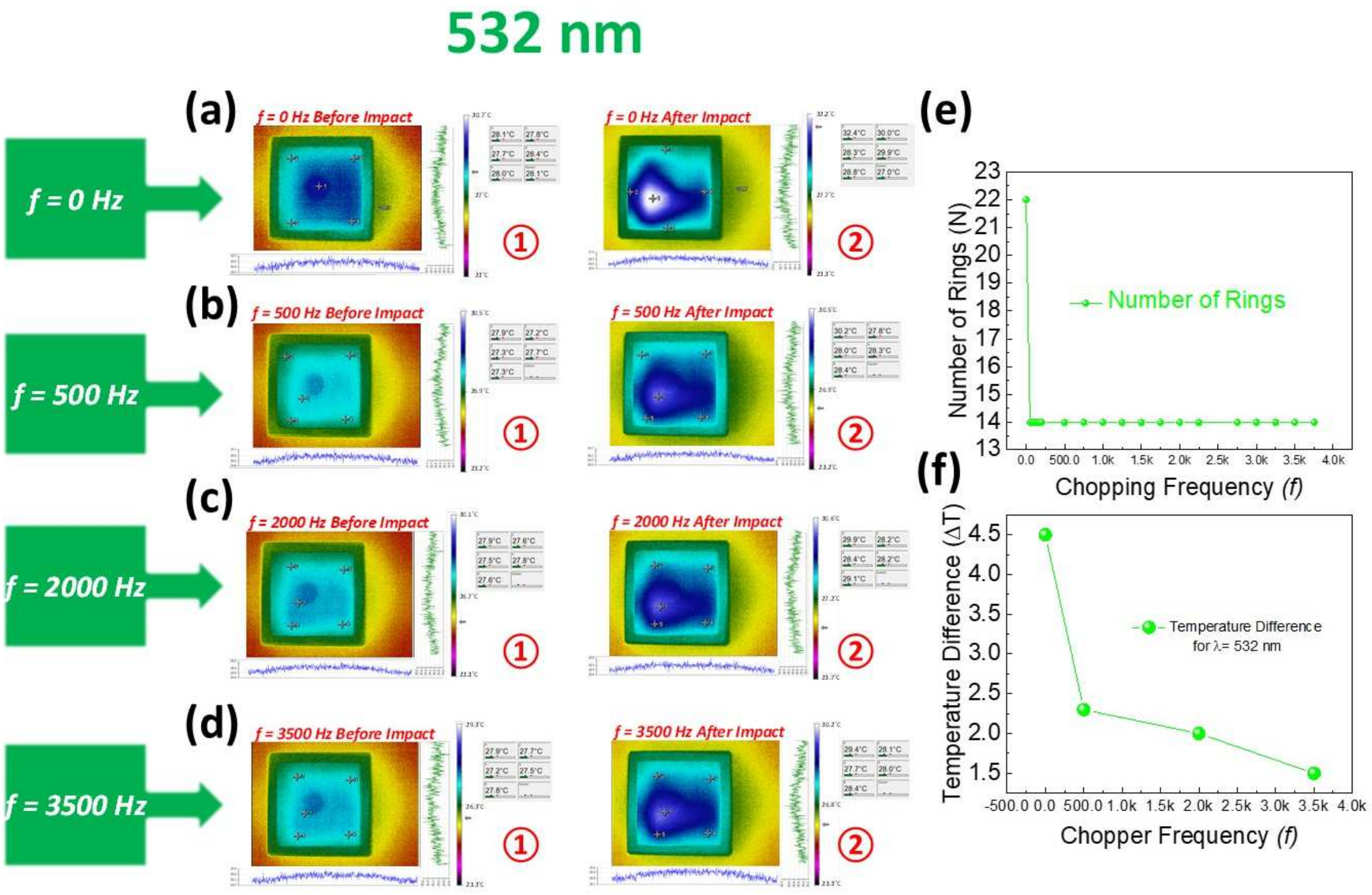

**Figure S11.** (a)-(b)- (c) -(d) Thermal profile of the cuvette before and after the laser impact for frequencies 0 kHz, 500 Hz, 2 kHz, and 3.5 kHz at wavelength 532 nm. (e) Number of Rings vs frequency of the mechanical chopper. (f) The temperature difference before and after laser impact of the incoming laser beam for wavelength 532 nm.

**Section S8: The bandgap calculation of 2D-hBN using Tauc method**

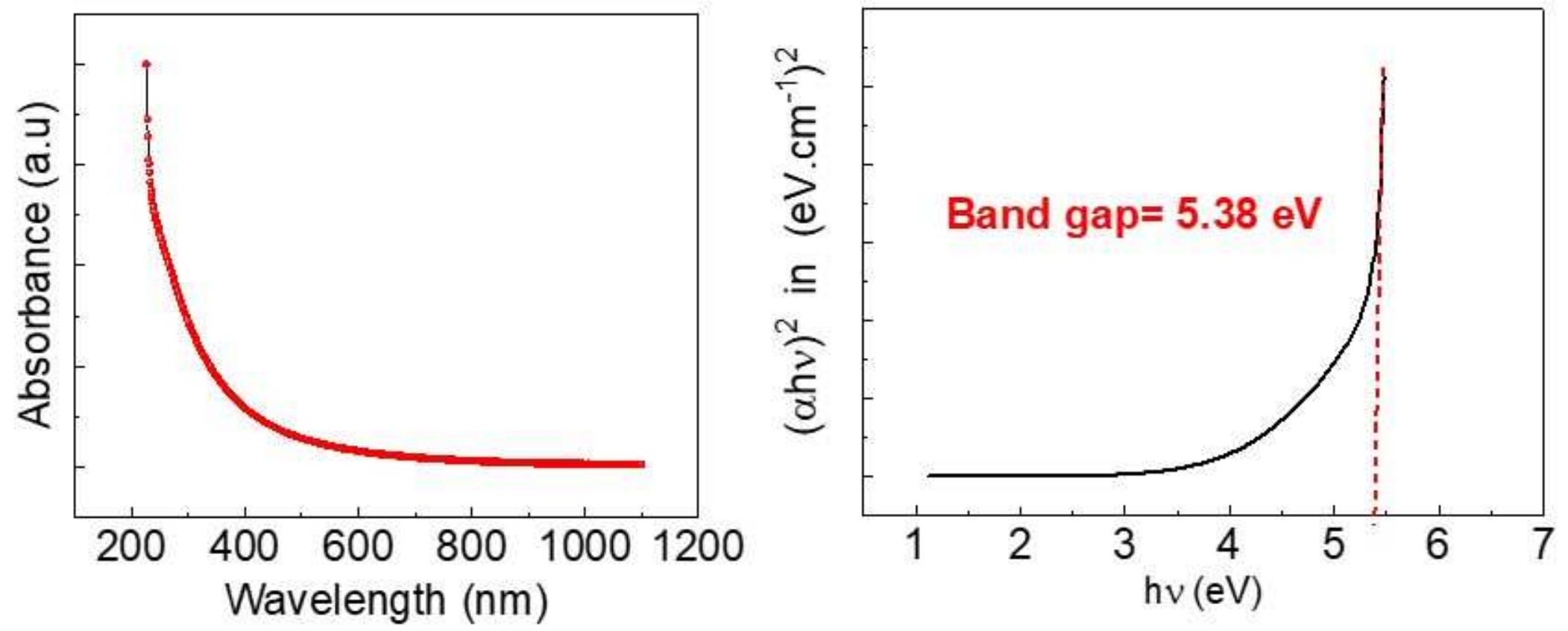

Figure S12. a) showing absorbance spectrum of the 2D-hBN derived from UV-vis Spectroscopy. Figure b) showing determined value of direct bandgap 5.38 eV using Tauc plot.

## Section S9: Laser Beam Solvent Control Experiment

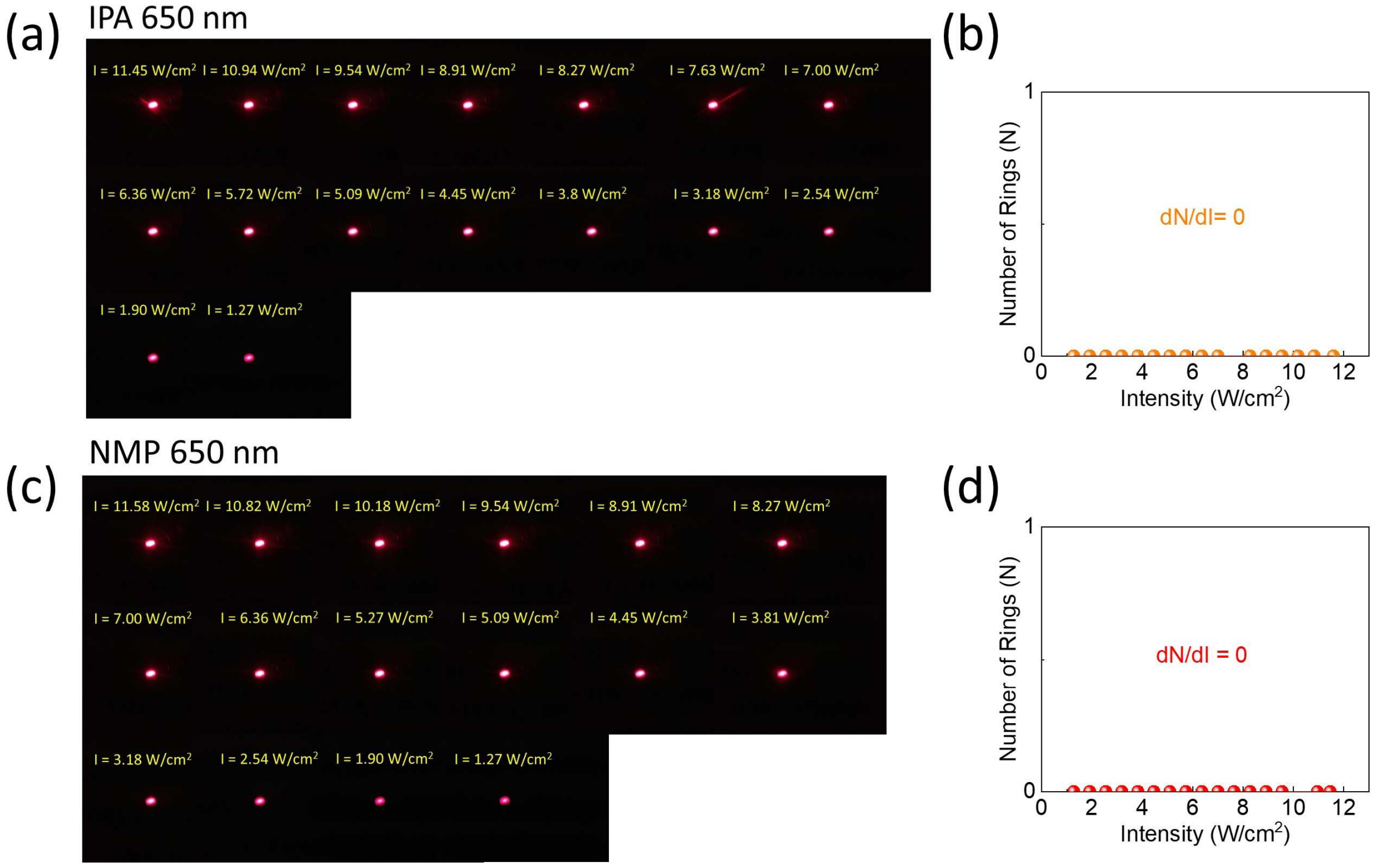

Figure S13. a) Outgoing laser beam recorded at far-screen for after laser beam and IPA interaction. b) corresponding value of dN/dI calculated through the experiment IPA-laser control experiment. c) Outgoing laser beam recorded at far-screen for after laser beam and NMP interaction. d) corresponding value of dN/dI calculated through the experiment NMP-laser control experiment.

## Section S10: The Similar Comparison Method- $n_2$ estimation

The Similar Comparison Method has been implemented to enhance the selection of specific wavelengths (650, 532, and 405 nm). Recent reports indicate the utilization of wavelengths 671, 532, 457, and 405 nm for a similar comparison method. A comparable method can be employed to determine the value of $n_2$ for $Bi_2Te_3$-based photonic diodes and other analogous semiconducting nanostructures, where nonreciprocal light propagation has been realized. The nonlinear refractive index can be defined as,

$$n_2 = \frac{\lambda}{2n_0 L_{eff}} . \frac{dN}{dI}$$

Here $\lambda/2n_0L_{eff}$ is a constant, λ, $n_0$, $L_{eff}$ are the waelength of the incoming laser beam, linear refractive index, and effective optical path length of the incoming laser beam inside the cuvette. Estimating the nonlinear refractive index of a 2D $Bi_2Te_3$-based photonic diode requires a comparative analysis with other materials that possess a well-established nonlinear refractive index.

Similar contrast (S) is defined as,

$$S = 1 - D = \frac{|n_{21} - n_{22}|}{n_{21}}$$

$$S = 1 - \frac{\left|\frac{\lambda}{2n_0L_{eff}}\frac{N_1}{I_1} - \frac{\lambda}{2n_0L_{eff}}\frac{N_2}{I_2}\right|}{\frac{\lambda}{2n_0L_{eff}}\frac{N_1}{I_1}} = 1 - \frac{\left|\frac{N_1}{I_1} - \frac{N_2}{I_2}\right|}{\frac{N_1}{I_1}}$$

D denotes the difference contrast, while $n_{21}$ and $n_{22}$ signify the nonlinear refractive index of the heterostructure acquired under forward bias and reverse bias conditions, respectively. The contrast is measured in relation to other 2D materials, with the coefficients of the 2D materials documented in Table S8.

Table S8: The value of material, nonlinear index, and Similar contrast calculated using SSPM Spectroscopy.

| Material | Nonlinear Refractive Index ($cm^2W^{-1}$) | Similar Contrast (%) | Reference |
|---|---|---|---|
| $MoS_2$ | $\approx 10^{-7}$ | 74% | [61] |
| $Bi_2Se_3$ | $\approx 10^{-9}$ | 58% | [31] |
| SnS | $\approx 10^{-5}$ | 90% | [35] |
| Sb | $\approx 10^{-6}$ | 85% | [26] |
| Graphdiyne | $\approx 10^{-5}$ | 92% | [21b] |
| CuPc | $\approx 10^{-6}$ | 75% | [23] |
| Graphene | $\approx 10^{-5}$ | 95% | [23] |
| $SnS_2$ | $\approx 10^{-9}$ | 54% | [21b] |
| $NiTe_2$ | $\approx 10^{-5}$ | 91.8% | [74] |
| $Bi_2Te_3$ | $\approx 10^{-4}$ | 98.3% | |

Figure S14 indicates that the 2D $NiTe_2$ demonstrates a nonlinear refractive index similar to that of Sb and SnS, with a $n_2$ range of $10^{-5}$ $cm^2$ $W^{-1}$. This was previously validated through experimentation using SSPM spectroscopy. This demonstrates the reliability and repeatability of the SSPM experiment. The 2D-$Bi_2Te_3$ demonstrates a significant contrast of approximately 98.3%.

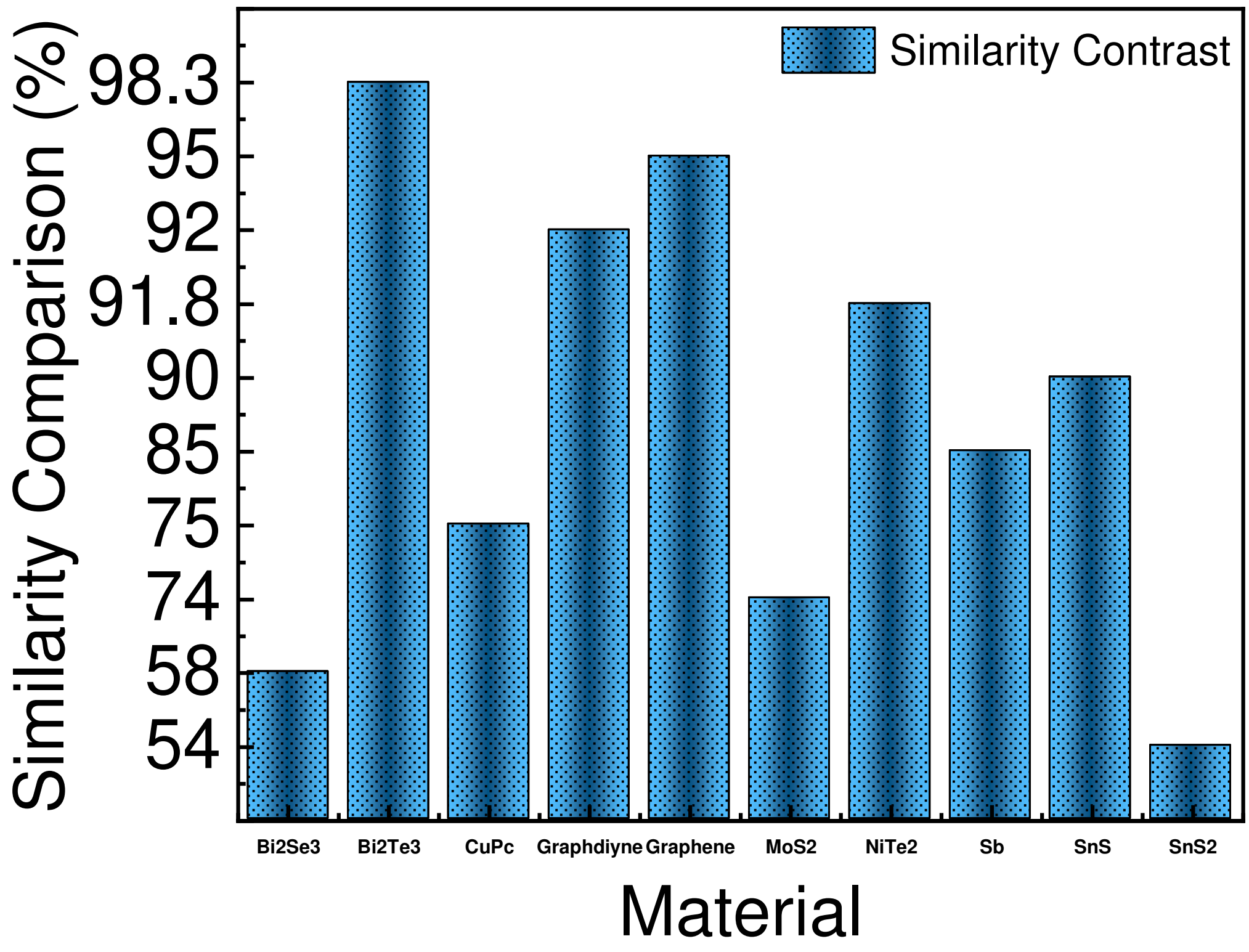

Figure S14: Similarity comparison method for $n_2$ estimation in all photonic diode

## References

[1] T. Sharifi, S. Yazdi, G. Costin, A. Apte, G. Coulter, C. Tiwary, P. M. Ajayan, *Chemistry of Materials* **2018**, 30, 6108.

[2] H. M. Rietveld, *Applied Crystallography* **1969**, 2, 65.

[3] a)P. Ngabonziza, R. Heimbuch, N. De Jong, R. Klaassen, M. Stehno, M. Snelder, A. Solmaz, S. Ramankutty, E. Frantzeskakis, E. Van Heumen, *Physical Review B* **2015**, 92, 035405; b)L. A. Reyes-Verdugo, M. d. J. Martinez-Carreon, C. Gutierrez-Lazos, F. J. Solis-Pomar, J. G. Quiñones-Galvan, E. Perez-Tijerina, *Journal of Nanotechnology* **2024**, 2024, 6623255; c)R. Galceran, F. Bonell, L. Camosi, G. Sauthier, Z. M. Gebeyehu, M. J. Esplandiu, A. Arrighi, I. Fernández Aguirre, A. I. Figueroa, J. F. Sierra, *Advanced Materials Interfaces* **2022**, 9, 2201997; d)P. Ngabonziza, M. P. Stehno, H. Myoren, V. A. Neumann, G. Koster, A. Brinkman, *Advanced electronic materials* **2016**, 2, 1600157; e)X. Zhang, X. Liu, C. Zhang, S. Peng, H. Zhou, L. He, J. Gou, X. Wang, J. Wang, *ACS nano* **2022**, 16, 4851.

[4] J. Yadav, M. Anoop, N. Yadav, N. S. Rao, F. Singh, T. Ichikawa, A. Jain, K. Awasthi, R. Singh, M. Kumar, *Journal of Materials Science: Materials in Electronics* **2023**, 34, 175.

[5] a)B. Trawiński, B. Bochentyn, M. Łapiński, B. Kusz, *Thermochimica Acta* **2020**, 683, 178437; b)H. Bando, K. Koizumi, Y. Oikawa, K. Daikohara, V. Kulbachinskii, H. Ozaki, *Journal of Physics: Condensed Matter* **2000**, 12, 5607; c)P. Kumar, P. Srivastava, J. Singh, R. Belwal, M. K. Pandey, K. Hui, K. Hui, K. Singh, *Journal of Physics D: Applied Physics* **2013**, 46, 285301.
[6] a)M. Ahmad, K. Agarwal, B. Mehta, *Journal of Applied Physics* **2020**, 128; b)S. M. Abzal, S. L. Janga, Y. Bhaskara Rao, S. Khatua, K. Kalyan, P. Maiti, R. Patel, L. N. Patro, J. K. Dash, *Journal of Materials Science* **2024**, 59, 6879.
[7] a)B. Long, Z. Qiao, J. Zhang, S. Zhang, M.-S. Balogun, J. Lu, S. Song, Y. Tong, *Journal of Materials Chemistry A* **2019**, 7, 11370; b)A. Ghosh, S. Shukla, M. Monisha, A. Kumar, B. Lochab, S. Mitra, *ACS Energy Letters* **2017**, 2, 2478.
[8] E. A. Hoffmann, T. Körtvélyesi, E. Wilusz, L. S. Korugic-Karasz, F. E. Karasz, Z. A. Fekete, *Journal of Molecular Structure: THEOCHEM* **2005**, 725, 5.
[9] R. Zheng, W. Cheng, E. Wang, S. Dong, *Chemical physics letters* **2004**, 395, 302.
[10] a)D. Teweldebrhan, V. Goyal, A. A. Balandin, *Nano letters* **2010**, 10, 1209; b)C. Wang, X. Zhu, L. Nilsson, J. Wen, G. Wang, X. Shan, Q. Zhang, S. Zhang, J. Jia, Q. Xue, *Nano Research* **2013**, 6, 688.
[11] W. Richter, C. Becker, *physica status solidi (b)* **1977**, 84, 619.
[12] H. Xu, Y. Song, Q. Gong, W. Pan, X. Wu, S. Wang, *Modern Physics Letters B* **2015**, 29, 1550075.
[13] V. Goyal, D. Teweldebrhan, A. A. Balandin, *Applied Physics Letters* **2010**, 97.
[14] a)G. D. Keskar, R. Podila, L. Zhang, A. M. Rao, L. D. Pfefferle, *The Journal of Physical Chemistry C* **2013**, 117, 9446; b)M. Hajlaoui, E. Papalazarou, J. Mauchain, G. Lantz, N. Moisan, D. Boschetto, Z. Jiang, I. Miotkowski, Y. Chen, A. Taleb-Ibrahimi, *Nano letters* **2012**, 12, 3532; c)Y. Zhao, X. Luo, J. Zhang, J. Wu, X. Bai, M. Wang, J. Jia, H. Peng, Z. Liu, S. Y. Quek, *Physical Review B* **2014**, 90, 245428.
[15] V. Wagner, G. Dolling, B. Powell, G. Landweher, *physica status solidi (b)* **1978**, 85, 311.
[16] Y. Liang, W. Wang, B. Zeng, G. Zhang, Y. Song, X. Zhang, J. Huang, J. Li, T. Li, *Solid State Communications* **2011**, 151, 704.
[17] W. G. Z. S. U. FA, *Appl. Phys. Lett* **2014**, 104, 141909.
[18] Y. Jia, Y. Shan, L. Wu, X. Dai, D. Fan, Y. Xiang, *Photonics Research* **2018**, 6, 1040.
[19] R. Wu, Y. Zhang, S. Yan, F. Bian, W. Wang, X. Bai, X. Lu, J. Zhao, E. Wang, *Nano letters* **2011**, 11, 5159.
[20] J. Zhang, X. Yu, W. Han, B. Lv, X. Li, S. Xiao, Y. Gao, J. He, *Optics letters* **2016**, 41, 1704.
[21] a)Y. Wu, Q. Wu, F. Sun, C. Cheng, S. Meng, J. Zhao, *Proceedings of the National Academy of Sciences* **2015**, 112, 11800; b)L. Wu, Y. Dong, J. Zhao, D. Ma, W. Huang, Y. Zhang, Y. Wang, X. Jiang, Y. Xiang, J. Li, *Adv. Mater.* **2019**, 31, 1807981; c)Y. Shan, L. Wu, Y. Liao, J. Tang, X. Dai, Y. Xiang, *Journal of Materials Chemistry C* **2019**, 7, 3811.
[22] Y. Liao, Y. Shan, L. Wu, Y. Xiang, X. Dai, *Advanced Optical Materials* **2020**, 8, 1901862.
[23] K. Sk, B. Das, N. Chakraborty, M. Samanta, S. Bera, A. Bera, D. S. Roy, S. K. Pradhan, K. K. Chattopadhyay, M. Mondal, *Advanced Optical Materials* **2022**, 10, 2200791.
[24] L. Wu, X. Jiang, J. Zhao, W. Liang, Z. Li, W. Huang, Z. Lin, Y. Wang, F. Zhang, S. Lu, *Laser & Photonics Reviews* **2018**, 12, 1870055.
[25] L. Wu, W. Huang, Y. Wang, J. Zhao, D. Ma, Y. Xiang, J. Li, J. S. Ponraj, S. C. Dhanabalan, H. Zhang, *Advanced Functional Materials* **2019**, 29, 1806346.
[26] L. Lu, X. Tang, R. Cao, L. Wu, Z. Li, G. Jing, B. Dong, S. Lu, Y. Li, Y. Xiang, *Advanced Optical Materials* **2017**, 5, 1700301.

[27] B. Shi, L. Miao, Q. Wang, J. Du, P. Tang, J. Liu, C. Zhao, S. Wen, *Applied Physics Letters* **2015**, 107.
[28] Y. Shan, J. Tang, L. Wu, S. Lu, X. Dai, Y. Xiang, *Journal of Alloys and Compounds* **2019**, 771, 900.
[29] L. Hu, F. Sun, H. Zhao, J. Zhao, *Optics Letters* **2019**, 44, 5214.
[30] Y. Jia, Y. Liao, L. Wu, Y. Shan, X. Dai, H. Cai, Y. Xiang, D. Fan, *Nanoscale* **2019**, 11, 4515.
[31] X. Li, R. Liu, H. Xie, Y. Zhang, B. Lyu, P. Wang, J. Wang, Q. Fan, Y. Ma, S. Tao, *Optics Express* **2017**, 25, 18346.
[32] Y. Liao, Q. Ma, Y. Shan, J. Liang, X. Dai, Y. Xiang, *Journal of Alloys and Compounds* **2019**, 806, 999.
[33] Y. Jia, Z. Li, M. Saeed, J. Tang, H. Cai, Y. Xiang, *Optics Express* **2019**, 27, 20857.
[34] C. Song, Y. Liao, Y. Xiang, X. Dai, *Science Bulletin* **2020**, 65, 1030.
[35] L. Wu, Z. Xie, L. Lu, J. Zhao, Y. Wang, X. Jiang, Y. Ge, F. Zhang, S. Lu, Z. Guo, *Advanced Optical Materials* **2018**, 6, 1700985.
[36] Y. Shan, Z. Li, B. Ruan, J. Zhu, Y. Xiang, X. Dai, *Nanophotonics* **2019**, 8, 2225.
[37] W. Wang, Y. Wu, Q. Wu, J. Hua, J. Zhao, *Scientific reports* **2016**, 6, 22072.
[38] Y. Huang, H. Zhao, Z. Li, L. Hu, Y. Wu, F. Sun, S. Meng, J. Zhao, *Advanced Materials* **2023**, 35, 2208362.
[39] S. Goswami, C. C. de Oliveira, B. Ipaves, P. L. Mahapatra, V. Pal, S. Sarkar, P. A. S. Autreto, S. K. Ray, C. S. Tiwary, *Laser & Photonics Reviews*, n/a, 2400999.
[40] X. Xu, Z. Cui, Y. Yang, Y. Zhang, Q. Li, L. Tong, J. Li, X. Zhang, Y. Wu, *Laser & Photonics Reviews* **2025**, 19, 2401521.
[41] Z. Xu, H. Wang, W. Niu, Y. Guo, X. Zhai, P. Li, X. Zeng, S. Gull, J. Liu, J. Cao, X. Xu, G. Wen, G. Long, Y. Wu, J. Li, *Laser Photonics Rev.*, n/a, 2401929.
[42] D. Weng, C. Ling, Y. Gao, G. Rui, L. Fan, Q. Cui, C. Xu, B. Gu, *Laser Photonics Rev.*, n/a, 2401587.
[43] S. Bera, S. Kalimuddin, A. Bera, D. S. Roy, T. Debnath, S. Das, M. Mondal, *Adv. Opt. Mater.* **2025**, 13, 2402318.
[44] Y. Wu, L. Zhu, Q. Wu, F. Sun, J. Wei, Y. Tian, W. Wang, X. Bai, X. Zuo, J. Zhao, *Appl. Phys. Lett.* **2016**, 108.
[45] X. Xu, Z. Cui, Y. Yang, Y. Zhang, Q. Li, L. Tong, J. Li, X. Zhang, Y. Wu, *Laser Photonics Rev.* **2025**, 19, 2401521.
[46] J. Ridout, M. R. Probert, *CrystEngComm* **2014**, 16, 7397.
[47] G. Müller, M. Lutz, S. Harder, *Structural Science* **1996**, 52, 1014.
[48] N. Mounet, M. Gibertini, P. Schwaller, D. Campi, A. Merkys, A. Marrazzo, T. Sohier, I. E. Castelli, A. Cepellotti, G. Pizzi, N. Marzari, *Nat. Nanotechnol.* **2018**, 13, 246.
[49] W. Kohn, L. J. Sham, *Physical review* **1965**, 140, A1133.
[50] a)J. M. Soler, E. Artacho, J. D. Gale, A. García, J. Junquera, P. Ordejón, D. Sánchez-Portal, *Journal of Physics: Condensed Matter* **2002**, 14, 2745; b)A. García, N. Papior, A. Akhtar, E. Artacho, V. Blum, E. Bosoni, P. Brandimarte, M. Brandbyge, J. I. Cerdá, F. Corsetti, *The Journal of chemical physics* **2020**, 152.
[51] J. Oroya, A. Martín, M. Callejo, M. García-Mota, F. Marchesin, *SIMUNE Atomistics* **2014**.
[52] J. P. Perdew, K. Burke, M. Ernzerhof, *Physical review letters* **1996**, 77, 3865.
[53] M. Dion, H. Rydberg, E. Schröder, D. C. Langreth, B. I. Lundqvist, *Physical review letters* **2004**, 92, 246401.
[54] J. Klimeš, D. R. Bowler, A. Michaelides, *Journal of Physics: Condensed Matter* **2009**, 22, 022201.
[55] W. Tang, E. Sanville, G. Henkelman, *Journal of Physics: Condensed Matter* **2009**, 21, 084204.

[56] H. J. Monkhorst, J. D. Pack, *Physical review B* **1976**, 13, 5188.
[57] a)C. D. Sherrill, *School of chemistry and biochemistry, Georgia Institute of Technology* **2010**, 130; b)S. Slathia, B. Ipaves, C. Campos de Oliveira, S. D. Negedu, S. Sarkar, P. A. Autreto, C. S. Tiwary, *Langmuir* **2024**, 40, 15731.
[58] a)A. Chakraborty, B. Ipaves, C. Campos de Oliveira, S. D. Negedu, S. Sarkar, B. Lahiri, P. A. Autreto, C. S. Tiwary, *ACS Applied Engineering Materials* **2024**, 2, 1935; b)J. Gomez Quispe, B. Ipaves, D. S. Galvao, P. A. d. S. Autreto, *ACS Omega* **2024**, 9, 39195.
[59] G. Wang, S. Zhang, F. A. Umran, X. Cheng, N. Dong, D. Coghlan, Y. Cheng, L. Zhang, W. J. Blau, J. Wang, *Appl. Phys. Lett.* **2014**, 104.
[60] J. Li, Z. Zhang, J. Yi, L. Miao, J. Huang, J. Zhang, Y. He, B. Huang, C. Zhao, Y. Zou, *Nanophotonics* **2020**, 9, 2415.
[61] G. Wang, S. Zhang, X. Zhang, L. Zhang, Y. Cheng, D. Fox, H. Zhang, J. N. Coleman, W. J. Blau, J. Wang, *Photonics Research* **2015**, 3, A51.
[62] J. Wang, Y. Hernandez, M. Lotya, J. N. Coleman, W. J. Blau, *Advanced Materials* **2009**, 21, 2430.
[63] L. Wu, X. Yuan, D. Ma, Y. Zhang, W. Huang, Y. Ge, Y. Song, Y. Xiang, J. Li, H. Zhang, *Small* **2020**, 16, 2002252.
[64] X. Sun, K. Zheng, M. Cai, J. Bao, X. Chen, *Applied Surface Science* **2019**, 491, 690.
[65] a)L. Cheng, Y. Liu, *Journal of the American Chemical Society* **2018**, 140, 17895; b)M. Orlita, C. Faugeras, P. Plochocka, P. Neugebauer, G. Martinez, D. K. Maude, A.-L. Barra, M. Sprinkle, C. Berger, W. A. de Heer, *Physical review letters* **2008**, 101, 267601.
[66] a)G. Long, D. Maryenko, J. Shen, S. Xu, J. Hou, Z. Wu, W. K. Wong, T. Han, J. Lin, Y. Cai, *Nano Letters* **2016**, 16, 7768; b)S.-L. Li, K. Tsukagoshi, E. Orgiu, P. Samorì, *Chemical Society Reviews* **2016**, 45, 118.
[67] S. Yu, H. D. Xiong, K. Eshun, H. Yuan, Q. Li, *Applied Surface Science* **2015**, 325, 27.
[68] A. Allain, A. Kis, *ACS nano* **2014**, 8, 7180.
[69] S. Kumar, U. Schwingenschlogl, *Chemistry of Materials* **2015**, 27, 1278.
[70] D. Ovchinnikov, A. Allain, Y.-S. Huang, D. Dumcenco, A. Kis, *ACS nano* **2014**, 8, 8174.
[71] R. Maiti, M. A. S. R. Saadi, R. Amin, V. O. Ozcelik, B. Uluutku, C. Patil, C. Suer, S. Solares, V. J. Sorger, *ACS Applied Electronic Materials* **2021**, 3, 3781.
[72] F. Dabby, T. Gustafson, J. Whinnery, Y. Kohanzadeh, P. Kelley, *Applied Physics Letters* **1970**, 16, 362.
[73] Y. Wang, Y. Tang, P. Cheng, X. Zhou, Z. Zhu, Z. Liu, D. Liu, Z. Wang, J. Bao, *Nanoscale* **2017**, 9, 3547.
[74] S. Goswami, C. C. de Oliveira, B. Ipaves, P. L. Mahapatra, V. Pal, S. Sarkar, P. A. Autreto, S. K. Ray, C. S. Tiwary, *Laser Photonics Rev.* **2025**, 2400999.